\documentclass[lonbibliography, reprint, aps, prd, superscriptaddress, nofootinbib, floatfix]{revtex4-2}  
\usepackage{amsmath, amssymb, mathtools,  microtype}   \allowdisplaybreaks    
\usepackage[dvipsnames]{xcolor}        
\usepackage{bm}                   
\usepackage{placeins}         
\usepackage{orcidlink}            
\usepackage{graphicx}          
\usepackage{hyperref}       
\graphicspath{{Figures/}}      
\usepackage{silence}
\allowdisplaybreaks  
\hypersetup{colorlinks=true, urlcolor=teal,  citecolor=teal, linkcolor=Blue}  
\newcommand{\mH}{\mathcal{H}}   
\newcommand{\Eq}[1]{Eq.~\eqref{#1}}    
\newcommand{\ogw} {\Omega_{\mathrm{GW}}}

\begin{document}

\title{Gauge Dependence of Scalar-Induced Gravitational Waves from \\ Isocurvature Perturbations: Analytical Results}   

\author{Arshad Ali\,\orcidlink{0000-0002-8744-2420}} 
\email{arshadali@suda.edu.cn}
\affiliation{Institute for Advanced Study \& School of Physical Science and Technology, Soochow University, Shizi Street 1, Suzhou 215006, P.R. China}  
\author{Yang Lei\,\orcidlink{0000-0002-8551-2608}}
\email{leiyang@suda.edu.cn}
\affiliation{Institute for Advanced Study \& School of Physical Science and Technology, Soochow University, Shizi Street 1, Suzhou 215006, P.R. China}  
\author{Mudassar Sabir\,\orcidlink{0000-0002-8551-2608}}  
\email{mudassar.sabir@uestc.edu.cn} 
\affiliation{School of Physics, University of Electronic Science and Technology of China, \\ 2006 Xiyuan Avenue, Chengdu, P.R. China}   

\begin{abstract}  
We analytically study the gauge dependence of scalar–induced gravitational waves (SIGWs) sourced by primordial isocurvature perturbations during radiation domination (RD), working across nine gauges. Through analytical integrations of the kernels supported by graphical comparison we identify a clear dichotomy. We find that in some gauges viz. the uniform-density (UD),   
total-matter (TM),    
uniform-curvature (UC),    
comoving-orthogonal (CO) and  
transverse-traceless (TT) 
gauges the energy density grows polynomially in conformal time $\eta^n$, where $n$ varies from $2$ to  $8$.           
While in rest of the gauges viz. the longitudinal (Long.), uniform-expansion  (UE), Newtonian-motion (Nm), and N-body  (Nb) gauges the late-time energy spectrum converges, and SIGWs behave as radiation. For subhorizon modes ($ k\eta \gg 1 $), the divergence becomes severe, showing that SIGWs are gauge-dependent observables in this regime. We resolve it through a kernel projection that isolates the luminal, freely propagating gravitational wave components (oscillating as $\sin(k\eta)$ and $\cos(k\eta)$), eliminating spurious contributions. The resulting kernel decays as $ (k\eta)^{-1} $ and yields a finite, gauge-independent late-time spectrum, confirming that only luminal modes represent physical SIGWs. 
\end{abstract} 

\maketitle   
\newpage

\tableofcontents   

\section{Introduction}\label{sec:Intro} 

In the standard cosmological framework, the origin of cosmic structure is attributed to tiny primordial perturbations generated in the early Universe. The characterization of primordial perturbations is fundamental to understanding the initial seeds of cosmic structure. On the largest cosmological scales, observations indicate that the primordial fluctuations were predominantly adiabatic \cite{WMAP:2003ivt, Akrami:2018odb, Planck:2018vyg}. In \emph{adiabatic} initial conditions, a specific spacetime slicing exists where the energy-density perturbation of each cosmic component vanishes. In this frame, the primordial fluctuations reside entirely in the geometry of spacetime (curvature perturbations) rather than in the relative number densities of different species. In contrast, \emph{isocurvature} perturbations correspond to spatial variations in the relative number densities between species (e.g., photons vs. cold dark matter), while the total energy density initially remains unperturbed \cite{Kodama:1984ziu, Bucher:1999re}. Adiabatic and isocurvature modes evolve differently as the Universe expands, leaving distinct imprints on the CMB. Current observations tightly constrain the isocurvature contribution, requiring it to be subdominant to the adiabatic mode. 

On large scales (approximately $10^{-3} \, \text{Mpc}^{-1} \lesssim k \lesssim 10^{-1} \, \text{Mpc}^{-1}$), the allowed isocurvature fraction is less than about 1--10\% \cite{Akrami:2018odb}. Given the measured amplitude of the adiabatic power spectrum of $\sim 10^{-9}$, this translates to an upper limit on the isocurvature power of $\lesssim 10^{-10}$ on these scales. The situation changes on sub–Mpc scales, where CMB bounds no longer apply. For $1\,{\rm Mpc}\!\gtrsim\!\lambda\!\gtrsim\!1\,{\rm pc}$, future CMB spectral–distortion measurements may probe isocurvature fluctuations \cite{Chluba:2013dna, Chluba:2019kpb}. 

In the literature, scalar–induced tensor perturbations are also called second order scalar induced gravitational waves (SIGWs) or secondary GWs. Primordial Black Hole (PBHs) may form from the collapse of large primordial \emph{adiabatic} fluctuations and have been discussed in several contexts  \cite{Carr:2020xqk, Carr:2020gox,  Sasaki:2016jop, Bird:2016dcv, Yang:2024ntt},  while induced GWs  arise from the nonlinear coupling of density  (scalar) perturbations to tensor modes  \cite{Matarrese:1992rp,  Matarrese:1993zf, Ananda:2006af,  Baumann:2007zm, Saito:2009jt}. Recent works have highlighted the phenomenology of induced GWs in a variety of settings \cite{Espinosa:2018eve, Inomata:2019yww, Cai:2021jbi, Kohri:2018awv}. Besides, \emph{isocurvature} fluctuations can also produce PBHs \cite{He:2024luf, Papanikolaou:2024kjb, Ruan:2018tsw, Luo:2019zal, TaijiScientific:2021qgx,  Papanikolaou:2022chm, Papanikolaou:2020qtd}, where non-Gaussianity plays a crucial role, leading to characteristic double-peak spectra with a low-frequency component scaling as $\tau_{\rm NL}^2$. 

On the observational side, scalar-induced gravitational waves constitute an important target for future gravitational-wave experiments. Currently, space-based interferometers such as LISA, TianQin, and Taiji are sensitive in the mHz band ~\cite{Danzmann:1997hm, LISA:2017pwj, Hu:2017mde, Luo:2025ewp}, while pulsar-timing arrays (EPTA, NANOGrav, PPTA) and the SKA probe the nHz frequency range \cite{Kramer:2013kea, Hobbs:2009yy, McLaughlin:2013ira, Hobbs:2013aka, Moore:2014lga}. However, the cosmological GWs, sourced after horizon reentry during the radiation-dominated era, predominantly peak at ultra-low frequencies of order $\sim 10^{-18}\text{--}10^{-16} \mathrm{Hz}$ \cite{Seto:2005tq}, well below the direct sensitivity of existing and planned interferometric or pulsar-timing experiments. As a result, direct detection of such signals is currently not feasible. Nevertheless, their physical effects may be indirectly constrained through their imprints on cosmological observables, such as CMB $B$-mode polarization, motivating continued theoretical investigations of SIGWs in this frequency regime. 

Unlike the first-order tensor perturbations, which correspond to linear GWs and are gauge invariant, the secondary GWs induced by first-order scalar perturbations are not gauge invariant because the nonlinear structure of Einstein's equations couples different perturbation modes \cite{Noh:2003yg}. Consequently, the predicted energy density spectrum of scalar-induced GWs can depend on the choice of gauge. This gauge dependence has been extensively investigated for curvature (adiabatic) perturbations, where the effects of gauge choice have been analyzed in detail under adiabatic initial conditions \cite{Hwang:2017oxa, Gong:2019mui, Tomikawa:2019tvi, DeLuca:2019ufz, Lu:2020diy, Domenech:2020xin, Ali:2020sfw, Ali:2023moi, Sabir:2023qyr, Lu:2020diy, Inomata:2019yww}. In contrast, the impact of gauge choice on induced GWs sourced by \emph{isocurvature} perturbations has not been systematically studied. A comprehensive analytic analysis of the gauge dependence of \emph{isocurvature}-induced GWs remains absent (cf. \cite{Domenech:2023jve} for related discussions). Addressing this gap is the main goal of the present work.

In this paper, we extend the study of gauge dependence to secondary GWs induced by primordial isocurvature perturbations. We present an analytic treatment during radiation domination across nine different gauges, namely, the longitudinal, comoving-orthogonal (CO), synchronous or transverse-traceless (TT), total-matter (TM), uniform-curvature (UC), uniform-density (UD), uniform-expansion (UE), Newtonian–motion (Nm), and N–body (Nb) gauges. For each gauge, we compute the kernel integrals in analytical form and systematically compare their behaviors. Specifically, we derive the corresponding energy density spectra of secondary GWs and organize the convolution integrals using the $(d, s)$ variables introduced in \cite{Espinosa:2018eve}. 

We find that the energy density of induced GWs exhibits a polynomial growth with conformal time, $\Omega_{\mathrm{GW}} \propto \eta^n$, where the exponent $n$ depends on the gauge choice. In particular, we obtain the following scaling behavior: in the uniform-density (UD) gauge, $\Omega_{\mathrm{GW}} \propto \eta^2$; in the total-matter (TM) and uniform-curvature (UC) gauges, $\Omega_{\mathrm{GW}} \propto \eta^4$; in the comoving-orthogonal (CO) gauge, $\Omega_{\mathrm{GW}} \propto \eta^6$; and in the synchronous or transverse-traceless (TT) gauge, $\Omega_{\mathrm{GW}} \propto \eta^8$. In contrast, in the longitudinal, uniform-expansion (UE), Newtonian–motion (Nm), and N–body (Nb) gauges, the late-time energy density spectra converge, and the secondary GWs behave as radiation. For subhorizon modes ($k\eta \gg 1$), however, the divergence of the induced GW amplitude becomes increasingly severe, clearly demonstrating that scalar-induced GWs are gauge-dependent observables in this regime.

The gauge dependence of secondary GWs induced by isocurvature perturbations exhibits stronger divergences than in the adiabatic case studied in Ref.~\cite{Ali:2023moi, Ali:2020sfw, Lu:2020diy}. Consequently, gauge choices that yield well-behaved and convergent results for adiabatic perturbations such as the uniform-curvature gauge can lead to unphysical divergences when applied to isocurvature sources. In particular, the TT and CO gauges display especially severe divergences in the case of isocurvature perturbations. It is important to note that such behavior may also signal a breakdown of perturbation theory due to the presence of growing modes in certain gauges. The enhanced gauge sensitivity originates from the differing evolution of isocurvature scalar modes, which amplifies the GW spectra and can cause divergence.

Resolving this strong gauge dependence is essential for developing a consistent theoretical framework for higher-order GWs and for ensuring reliable observational predictions. We show that the apparent gauge dependence of secondary GWs from isocurvature perturbations arises mainly from unphysical, non-radiative tensor modes that contaminate the metric perturbations in certain gauges. Within our nine-gauge analysis, we isolate the physical tensor component by retaining only the freely propagating oscillatory terms, ${\sin(k\eta)}$ and ${\cos(k\eta)}$, which represent genuine gravitational radiation. This \emph{radiative projection} eliminates gauge artifacts and yields a gauge-independent late-time spectrum for the physically observable induced GWs. Our results clarify the origin of gauge dependence in scalar-induced GWs and establish a consistent framework for higher-order gravitational radiation from both adiabatic and isocurvature sources, providing a foundation for connecting theoretical predictions with future observations in the mHz–Hz range.

The structure of this paper is as follows. In Sec.~\ref{sec.2}, we review the basic formalism for calculating secondary (scalar-induced) GWs and discuss the relevant gauge transformations. We employ the \emph{Mathematica} package \emph{xPand}~\cite{Pitrou:2013hga} to derive several key relations. We also provide a general prescription to obtain the results in arbitrary gauges from the longitudinal-gauge expressions by applying the gauge transformation of the second-order tensor perturbation. 
In Sec.~\ref{sec3}, we apply this prescription to derive the kernels in nine different gauges, namely, the longitudinal, CO, synchronous (TT), TM, UC, UD, UE, Nm, and Nb gauges. For the CO, UE, and Nm gauges, the kernels are explicitly obtained through coordinate (gauge) transformations. We then analyze the late-time behavior of secondary GWs in all these gauges. Interestingly, we find notable gauge-dependent variations in both the kernels and the resulting spectra: while the late-time kernels remain finite in the longitudinal, UE, Nm, and Nb gauges, five other gauges exhibit growing modes. 
In Sec.~\ref{sec:gauge-independent}, we compare the energy density spectra obtained in different gauges and present a simple resolution of the divergences that arise in some of them, consistent with the physical requirement that the GW energy density be gauge invariant. Finally, our discussion and conclusions are given in Sec.~\ref{summary}.  

\section{Basics of Secondary GWs}\label{sec.2} 
In this pivotal section, we provide a comprehensive analysis of the fundamental formulas essential for calculating the kernel functions that describe the energy density of secondary gravitational waves\footnote{ Also referred to as second-order GWs, scalar-induced secondary GWs, or simply scalar-induced GWs, as discussed in \cite{Inomata:2019yww}.}. These waves are particularly significant in modern astrophysical studies, as they are generated by primordial isocurvature perturbations. This research domain sheds light on the complex dynamics of the early Universe, where second-order tensor perturbations arise from quadratic combinations of linear scalar perturbations.  
To tackle this intricate phenomenon, it is crucial to develop the general formula for secondary GWs across various gauges.

To accurately encapsulate secondary GWs within the stochastic GW background, we model the perturbed metric around the Friedmann-Lemaître-Robertson-Walker (FLRW) framework, expressed by the following equations:
\begin{align}
g_{00} &= -a^2(\eta)(1 + 2 \phi(\eta, x)),\nonumber \\
g_{0i} &= 2a^2(\eta)\partial_i B(\eta, x) , \nonumber\\
g_{ij} &= a^2(\eta)\delta_{ij} + a^2(\eta)\Big(\frac{1}{2} h_{ij}^{\text{TT}}(\eta, x) - 2\delta_{ij}\psi(\eta, x) \nonumber\\
&\quad + 2\partial_i \partial_j E(\eta, x)\Big)\label{rdmetric},
\end{align}
where $a(\eta)$ denotes the scale factor of the Universe. The scalar perturbations $\phi$, $\psi$, $B$, and $E$ are of first order. At the same time, the traceless transverse component $h_{ij}^{\text{TT}}$ represents the significant second-order tensor mode essential to calculate scalar-induced tensor perturbations. This component satisfies conditions $h^{\text{TT}}_{ii} = 0$ and $\partial_i h^{\text{TT}}_{ij} = 0$. 

To investigate secondary GWs, we apply the spacetime \eqref{rdmetric} to the general Einstein equations:
\begin{equation}
G_{\mu\nu} = T_{\mu\nu},\label{Einsteq}
\end{equation}
where $G_{\mu\nu}$ represents the Einstein tensor. For simplicity, we use reduced Planck units, setting $M_\text{pl} = (8\pi G)^{-1/2} = 1$. In the subsequent subsections, we will perturb the equation \eqref{Einsteq} into scalar and tensor parts. 

\subsection{Metric Perturbations and Scalar Modes}
To develop a general framework for isocurvature-induced GWs, we consider that isocurvature fluctuations are due to differences in relative number densities. We hypothesize the existence of at least two types of fluid in the primordial Universe. For simplicity, we assume that after cosmic inflation, the Universe is predominantly composed of relativistic particles (radiation), with a small portion of non-relativistic particles (matter). The energy-momentum tensors for radiation and matter are expressed as follows \cite{Domenech:2021and,   Domenech:2023jve}:   
\begin{align}
T^r_{\mu \nu}       & =\left(\rho^{\text{r}}+P^{\text{r}}\right) U^{\text{r}}_{\mu} U^{\text{r}}_{\nu}+P^{\text{r}} g_{\mu \nu}, \nonumber \\ 
\quad T^m_{\mu \nu} & =\rho^{\text{m}} U^{\text{m}}_{ \mu} U^{\text{m}}_{ \nu},      
\end{align} 
where the subscripts  $r$  and $m$ correspond to the radiation and matter components, respectively, and it is assumed that the background anisotropic stress ${\Pi}^0_{\mu\nu}$ is absent.  First-order perturbations in velocity $U_\mu$, energy density, pressure, and anisotropic stress are indicated by $\delta U_\mu$, $\delta\rho$, $\delta P$, and $\delta\Pi_{ij}$, respectively. The first-order four-velocity perturbation $\delta U_\mu$ is decomposed according to $\delta U_\mu=a(\eta)(\delta V_0,\delta V_{,i}+\delta V_i)$ with $\delta V_{i,i}=0$. 
During the radiation domination (RD) phase, for energy density $\rho$ and pressure $P$, we utilize $P^{\text{r}} =(1/3)\rho^{\text{r}}$. 
In our subsequent analysis, we adopt the notations $\delta \rho$ and $\delta P$ to signify perturbations in energy density and pressure, respectively.

\subsubsection{Metric perturbations}\label{metricpert}   
We present first-order metric perturbations and the relation between the conformal Hubble rate and the background matter and radiation energy densities. We derive $a(\eta)$ and note that first-order energy–momentum conservation governs the evolution of densities, velocities, and gravitational potentials. At leading order, the Einstein equation and energy conservation yield:
\begin{equation}
\begin{aligned}
&3\mH^2=8\pi a^2 (\rho_{0m}+\rho_{0r}), \\
& \mH^2+2\mH'=-\frac{8\pi}{3} a^2 \rho_{0r},\\
&\rho_{0m}'+3\mH \rho_{0m}=0, \\
&\rho_{0r}'+4\mH\rho_{0r} =0 .
\end{aligned}\label{eq4}
\end{equation}

The scale factor can be determined from the solution
\begin{equation}
{a(\eta)\over a_{\mathrm{eq}}}=2\left(\eta \over \eta_*\right) + \left(\eta \over \eta_*\right)^2,\label{eq5}
\end{equation}
where $\eta_*=\frac{\eta_{\mathrm{eq}}}{\sqrt{2}-1}$.

Here, in Eqs.~\eqref{eq4} and \eqref{eq5}, $\mathcal{H} \equiv a'/a$ is the conformal Hubble rate, where a prime ($'$) denotes a derivative with respect to conformal time $\eta$, and the subscript `eq' indicates evaluation at matter–radiation equality.
Based on the equations mentioned above, we can derive a solution so that
\begin{align}
\rho_m (\eta)&={1
\over 2}\rho_{\mathrm{eq}}\left(a \over a_{\mathrm{eq}}\right)^{-3},\nonumber\\ 
\rho_r (\eta)&= {1
\over 2}\rho_{\mathrm{eq}}\left(a \over a_{\mathrm{eq}}\right)^{-4}. 
\end{align}

In the matter-radiation equality, the total energy density is expressed as $\rho_{\mathrm{eq}}=\rho_m(\eta_{\mathrm{eq}})+\rho_r(\eta_{\mathrm{eq}})$. Energy conservation at first order yields  
\begin{equation} 
\begin{aligned}
&\delta\rho_m'+3\mH\delta\rho_m+\rho_m\left( -3\psi' + \nabla^2 E'+\nabla^2 V_m \right)=0,\\
&\delta\rho_r'+4\mH\delta\rho_r+{4\over 3}\rho_r\left( -3\psi' + \nabla^2 E'+\nabla^2 V_r \right)=0,\\
&V_m'+\mH V_m+(\phi+\mH B+B') = 0, \\
&V_r'+{1\over 4}\frac{\delta\rho_r}{\rho_r} +(\phi+B') = 0.
\label{eq:background01}
\end{aligned}
\end{equation}

The Einstein equations with first-order perturbations are expressed as
\begin{equation}
\begin{aligned}
&3\mH^2\phi+\mH(3\psi'+\nabla^2 \sigma)-\nabla^2 \psi = -4\pi a^2 (\delta\rho_m+\delta\rho_r),\\
&\mH^2 B-\mH'B+\mH\phi+\psi'=-4\pi a^2\left(\rho_{0m} V_m +{4\over 3}\rho_{0r} V_r\right),\\
&\psi'' + 3\mathcal{H}\psi' + \left(2\mathcal{H}' + 3\mathcal{H}^2\right)\phi - \frac{1}{2}\nabla^2 \phi = 4\pi a^2 (\delta P_m+\delta P_r),\\  
&\mathcal{H}^2\left(\phi + \frac{1}{2}\nabla^2 B\right) + \mathcal{H}\left(\psi' + \nabla^2 \sigma\right) = -4\pi a^2\left(\delta\rho_m + \frac{4}{3}\delta\rho_r\right).
\label{eq:background02} 
\end{aligned}
\end{equation} 
These equations relate the metric potentials $\phi$, $\psi$ and the shift/shear ($B,\sigma$) to the matter and radiation perturbations during RD. Radiation pressure, with $\delta  P_r=\tfrac13\,\delta\rho_r$, supports with sound speed $c_s^2=1/3$, while cold matter is pressureless and follows the flow. Adiabatic initial conditions tie all species to a common time slicing so that curvature tracks the total density; isocurvature instead keeps the total density unperturbed while relative number-density contrasts source the potentials via  pressure/velocity terms. A gauge choice fixes the slicing/threading and redistributes non-propagating pieces among $(\phi,\psi,B,E)$, a point that will be important for the apparent late-time growth of the quadratic tensor source. 
\subsubsection{Scalar modes}\label{scalr} 
Under an infinitesimal coordinate transformation $x^\mu\rightarrow \tilde{x}^\mu=x^\mu+\epsilon^\mu(x)$ where $\epsilon^\mu=(\alpha,\partial^i\beta)$ \cite{Malik:2008im}, the scalar components of the perturbations undergo transformation as  
\begin{equation} 
\begin{aligned}
\tilde{\phi}         & = \mathcal{H}\alpha +\alpha^\prime +\phi, \\
\tilde{\psi}         & = -\mathcal{H}\alpha +\psi,               \\
\tilde{B}            & = -\alpha +\beta^\prime +B,               \\
\tilde{E}            & = \beta +E,                               \\
\tilde{\sigma}       & = \alpha +\sigma,                         \\
\delta\tilde{\rho}\,^{\text{r}} & = {\rho}_0^{\text{r}\prime}\alpha +\delta\rho^{\text{r}},    \\
\delta\tilde{\rho}^{\text{m}} & = {\rho}_0^{\text{m}\prime}\alpha +\delta\rho^{\text{m}},    \\
\delta\tilde{P}^\text{r}      & = P_0^{\prime\,r}\alpha +\delta P^r,             \\
\delta\tilde{V}^r      & = -\alpha +\delta V^r, \\
\delta\tilde{V}^m      & = -\alpha  +\delta V^m,  
\end{aligned}\label{gauge:etransf}
\end{equation} 
where quantities in tilde are transformed quantities, a prime denotes the derivative with respect to conformal time,  $\mH \equiv a'/a$ is identified as the conformal Hubble parameter and $\sigma=E'-B$ is the shear potential. Using the aforementioned gauge transformation, we derive two gauge-invariant Bardeen potentials  \cite{Bardeen:1980kt,Ali:2020sfw},
\begin{align}
\label{bardeenphi}
\Phi &=-\mathcal{H}\sigma+\phi-\sigma^\prime,\\
\label{bardeenpsi}
\Psi &= \mathcal{H}\sigma+\psi.
\end{align}  
By selecting specific $\alpha$ and $\beta$, it is possible to eliminate two degrees of freedom from among the four scalar modes. Furthermore, the first-order $ij$ component of the Einstein equation removes an additional degree of freedom, thereby leaving only one scalar mode, governed by the equations\footnote{When anisotropic stress is absent, in Longitudinal gauge, the potentials satisfy $\Psi = \Phi$. Pure isocurvature perturbations correspond to an initially uniform total energy density, meaning $\delta \rho = \delta \rho_m + \delta \rho_r = 0$. Additionally, the time derivative of the total density contrast must also vanish initially \cite{Kodama:1986ud}. Under these conditions, the Einstein equations imply an unperturbed metric at the initial time, leading to $\Phi = 0$.}: 
\begin{gather}
\psi''+\mH\left( \phi'+(2+3c_s^2)\psi'\right)+\left( (1+3c_s^2)\mH^2 +2\mH' \right) \phi \nonumber\\
- c_s^2\left(\mH \nabla^2\sigma+\nabla^2\psi\right)= 4\pi a^2 \, \tau \, \delta s, \nonumber\\
\psi - \phi + \sigma' + 2\mH\sigma = 0.\label{eom1}
\end{gather}
The first line represents the equation of motion for the scalar modes, where $c_s^2$ and $\delta s$ are the sound speed and entropy perturbation, respectively, arising from $\delta P = c_s^2 \delta \rho +\tau \delta s$, which takes the form 
\begin{align}
c_s^2                       & = {1\over 3}\left(1+{3\over 4}{\rho^{\text{m}}\over \rho^{\text{r}}}\right)^{-1}, \qquad \\  
\tau                        & ={c_s^2 \rho^{\text{m}} \over s}, \qquad                                        \\
S& \equiv {3\over 4}{\delta\rho^{\text{r}}\over \rho^{\text{r}}}-{\delta\rho^{\text{m}}\over \rho^{\text{m}}}.  
\end{align} 
Here, the isocurvature perturbations $S$ is a gauge-invariant quantity \cite{Domenech:2021and,  Domenech:2021ztg, Domenech:2025ffb, Domenech:2023jve, Yuan:2024qfz}.  Moreover, by defining the relative velocity as $V\equiv V^{\text{m}}-V^{\text{r}}$, one can obtain the relation $ S' = \nabla^2  V$. Combining the Einstein equation and energy conservation, $\nabla_{\mu} T^{\mu\nu}=0$, up to first order, one obtains the equation of motion for the entropy as follows: 
\begin{equation}\label{eom2}
S''+3\mH c_s^2 S'+{3\rho^{\text{m}} \over 4\rho^{\text{r}}}c_s^2 k^2 S - {3\over 16\pi a^2 \rho^{\text{r}}}c_s^2 k^4 (\mH \sigma+\psi) =0. 
\end{equation} 

On scales larger than $10\,\mathrm{Mpc}$, the Cosmic Microwave Background (CMB) informs us that primordial isocurvature fluctuations can contribute between 1-10{\%} to the total fluctuations \cite{Akrami:2018odb}. Since the observed amplitude of the power spectrum of primordial adiabatic  fluctuations is approximately $10^{-9}$, the corresponding power spectrum of isocurvature fluctuations might be smaller than $10^{-10}$ on larger scales. However, the situation changes at scales smaller than $1\,\mathrm{Mpc}$, where CMB constraints are not applicable. For scales between $1\,\mathrm{Mpc}$ and $1\,\mathrm{pc}$, future CMB spectral distortions have the potential to probe isocurvature fluctuations \cite{Chluba:2013dna, Chluba:2019kpb}.
In the adiabatic case, the initial values of the scalar modes are not detailed here but are elaborated in \cite{Domenech:2021and}. Conversely, in the isocurvature scenario, metric perturbations originate from the entropy. We determine the initial value of $S$ in Fourier space using the relation:
\begin{align}
S = S_{\bm{k}} T_{S}(k\eta),   
\end{align} 
where $T_S$ represents the transfer function of entropy, describing its temporal evolution, and is normalized such that $T_S(0)=1$. The initial value $S_{\bm{k}}$ is connected to the dimensionless primordial entropy spectrum, and can be evaluated in a two-point correlator of the form as:
\begin{equation}
\left\langle S_{\bm{k}}S_{\bm{k}}' \right\rangle={2\pi^2\over k^3}\mathcal{P}_S(k) \delta^{(3)}(\bm{k}+\bm{k'}),
\end{equation}
where $\mathcal{P}_S(k)$ is the power spectrum that characterizes the statistical properties of isocurvature perturbations, and  $\delta^{(3)}$ denotes the three-dimensional Dirac delta function. 

For convenience, we introduce dimensionless parameters 
\begin{equation}
x = k\eta, \qquad \kappa = \frac{k}{k_{\text{eq}}}. 
\end{equation} 
The time coordinate $x$ distinguishes the superhorizon ($x \ll 1$) and subhorizon ($x \gg 1$) regimes of the scalar perturbation. Meanwhile, $\kappa \gg 1$ governs how deeply the mode enters the horizon during radiation domination. In this phase, $x/\kappa = k_{\text{eq}} \eta \ll 1$ always holds \cite{Domenech:2021and}, which simplifies the equations of motion together with \Eq{eom1} and \Eq{eom2}.
    
In various gauges \cite{Clifton:2020oqx}, we impose the following gauge conditions: 
\begin{equation}
\begin{aligned}
& \text{Longitudinal:} && B = 0, && E = 0, \\
& \text{Comoving Orthogonal:} && \delta V = 0, && B = 0, \\
& \text{Synchronous:} && \phi = 0, && B = 0, \\
& \text{Total Matter:} && \delta V = 0, && E = 0, \\
& \text{Uniform Curvature:} && \psi = 0, && E = 0, \\
& \text{Uniform Density:} && \delta \rho = 0, && E = 0, \\
& \text{Uniform Expansion:} && E = 0, && \nabla^2 \sigma = 3(\mathcal{H}\phi + \psi'), \\
& \text{Newtonian-motion:} && B = 0, && E'' = - \mathcal{H} E', \\
& \text{N-body:} && \psi = \tfrac{1}{3}\nabla^2 E, && \delta V = -B.  
\end{aligned}
\label{eq:gauge-conditions}
\end{equation} 
We fix the scalar part of the infinitesimal gauge transformation (sometimes called a scalar diffeomorphism), generated by $\epsilon^\mu=(\alpha,\partial^i\beta)$, by imposing the two conditions listed in Eq.~\eqref{eq:gauge-conditions}. In particular, $B=0$ removes the scalar shift ($g_{0i}=a^2\partial_i B=0$) and $E=0$ removes the scalar shear, while conditions such as $\delta V=0$, $\psi=0$, or $\delta\rho=0$ choose the time slicing (cf.\ Eq.~\eqref{gauge:etransf}; see \cite{Ma:1995ey,Malik:2008im}). In the N-body gauge, $\psi=\tfrac{1}{3}\nabla^2E$ and $\delta V=-B$ eliminate relativistic volume deformation and align cold-matter trajectories with Newtonian $N$-body evolution~\cite{Clifton:2020oqx}. The choices above then have the following physical interpretation~\cite{Malik:2008im, Clifton:2020oqx}:
\begin{itemize}
\item \textbf{Longitudinal (Long.):} $B=E=0$,
(no scalar shear, $\sigma$).  The remaining potentials are the Bardeen pair $(\phi,\psi)$.   
\item \textbf{Comoving Orthogonal (CO):} $\delta V=0$, $B=0$  comoving slicing  
with hypersurfaces orthogonal to the 4-velocity; momentum density vanishes.
\item \textbf{Synchronous (TT):} $\phi=0$, $B=0$  proper-time slicing with
vanishing shift (noting the usual residual freedom). 
\item \textbf{Total Matter (TM):} $\delta V=0$, $E=0$  comoving slicing with
shear-free threading.
\item \textbf{Uniform Curvature (UC):} $\psi=0$, $E=0$ flat spatial slices,
shear-free threading.
\item \textbf{Uniform Density (UD):} $\delta\rho=0$, $E=0$  constant-density slices,
shear-free threading.
\item \textbf{Uniform Expansion (UE):} $E=0$, $\nabla^2\sigma=3(\mathcal H\phi+\psi')$ shear-free threading and uniform perturbation of the local expansion (trace of the
extrinsic curvature).
\item \textbf{Newtonian-motion (Nm):} $B=0$, $E''=-\mathcal H E'$  coordinates chosen
so the relativistic Euler equation reduces to its Newtonian form; matter follows
Newtonian trajectories.
\item \textbf{N–body (Nb):} $\psi=\tfrac13\nabla^2E$, $\delta V=-B$. The conditions remove volume
deformation and align particle trajectories and continuity/Euler equations with those
used in Newtonian $N$-body simulations.
\end{itemize}
These gauge choices fix the scalar diffeomorphisms $(\alpha,\beta)$ via 
Eqs. \eqref{eq:background01}--\eqref{gauge:etransf}; we apply them to the transformed scalar variables  
when constructing the sources and kernels for $\Omega_\text{GW}$ in the
sections that follow. For clarity, the velocity potential used in the gauge conditions is the
momentum–weighted total scalar velocity,
\begin{equation}
(\rho_\text{tot}+P_\text{tot})\,\delta V \;\equiv\; \rho_m\,V_m \;+\; \frac{4}{3}\,\rho_r\,V_r,
\end{equation}
so that $T^{0}{}_{i}\propto(\rho_\text{tot}+P_\text{tot})\,\partial_i\delta V$ vanishes when $\delta V=0$.

By “relativistic Euler equation’’ we mean the spatial component of energy–momentum
conservation, $\nabla_\mu T^{\mu}{}_{i}=0$. In our variables (cf.\ Eq.~\eqref{eq:background01}) its linear form reads
\begin{equation}
\begin{aligned}
V_m' + \mathcal{H} V_m + (\phi + \mathcal{H} B + B') &= 0,\\
V_r' + \dfrac{1}{4}\frac{\delta\rho_r}{\rho_r} + (\phi + B') &= 0.  
\end{aligned}
\end{equation}
 The Newtonian–motion gauge $B=0$ with $E''=-\mathcal{H}E'$ makes the matter equation take its Newtonian gauge, while the N–body choice $\psi=\tfrac{1}{3}\nabla^2E$ and $\delta V=-B$ removes relativistic volume deformation and aligns the continuity/Euler system with Newtonian $N$-body gauge \cite{Clifton:2020oqx}.

We then use the first-order background Einstein equations and introduce the transfer functions of the scalar modes:
\begin{equation}
T_Y(x) \equiv \frac{Y}{S_{\bm k}}, \qquad
Y \in \{\alpha,\beta,\sigma,\phi,\psi,B,E\}, 
\label{transferfun}
\end{equation}
which relate scalar perturbations to the isocurvature amplitude $S_{\bm k}$.\footnote{For brevity we also use $T_S \equiv S/S_{\bm k}$. No additional normalization is assumed for $T_Y$; only $T_S(0)=1$ was fixed earlier.}
This permits an expansion of the equations of motion to order $\mathcal{O}(\kappa^{-1})$, resulting in: 
\begin{widetext}
\begin{equation}\label{expansionequation}  
\begin{aligned}
& \text{Longitudinal (Long.):}        &&\quad
\begin{aligned} 
T_{\psi}^{**} + \frac{3}{x} T_{\psi}^* + \left( \frac{1}{x} + \frac{1}{4\sqrt{2}\kappa} \right) T_{\phi}^*  + \left( \frac{1}{3x} - \frac{1}{6\sqrt{2}\kappa} \right) T_{\sigma} \\
+ \frac{1}{4\sqrt{2}x\kappa}\, T_{\phi} + \left( \frac{1}{3} - \frac{x}{4\sqrt{2}\kappa} \right) T_{\psi} - \frac{1}{2\sqrt{2}x\kappa}\, T_S \simeq 0,
\end{aligned}  \\
& \text{Comoving Orthogonal (CO):} &&\quad T_{\phi} + \frac{x}{2} T_{\phi}^* - \frac{3}{2x\sqrt{2}\kappa}\, T_{\beta}^* \simeq 0,  \\
& \text{Synchronous (TT):}         &&\quad T_S^{**} + \left( \frac{1}{x} - \frac{1}{2\sqrt{2}\kappa} \right) T_S^*  + \frac{x}{4\sqrt{2}\kappa}\, T_S  - \frac{x}{6}\bigl(T_{\sigma} + x T_{\psi}\bigr) \simeq 0, \\
& \text{Total Matter (TM):}        &&\quad T_{B} + \frac{x}{6} T_{\phi} - \frac{x}{6\sqrt{2}\kappa}\, T_{\psi}^* \simeq 0, \\
& \text{Uniform Curvature (UC):}   &&\quad T_{\phi} - T_{\psi} - T_{\sigma}^* + \left( \frac{2}{x} - \frac{1}{2\sqrt{2}\kappa} \right) T_{\sigma} \simeq 0,  \\
& \text{Uniform Density (UD):}   && \quad T_{B} - 2x^2\!\left( \frac{1}{x^2} - \frac{1}{2\sqrt{2}\kappa} \right) T_{\phi} - 2x\, T_{\phi}^* \simeq 0,  \\
& \text{Uniform Expansion (UE):}  && \quad T_{\alpha} + \frac{3}{x} T_{\phi} + \frac{3}{2\sqrt{2}\kappa}\, T_{\phi}^* \simeq 0,\\
& \text{Newtonian-motion (Nm):}   &&\quad
T_{\beta}^{**}+\;\Bigl(\frac{1}{x}+\frac{1}{4\sqrt{2}\,\kappa}\Bigr)\,T_{\beta}^{*}\;\simeq 0 , \\
& \text{N-body (Nb):}   &&\quad 
T_B^{**}+\Bigl(\tfrac{2}{x}+\tfrac{1}{4\sqrt{2}\,\kappa}\Bigr)T_B^{*}
+\tfrac{1}{3}\,T_B\simeq 0. \\\; 
\end{aligned}  
\end{equation}
\end{widetext}
Here the superscript $^{*}$ on transfer functions denotes differentiation with respect to their argument $x$ (and $^{**}$ the second derivative).
The transfer functions encapsulate the dynamic behavior of scalar  perturbations during the radiation-dominated era, allowing us to gain insights into the relationship between initial scalar fluctuations and their evolution over time. They will be utilized in subsequent sections to evaluate the kernels of the energy-density spectra of secondary GWs. 
	
\subsection{General Formalism of Isocurvature Secondary GWs}
In this section, we begin by deriving the general formalism applicable to any gauge for isocurvature secondary GWs. Isocurvature fluctuations, characterized by variations in relative number density, require the consideration of at least one fluid present in the primordial Universe. For the sake of simplicity, it is assumed that after cosmic inflation, the Universe is predominantly filled with relativistic particles, commonly referred to as radiation\footnote{A small but non-zero fraction of non-relativistic particles, referred to here as `matter', will be explored in detail in our next project with a focus on general gauges during matter domination (MD) phase and RD-to-MD transition.}. 
Before delving into the computational aspects, an understanding of the primary sources of secondary gravitational waves in a general gauge is essential. 
For understanding the source term of secondary GWs, we study the spatial component of the second-order Einstein tensor, specifically the transverse, trace-free part. This is done by first applying the projection tensor\footnote{In the Fourier space, the projection tensor is expressed as $\mathcal{T}_{ij}^{lm}=[\mathbf e_{ij}^+ \mathbf e^{+lm}+\mathbf e_{ij}^\times \mathbf e^{\times lm}],$ see Refs. \cite{Ali:2020sfw}, for more detail.},  $\mathcal{T}_{ij}$, to the spatial part of the field equation \eqref{Einsteq}, leading to an expression that relates the spatial Einstein tensor to the energy-momentum tensor as follows:
\begin{align}
\label{theeq}                            
\mathcal{T}_{ij}\,^{lm}G_{lm}= \mathcal{T}_{ij}\,^{lm} T_{lm}, \qquad\\\nonumber \\
h_{ij}^{\mathrm{TT}\prime\prime}+2\mathcal{H}h_{ij}^{\mathrm{TT}\prime}-\nabla^2h_{ij}^{\mathrm{TT}}=4\mathcal{T}_{ij}^{lm}s_{lm}. 
\end{align}

For any gauge, after some simplifications, the source term $s_{ij}$ is given as follows:  
\begin{widetext}
\begin{align}
-s^{\mathrm{}}_{ij}=&\,{\partial_i}\psi{\partial_j}\psi
+{\partial_i}\phi{\partial_j}\phi
-{\partial_i}{\partial_j}\sigma\left(\phi^\prime+\psi^\prime-\nabla^2\sigma\right)
+\left({\partial_i}\psi^\prime\sigma{\partial_j}+{\partial_j}\psi^\prime{\partial_i}\sigma\right)
-{\partial_i}{\partial_k}\sigma{\partial_j}{\partial_k}\sigma
\nonumber\\
&+2{\partial_i}{\partial_j}\psi\left(\phi+\psi\right)
-8\pi Ga^2({\rho^r}+{P^r}){\partial_i}\delta  \mathcal{U}{\partial_j}\delta  \mathcal{U}
-2{\partial_i}{\partial_j}\psi\nabla^2E
-{\partial_i}{\partial_k}E^\prime {\partial_j}{\partial_k}E^\prime
\nonumber\\
&+2{\partial_i}{\partial_j}E\left(\psi^{\prime\prime}+2\mathcal{H}\psi^\prime-\nabla^2\psi\right)
+{\partial_i}{\partial_k}{\partial_l}E{\partial_j}{\partial_k}{\partial_l}E
+2\left({\partial_j}{\partial_k}\psi{\partial_i}{\partial_k}E
+{\partial_i}{\partial_k}\psi{\partial_j}{\partial_k}E\right)
\nonumber\\
&-2\mathcal{H}({\partial_i}\psi{\partial_j}E^\prime+{\partial_j}\psi{\partial_i}E^\prime)
-\left({\partial_i}\psi^\prime {\partial_j}E^\prime+{\partial_j}\psi^\prime {\partial_i}E^\prime\right)
-\left({\partial_i}\psi{\partial_j}E^{\prime\prime}
+{\partial_j}\psi{\partial_i}E^{\prime\prime}\right)
\nonumber\\
&+2{\partial_i}{\partial_j}E^\prime\psi^\prime 
+{\partial_i}{\partial_j}{\partial_k}E{\partial_k}\left(E''+2\mathcal{H}E'-\nabla^2E\right).
\label{source}
\end{align}
\end{widetext}
It should be noted that $\mathcal{U}$ represents the scalar component of the fluid velocity perturbation, and $\rho^{\text{r}}$ and $P^{\text{r}}$ refer to the background values of energy density and pressure during the radiation-dominated phase. In gauges characterized by $B=E=0$, the aforementioned equation \eqref{source} simplifies to the results provided in \cite{Nakamura:2004rm}, with the anisotropic stress absent. Generally, Equation \eqref{source} should be used instead. Specifically, all terms involving $B=0$ and $E\ne 0$ should be incorporated into the TT gauge. 
	
For gravitational waves propagating in the direction indicated by $\bm{k}$, we establish normal bases as denoted in ${\mathbf e}$ and $\bar{\mathbf e}$, incorporating ${\bm k}\! \cdot\!{\mathbf e}={\bm k}\! \cdot\!\bar{\mathbf e}={\mathbf e}\!\cdot\!\bar{\mathbf e}=0$ and $|{\mathbf e}|=|\bar{\mathbf e}|=1$. Subsequently, the plus and cross–polarization tensors are articulated as
\begin{align}
\mathbf e^+_{ij}&=\frac{1}{\sqrt{2}}\!\left(\mathbf e_i \mathbf e_j-\bar{\mathbf e}_i \bar{\mathbf e}_j\right),\notag\\
\mathbf e_{ij}^\times&=\frac{1}{\sqrt{2}}\!\left(\mathbf e_i\bar{\mathbf e}_j+\bar{\mathbf e}_i \mathbf e_j\right),
\label{poltensor1}
\end{align}

which are transverse and traceless: $k_i \mathbf{e}^{+ij} = k_i \mathbf{e}^{\times ij} = 0$ and $\mathbf{e}^+_{ij}\mathbf{e}^{\times ij}=0$. Indices are raised/lowered with the flat spatial metric $\delta_{ij}$, hence $e^{\lambda ij}\equiv\delta^{ia}\delta^{jb}e^\lambda_{ab}=e^\lambda_{ij}$.
	
These tensors enable the expansion of $h^{\mathrm{TT}}_{ij}$:
\begin{equation}\label{h_ft}
h^{\mathrm{TT}}_{ij}(\eta,\bm{k})=h^{(+)}(\eta,\bm{k})e_{ij}^{(+)}(\bm{k})+h^{(\times)}(\eta,\bm{k})e_{ij}^{(\times)}(\bm{k}),
\end{equation}
where the mode functions are obtained via the Green’s function method:
\begin{equation}\label{solh}
h^+_{\bm{k}}(\eta)=\frac{1}{a(\eta)}\int_0^\eta G_k(\eta;\tilde{\eta})\,a(\tilde{\eta})\,s^+(\tilde{\eta},\bm{k})\,\mathrm{d}\tilde{\eta},
\end{equation}
with
$G_k(\eta;\tilde{\eta})=\sin(k\eta-k\tilde{\eta})/k.$
For convenience, we set $x=k\eta$ and introduce $u\equiv p/k$ and $v\equiv|\bm k-\bm p|/k$. The plus–polarization source in Fourier space is then
\begin{equation}\label{plussour}
s^+_{\bm{k}}(\eta)=-4\int\frac{\mathrm{d}^3p}{(2\pi)^{3}}\; e^{ij}_{+}(\hat{\bm k})\,p_i p_j\; S_{\bm p}\,S_{|\bm k-\bm p|}\, f(u,v,x).
\end{equation}
The source function $ f(u, v, x)$ in terms of the transfer functions related to the source term is derived by extracting $p_i$, $p_j$, $S_p$, and $S_{ |\bm k-\bm p|}$ from the Fourier transformation. The formulation for $ f(u, v, x)$, absent the imposition of gauge conditions,  is particularly intricate and is expressed as:
\begin{widetext}
\begin{align}   
{f}_{\mathrm{}}(u,v,x) & =T_\psi(ux)T_\psi(vx)-T_\phi(ux)T_\phi(vx)-\frac{v}{u}T_\sigma(ux)\left[T_\phi^*(vx)+T_\psi^*(vx)+T_\sigma(vx)\right] -2\frac{u}{v}T_\psi^*(ux)T_\sigma(vx)+2T_\psi(ux)T_E^{**}(vx)\nonumber\\
&\quad-\frac{1-u^2-v^2}{2uv}T_\sigma(ux)T_\sigma(vx)+2T_\psi(ux)T_\phi(vx)+\frac{2}{\mathcal{H}^2-\mathcal{H}^\prime}\left[kuT_\psi^*(ux)+\mathcal{H}T_\phi(ux)\right]\left[kvT_\psi^*(vx)+\mathcal{H}T_\phi(vx)\right]\nonumber\\
&\quad + 2T_\psi(ux)T_E(vx)+2\frac{u^2}{v^2}T_E(vx)\left[T_\psi^{**}(ux)+\frac{2\mathcal{H}}{ku}T_\psi^*(ux)+T_\psi(ux)\right] -\frac{1-u^2-v^2}{2uv}T_E^*(ux)T_E^*(vx)\nonumber\\
&\quad  +2\left(\frac{1-u^2-v^2}{v^2}\right)T_\psi(ux)T_E(vx)-\frac{1-u^2-v^2}{2u^2}T_E(ux)\left[T_E^{**}(vx)+2\frac{\mathcal{H}}{kv}T_E^*(vx)+T_E(vx)\right]\nonumber\\
&\quad-\left(\frac{1-u^2-v^2}{2uv}\right)^2T_E(ux)T_E(vx) +4\frac{\mathcal{H}}{kv}T_\psi(ux)T_E^*(vx)        +4\frac{u}{v}T_\psi^*(ux)T_E^*(vx),  \end{align}
\end{widetext} 
where $T^*(y)=\mathrm{d}T(y)/\mathrm{d}y$. However, computations based solely on specific gauges will be addressed in the following sections.  
	
A fundamental parameter in the observation of secondary GWs is the energy density parameter, denoted as $\Omega_{\mathrm{GW}}(f)$. This parameter is defined as the energy density of gravitational waves per logarithmic frequency interval (or wavelength, as per $k=2\pi f$), normalized by the critical energy density of the Universe. It is mathematically represented by \cite{Ali:2023moi, Lu:2020diy, Domenech:2021and} 
\begin{equation}\label{ogwd}
\Omega_{\mathrm{GW}}(k,\eta) 
\equiv \frac{1} {\rho_{\mathrm{c}}}\,\frac{\mathrm{d}\rho_{\mathrm{GW}}}{\mathrm{d}\ln k}
= \frac{1}{24}\left(\frac{k} {\mathcal{H}}\right)^{2}\,
\overline{\mathcal{P}_h(k,\eta)}.
\end{equation} 
During radiation domination, $\mathcal{H}\propto\eta^{-1}$, and the overline denotes an average over oscillations, e.g., $\sin^2 x=\cos^2 x\to 1/2$. 
We include both tensor polarizations in Eq.~(\ref{ogwd}), i.e. $\mathcal{P}_h \equiv \sum_{\lambda=+,\times}\mathcal{P}_{h,\lambda}$. 
The dimensionless tensor power spectrum is defined by
\begin{equation}
\big\langle h_{\lambda}(\eta,\bm{k})\,h_{\lambda'}(\eta,\bm{k}')\big\rangle
= (2\pi)^{3}\,\delta_{\lambda\lambda'}\,\delta^{(3)}(\bm{k}+\bm{k}')\,
\frac{2\pi^{2}}{k^{3}}\,\mathcal{P}_{h,\lambda}(k,\eta)\,.
\end{equation}

The calculation of secondary GW during the radiation-dominated era necessitates an evaluation of $\ogw(k,\eta)$ within the sub-horizon regime $\eta\to\eta_{c}$, wherein the source term becomes negligible at $\eta_c$, indicating the stabilization of the  secondary GW signal. At the juncture of radiation-matter equality, the energy density parameter is defined by $\ogw(k) \equiv \ogw(k,x_c)$ for $x_c\gg 1$. Utilizing the aforementioned expressions, one can establish a connection between $\ogw(k)$ and the primordial power spectrum as described by \cite{Domenech:2021and}, 
\begin{align}\label{Omegagw}
\ogw(k) &= {1\over 6}\int_0^\infty\mathrm{d}u\int_{|1-u|}^{1+u}\mathrm{d}v~{v^2\over u^2}\Bigg[1-\left({1+v^2-u^2\over 2v}\right)^2\Bigg]^2 \nonumber\\
&\quad \times \mathcal{P}_S(uk)  \mathcal{P}_S(vk)\overline{I^2(u,v,x\to \infty)}.
\end{align}  
The kernel function is articulated by  
\begin{align} 
\label{iuv} 
I(u,v,x)&=x\int_{x_i}^{x}\!\mathrm{d}\tilde x\; G(x,\tilde x)\,f(u,v,\tilde{x}), 
\end{align} 
where, the source function $f(u,v,\tilde{x})$ can be written in the symmetrize form as follows:  
\begin{widetext} 
\begin{align}    
f(u,v,\tilde{x})
&= -\tfrac{1}{2}\Bigl(
\frac{u} {v}\,T_E(u\tilde{x})\,T^{*}_B(v\tilde{x}) 
+\frac{v}  {u}\,T_E(v\tilde{x})\,T^{*}_B(u\tilde{x}) 
\Bigr) 
-\frac{u^{2}}{v^{2}}
\Bigl[\frac{1}{2v^{2}}\,T_\sigma(u\tilde{x})\bigl(T^{*}_E(v\tilde{x})-T^{*}_\phi(v\tilde{x})\bigr)
+\frac{1}{2u^{2}}\,T_\sigma(v\tilde{x})\bigl(T^{*}_E(u\tilde{x})\nonumber\\
&-T^{*}_\phi(u\tilde{x})\bigr)
\Bigr]
-\frac{16}{3\tilde{x}}
\Bigl(
\frac{u}{v^{2}}\,T^{*}_E(u\tilde{x})\,T_E(v\tilde{x})
+\frac{v} {u^{2}}\,T^{*}_E(v\tilde{x})\,T_E(u\tilde{x})
\Bigr)
-\tfrac{1}{2}\Bigl[
(3+2u^{2}-3v^{2})\,T_E(u\tilde{x})\,T_\psi(v\tilde{x})
+(3+2v^{2}\nonumber\\
&-3u^{2})\,T_E(v\tilde{x})\times\,T_\psi(u\tilde{x})
\Bigr]
+\frac{1}{uv}\,(1+u^{2}-v^{2})\,T^{*}_E(u\tilde{x})\,T^{*}_E(v\tilde{x})    
+\frac{(u^{2}+v^{2}-1)\bigl(3(u^{2}+v^{2})-1\bigr)}{4u^{2}v^{2}}\,T_E(u\tilde{x})\,T_E(v\tilde{x})
\nonumber\\ 
&
+\frac{2}{uv}\,T^{*}_E(u\tilde{x})\,T^{*}_E(v\tilde{x})
+2\,T_\psi(u\tilde{x})\,T_\psi(v\tilde{x})
+\tfrac{1}{2}\Bigl(
\frac{1}{v^{2}}\,T_B(u\tilde{x})\,T_\sigma(v\tilde{x})
+\frac{1}{u^{2}}\,T_B(v\tilde{x})\,T_\sigma(u\tilde{x})
\Bigr)
-\tfrac{1}{2}\Bigl(
\frac{u^{2}}{v^{2}}\,T^{**}_E(u\tilde{x})\,T_E(v\tilde{x})
\nonumber\\
&+\frac{v^{2}}{u^{2}}\,T^{**}_E(v\tilde{x})\,T_E(u\tilde{x})
\Bigr)
-\frac{8}{\tilde{x}}
\Bigl(
\frac{u}{v^{2}}\,T^{*}_\psi(u\tilde{x})\,T_E(v\tilde{x})
+\frac{v}{u^{2}}\,T^{*}_\psi(v\tilde{x})\,T_E(u\tilde{x})
\Bigr)
-\frac{3}{uv}\Bigl(
\frac{u^{2}}{v^{2}}\,T^{**}_\psi(u\tilde{x})\,T_E(v\tilde{x})
+\frac{v^{2}}{u^{2}}\,T^{**}_\psi(v\tilde{x})\nonumber\\
&\times\,T_E(u\tilde{x})
\Bigr)
+\frac{8}{\tilde{x}}
\Bigl(
\frac{u}{v^{2}}\,T_\phi(u\tilde{x})\,T_\psi(v\tilde{x})
+\frac{v}{u^{2}}\,T_\phi(v\tilde{x})\,T_\psi(u\tilde{x})
\Bigr)
-\tfrac{1}{2}\Bigl(
\frac{u}{v}\,T^{*}_\sigma(u\tilde{x})\,T_\psi(v\tilde{x})
+\frac{v}{u}\,T^{*}_\sigma(v\tilde{x})\,T_\psi(u\tilde{x})
\Bigr)
\nonumber\\
&
-\frac{4}{3}\Bigl[
\frac{6-13u^{2}+3v^{2}}{v^{2}}\,T_\psi(u\tilde{x})\,T_E(v\tilde{x}) 
+\frac{6-13v^{2}+3u^{2}} {u^{2}}\,T_\psi(v\tilde{x})\times\,T_E(u\tilde{x})\Bigr]
-\frac{3}{uv}\Bigl(
T_\psi(u\tilde{x})\,T_B(v\tilde{x})
\nonumber\\
&+\tfrac{1}{2}\Bigl(
\frac{u^{2}}{v^{2}}\,T_B(u\tilde{x})\,T_B(v\tilde{x})
+\frac{v^{2}}{u^{2}}\,T_B(v\tilde{x})\,T_B(u\tilde{x})
\Bigr)
-\frac{2\mH}{\tilde{x}}
\Bigl(
\frac{u}{v}\,T_B(u\tilde{x})\,T_E(v\tilde{x})
+\frac{v}{u}\,T_B(v\tilde{x})\,T_E(u\tilde{x})
\Bigr)
\nonumber\\
&+T_\psi(v\tilde{x})\,T_B(u\tilde{x})
\Bigr)-\frac{8}{3\tilde{x}}
\Bigl(
\frac{u}{v^{2}}\,T^{*}_E(u\tilde{x})\,T_E(v\tilde{x})
+\frac{v}{u^{2}}\,T^{*}_E(v\tilde{x})\,T_E(u\tilde{x})
\Bigr).\label{sourcesymmetric}
\end{align}
\end{widetext}
We wish to compute the integration kernel $I(u,v,x)$ defined in Eq.~\eqref{iuv}.
This requires the source $f(u,v,x)$ in Eq.~\eqref{sourcesymmetric}, constructed from
the scalar transfer functions in a generic scalar gauge. No specific slicing is assumed
at this stage; the explicit gauge choices and their relations via second–order
gauge transformations are introduced next.

We now turn our attention to the gauge transformation. The infinitesimal coordinate transformation is characterized by $x^\mu\to x^\mu+\epsilon^\mu$ with $\epsilon^\mu=[\alpha,\partial^i \beta]$. In the context of studying secondary GW, we exclude the vector degrees of freedom from the coordinate transformation, while the scalars $\alpha$ and $\beta$ are treated as first-order quantities. Considering that the transformation of tensor modes under gauge changes is independent of coordinate transformations at the same order, it becomes unnecessary to account for second-order coordinate transformations. This is elaborated extensively in the literature \cite{Malik:2008im,  Bruni:1996im} regarding curvature perturbation. The expression for the transformation of second-order tensor perturbations in the case of isocurvature perturbations is given by  
\begin{equation} 
\label{gaugetranf1}
h_{ij}^{\mathrm{TT}}\to h_{ij}^{\mathrm{TT}}+\chi_{ij}^{\mathrm{TT}}. 
\end{equation}
where 
\begin{equation} \label{hgaugetrans} 
\begin{split} \chi_{ij}=&\,\,2\left[\left(\mathcal{H}^2+\frac{a^{''}}{a}\right)\alpha^{2}+\mathcal{H}\left(\alpha\alpha'+\alpha_{,k}\epsilon^{k}\right)\right]\delta_{ij}\\&+4\left[\alpha\left(C'_{ij}+2\mathcal{H}C_{ij}\right)+C_{ij,k}\epsilon^k+C_{ik}\epsilon^k_{,j}+C_{jk}\epsilon^k_{,i}\right]\\&+2\left(B_{i}\alpha_{,j}+B_{j}\alpha_{,i}\right)+4\mathcal{H}\alpha\left(\epsilon_{i,j}+\epsilon_{j,i}\right)-2\alpha_{,i}\alpha_{,j}\\&+\left(\epsilon_{i,jk}+\epsilon_{j,ik}\right)\epsilon^k+\epsilon_{i,k}\epsilon^{k}_{,j}+\epsilon_{j,k}\epsilon^{k}_{,i}+\epsilon'_{i}\alpha_{,j}+\epsilon'_{j}\alpha_{,i},\\&+2\epsilon_{k,i}\epsilon^{k}_{,j}+\alpha\left(\epsilon'_{i,j}+\epsilon'_{j,i}\right)
\end{split}
\end{equation} 
and  $C_{ij}=-\psi\delta_{ij}+E_{,ij}$. 
Here  $\chi_{ij}^{\mathrm{TT}}$ represents the transverse--traceless tensor contribution generated by a second--order gauge transformation, sourced by  scalar perturbations. This term accounts for the gauge transformation of $h_{ij}^{\mathrm{TT}}$ and is required to relate results between different gauges,
and further we have  
\begin{equation}
\label{xijtt}
\chi_{ij}^{\mathrm{TT}}(\bm x,\eta)=\mathcal{T}_{ij}^{lm}\chi_{lm}=\int\frac{\mathrm{d}^3 k}{(2\pi)^{3/2}}e^{i\bm k\cdot \bm x}[\chi^+_{\bm k}(\eta) \mathbf e_{ij}^++{\chi}^\times_{\bm k}(\eta)\mathbf e_{ij}^\times],
\end{equation}
\begin{align}	\label{chi_fourier}
\chi^+_{\bm k}(\eta)= & -\int\frac{\mathrm{d}^3p}{(2\pi)^{3/2}} \mathbf e^{+ij}p_ip_j\left(4\alpha(\bm p)\sigma(\bm k-\bm p)+8\mathcal{H}\alpha(\bm p)\right. \nonumber \\
& \left.\times [E(\bm k-\bm p)+\beta(\bm k-\bm p)]                     
+\bm{p}\cdot
(\bm k-\bm p)\beta(\bm p)\right.\nonumber \\
& \left.\times[4E(\bm k-\bm p)+2\beta(\bm k-\bm p)]-8\psi(\bm p)\beta(\bm k-\bm p)\right.\nonumber \\
& \left.+2\alpha(\bm p)\alpha(\bm k-\bm p)\right),\nonumber\\
=& 4\int\frac{\mathrm{d}^3p}{(2\pi)^{3/2}}\mathbf e^{+ij}p_ip_j S_p S_{|\bm k-\bm p|} \frac{1}{k^2}I_\chi(u,v,x),                      \end{align}
\begin{align}
I_\chi(u,v,x)
&= -\frac{1}{9\,u v}\Bigl[
2\,T_\alpha(ux)\,T_\sigma(vx)
+2\,T_\alpha(vx)\,T_\sigma(ux)
\notag\\
&\quad+2\,T_\alpha(ux)\,T_\alpha(vx)
-4\Bigl(\tfrac{u}{v}\,T_\psi(ux)\,T_\beta(vx)
\notag\\
&\quad+\tfrac{v}{u}\,T_\psi(vx)\,T_\beta(ux)\Bigr)
+\tfrac{1-u^{2}-v^{2}}{u v}\Bigl(
T_\beta(ux)\,\notag\\
&\quad\times T_E(vx)
+T_\beta(vx)\,T_E(ux)
+T_\beta(ux)\,T_\beta(vx)\Bigr)
\notag\\
&\quad+4\,\tfrac{\mathcal{H}}{k}\Bigl(
\tfrac{1}{v}\,T_\alpha(ux)\,T_E(vx)
+\tfrac{1}{u}\,T_E(ux)\,T_\alpha(vx)
\notag\\
&\quad+\tfrac{1}{v}\,T_\alpha(ux)\,T_\beta(vx)
+\tfrac{1}{u}\,T_\beta(ux)\,T_\alpha(vx)
\Bigr)
\Bigr].
\label{Ichi}
\end{align} 
We have symmetrized $I_\chi(u,v,x)$ under $u\leftrightarrow v$. Note that the first-order scalar coordinate transformation appears in the transformed second-order tensor perturbations.
With the gauge transformation \eqref{gaugetranf1} and the result for secondary GWs in the Longitudinal gauge, it is straightforward to derive the semianalytic expression for secondary GWs in other gauges without performing the detailed calculation in that gauge.
	
Combining Eqs. \eqref{h_ft}--\eqref{plussour}, \eqref{iuv}, \eqref{gaugetranf1}, \eqref{xijtt} and \eqref{chi_fourier}, we get the following gauge transformation:
\begin{align}\label{hchi}
h^+_{\bm k}\to h^+_{\bm k}+\chi^+_{\bm k} 
& = 4\int\frac{\mathrm{d}^3p}{(2\pi)^{3/2}}\mathbf e^{+ij}(\bm k)p_ip_jS_p S_{|\bm k-\bm p|} \nonumber\\
&\quad \times \frac{1}{k^2} \left[I(u,v,x)+I_\chi(u,v,x)\right]. 
\end{align}
This gauge transformation \eqref{hchi} is an important result of our paper. 
It shows how the solution and the power spectrum of secondary GWs transforms under a change of gauge, which is especially useful in gauges where the kernels cannot be obtained directly from the transfer function. For general gauge choices, for example, starting from the longitudinal gauge, we can obtain the solution in any other gauge by replacing the longitudinal gauge kernel $I_{\mathrm{Long.}}(u,v,x)$ in Eq.~\eqref{iuv} according to the following rule:  
\begin{align}\label{IN}
I_\mathrm{Long.}(u,v,x) \to I_\mathrm{Long.}(u,v,x)+I_\chi(u,v,x),  
\end{align}
where
\begin{align}
I_\chi(u,v,x)
&= -\frac{1}{9\,u v}\Bigl[
-4\bigl(\tfrac{u}{v}\,T_{\mathrm{Long.}}(ux)\,T_\beta(vx)
\notag\\
&\qquad+\tfrac{v}{u}\,T_{\mathrm{Long.}}(vx)\,T_\beta(ux)\bigr)
\notag\\&\qquad+2\,T_\alpha(ux)\,T_\alpha(vx)
\notag\\
&\qquad+4\,\tfrac{\mathcal{H}}{k}\bigl(\tfrac{1}{v}\, T_\alpha(ux)\,T_\beta(vx)
\notag\\
&\qquad+\tfrac{1}{u}\,T_\beta(ux)\,T_\alpha(vx)\bigr)\notag\\
&\qquad+\tfrac{1-u^{2}-v^{2}}{u v}\,T_\beta(ux)\,T_\beta(vx)
\Bigr].\label{xitransfer}
\end{align}  
Here $T_{\mathrm{Long.}}(z)$ denotes the longitudinal gauge transfer function (unit early–time normalization $T_{\mathrm{Long.}}(0)=1$),  
while $T_{\alpha}(z)$ and $T_{\beta}(z)$  are the transfer functions of the gauge generators $(\alpha,\beta)$ that map the longitudinal gauge to the target gauge (e.g. UE, CO, Nm). Equation~\eqref{xitransfer} follows from Eq.~\eqref{Ichi} upon inserting $T_{\sigma}=T_{E}=0$ and $T_{\psi}=T_{\mathrm{Long.}}$ in the longitudinal gauge and expressing $T_{\alpha}$, $T_{\beta}$ via the corresponding coordinate transformation.   

\subsection{Analytical kernel functions in the $(d,s)$ domain}\label{sec.2.ds} 

In this work, we provide a comprehensive analytical treatment of the kernel function relevant for computing
the energy density of secondary GWs in multiple gauge choices. Previous studies have
often focused on a subset of gauges or presented numerical results without a unified analytical framework. 
By contrast, in the following subsections we derive explicit analytical expressions of the kernel function
in all nine commonly used gauges, enabling a systematic investigation of potential gauge dependencies.

To achieve this, we adopt a transformation of the integration domain from the conventional variables $(u,v)$
to a pair of dimensionless auxiliary variables $(d,s)$, defined as
\begin{gather}
d = \frac{1}{\sqrt{3}}|u - v|, \qquad s = \frac{1}{\sqrt{3}}(u + v), \nonumber\\
(d,s) \in \left[0, \tfrac{1}{\sqrt{3}}\right] \times \left[\tfrac{1}{\sqrt{3}}, +\infty \right),
\end{gather}
as introduced in~\cite{Espinosa:2018eve} and employed here for isocurvature perturbations.
This redefinition streamlines the kernel structure, aligns the radiative (luminal) condition with simple lines
in the $(d,s)$ plane, and facilitates exact integration.

Assuming Gaussian isocurvature fluctuations, the spectral density of induced GWs can be written directly in
the $(d,s)$ domain:
\begin{widetext}
\begin{align}
\label{eq:OmegaGW-dsg2}
\Omega_{\mathrm{GW},c}(k)
= \int_{0}^{1/\sqrt{3}}\!\! \mathrm{d}d\,
\int_{1/\sqrt{3}}^{\infty}\!\! \mathrm{d}s\;
\left(\frac{3(s-d)^2-(1-3sd)^2}{3(s^2-d^2)}\right)^{\!2}
\,\overline{I^2\!\big(d,s\big)}\;
{\cal P}_S\!\!\left(k\,\tfrac{\sqrt{3}}{2}(s+d)\right)
{\cal P}_S\!\!\left(k\,\tfrac{\sqrt{3}}{2}(s-d)\right),
\end{align}
\end{widetext}
with the primordial isocurvature spectrum defined by
\begin{equation}
\langle S_{\mathbf{k}}(0)S_{\mathbf{k}'}(0)\rangle
= \frac{2\pi^2}{k^3}\,{\cal P}_S(k)\,(2\pi)^3\delta^{(3)}(\mathbf{k}+\mathbf{k}').
\end{equation}

The kernel encapsulates the scalar transfer functions via the Green–function solution,
\begin{align}
\label{eq:kernel-rad2}
I(d,s,x)&=x\int_{x_i}^{x}\!\mathrm{d}\tilde x\; G(x,\tilde x)\,f(\tilde x;d,s),\notag\\
G(x,\tilde x)&=\frac{a(\tilde x)}{a(x)}\big(\sin x\cos\tilde x-\cos x\sin\tilde x\big),
\end{align}
and can be decomposed as
\begin{align}
I(d,s,x)&=I_c(d,s,x)\,(-\sin x) + I_s(d,s,x)\,\cos x, \label{ICS} \\
\intertext{where,}
I_c(d,s,x)&\equiv \int_{0}^{x}\!\mathrm{d}\tilde x\;\tilde x\,
\big(-\sin\tilde x\big)\,f(\tilde x;d,s),\notag\\
I_s(d,s,x)&\equiv \int_{0}^{x}\!\mathrm{d}\tilde x\;\tilde x\,
\cos\tilde x\,f(\tilde x;d,s). \nonumber
\end{align} 
Averaging over many oscillations ($x\to\infty$) yields
\begin{equation}
\label{eq:Ibar2}
\overline{I^2(d,s)}\simeq \tfrac12\big(I_{c,\infty}^2(d,s)+I_{s,\infty}^2(d,s)\big).
\end{equation}
Here we take the limit $x \to \infty$, since our interest lies in GWs that are well inside the horizon,
and Eq.~\eqref{eq:OmegaGW-dsg2} gives the expression for the energy density. 
The integrals $I_s$ and $I_c$ in \eqref{ICS} can be evaluated analytically for isocurvature fluctuations, as they involve trigonometric kernels; their explicit forms are lengthy (see Sec.~\ref{sec3})\footnote{Exact analytical results for $I_{c}(d,s,x)$ and $I_{s}(d,s,x)$ are given for all gauges. TT and N–body (Nb) formulas span several pages; a unified \texttt{Mathematica} notebook is provided at Ref.~\cite{Ali_SIGW_TT_Nb_results_2025}.} where we present them gauge by gauge, together with compact limits and figures obtained by direct evaluation.

Throughout, we work in the $(d,s)$ domain. We can find that the difference between the longitudinal gauge, and rest of the gauges comes from the terms of 
the form $(\cos(d\pm s)k\eta)$ or $(\sin(d\pm s)k\eta)$ which do not correspond to gravitational waves, i.e. $\cos(k\eta)$ or  $\sin(k\eta)$ \cite{Tomikawa:2019tvi}. This indicates that the gauge dependence appears only in the tensor perturbations coupling with scalar perturbation, not in the gravitational 
wave. When forming late-time observables, we evaluate the radiative sector by restricting to the luminal lines $d\pm s=1$ in the $(d,s)$ plane, which amounts to retaining only the oscillatory $\sin x$ and $\cos x$ pieces of the tensor solution. These oscillatory components are precisely the freely propagating tensor modes (free GWs). This selection is used in the following to construct the gauge-independent spectrum in
Sect.~\ref{sec:gauge-independent}. 

\section{Results in various gauges}\label{sec3}
We now present a gauge-by-gauge analytic treatment of the source $f(d,s,x)$ and of the kernel components $I_c(d,s,x)$ and $I_s(d,s,x)$ in the $(d,s)$ representation introduced above. In each slicing we impose the standard gauge conditions and, when needed, fix residual freedom so that pure gauge pieces do not enter the source. Our objective is to provide the transfer kernels that control the second-order tensor mode induced by isocurvature fluctuations during radiation domination.  

Each subsection shows the source $f(d,s,x)$, the kernels $I_c$ and $I_s$, and the corresponding energy density spectrum $\Omega_{\mathrm{GW}}(k)$. All expressions are obtained analytically. We evaluate these analytical results at $(d,s)=(0,1/\sqrt{3})$ to display their late-time evolution. While the late-time construction of the gauge-independent spectrum is provided in Sec.~\ref{sec:gauge-independent}.

\subsection{Secondary GWs  in Longitudinal  gauge} \label{n-gauge}
\begin{figure}[th]  
\centering \includegraphics[width=\linewidth]{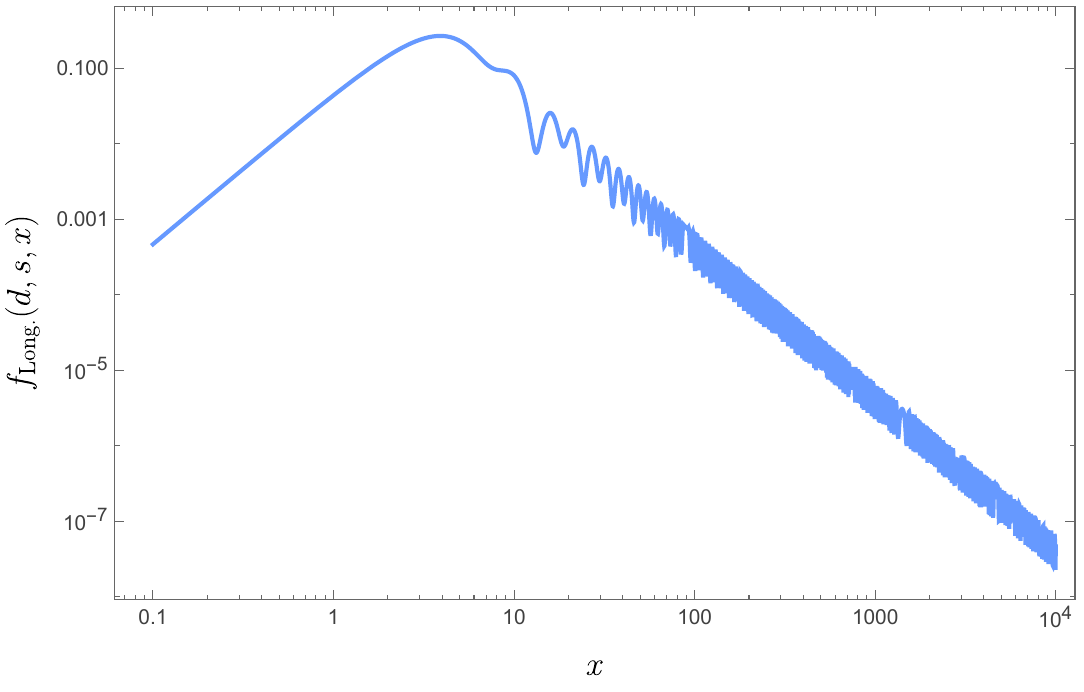} 
\caption{Source term in longitudinal gauge, 
  $f_{\rm long.}(d,s,x)$, evaluated at $d=0$ and $s=1/\sqrt{3}$ as a  
  function of the dimensionless time variable $x=k\eta$.
  The growth around horizon entry is followed by an oscillatory decay 
  $f_{\rm long.}\propto x^{-2}$ (up to trigonometric factors),
  signalling the transition from causal scalar sourcing to free GW 
  propagation at late times.}  
\label{fig:source-kernel} 
\end{figure} 
Secondary GWs generated by isocurvature perturbations emerge within the framework of the Longitudinal gauge, as delineated in $B=E=0$. Within this gauge, the governing equations are specified by ${\psi}={\phi}$, and the resultant solution is provided by 
\begin{align}
T_\phi(x)=T_{\mathrm{Long.}}(x)
&= \frac{3}{2\sqrt{2}\,\kappa\,x^{3}}\left[
6 + x^{2} - 2\sqrt{3}\,x \sin\!\left(\frac{x}{\sqrt{3}}\right)
\right]\notag\\
&\quad - \frac{3}{2\sqrt{2}\,\kappa\,x^{3}}\,6 \cos\!\left(\frac{x}{\sqrt{3}}\right).	\label{trealg}
\end{align}
The subsequent expression for $T_S(x)$ is articulated as follows: 
\begin{align}
T_S(x)
&= 1+\frac{3}{2\sqrt{2}\,\kappa}
\Big[
x+\sqrt{3}\,\sin\!\big(\tfrac{x}{\sqrt{3}}\big)
-2\sqrt{3}\,\mathrm{Si}\!\big(\tfrac{x}{\sqrt{3}}\big)
\Big],
\label{eq:TSiso}
\end{align}  
where $\text{Si}(z)$ appears in \eqref{eq:TSiso} and both $\text{Si}(z)$ and $\text{Ci}(z)$ appear in the subsequent expressions, representing the Sine and Cosine Integral functions.
    
Because the isocurvature $S$ is gauge invariant, the isocurvature transfer function $T_S(x)\equiv S/S_{\bm k}$ is identical in all gauges; we present it once (computed in the Longitudinal gauge) and use it throughout. 
\begin{figure*}[th]
\centering
\begin{minipage}[t]{0.48\linewidth}
\centering
\includegraphics[width=\linewidth]{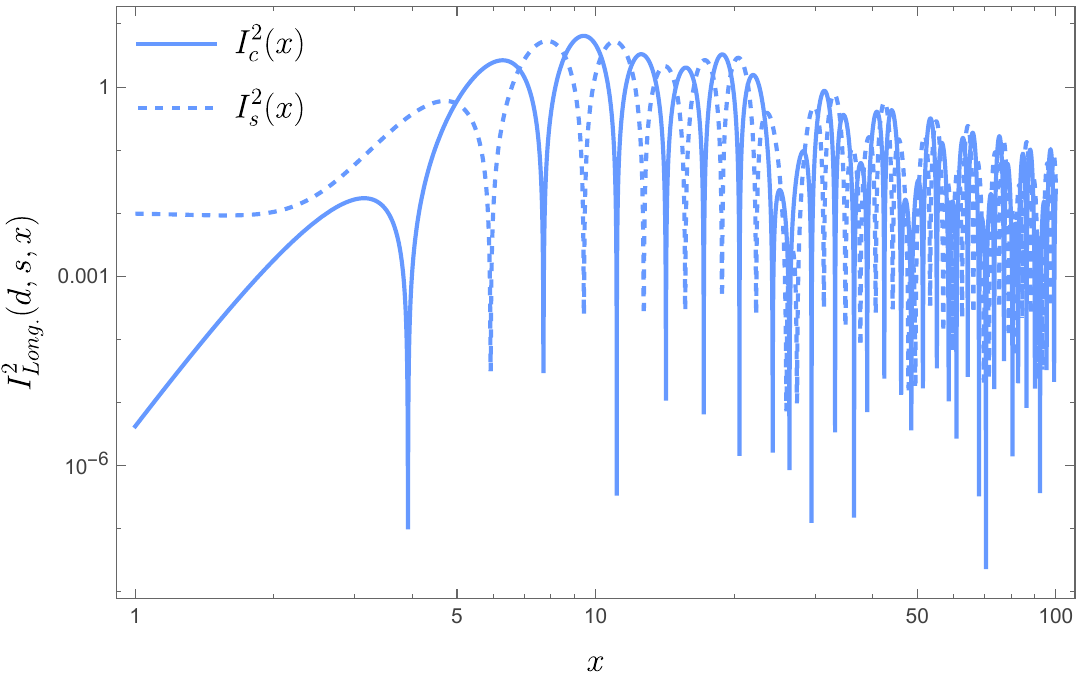}
\caption{Squared Green–function kernels in longitudinal gauge at
$(d,s)=(0,1/\sqrt{3})$:
$I_c^2(x)$ (solid) and $I_s^2(x)$ (dashed).
Both exhibit a decaying evolution, 
$I_{c/s}^2\propto x^{-2}$.
This ensures a convergent tensor solution and a time-independent late-time energy density.}
\label{fig:L_kernels}
\end{minipage}
\hfill
\begin{minipage}[t]{0.48\linewidth}
\centering
\includegraphics[width=\linewidth]{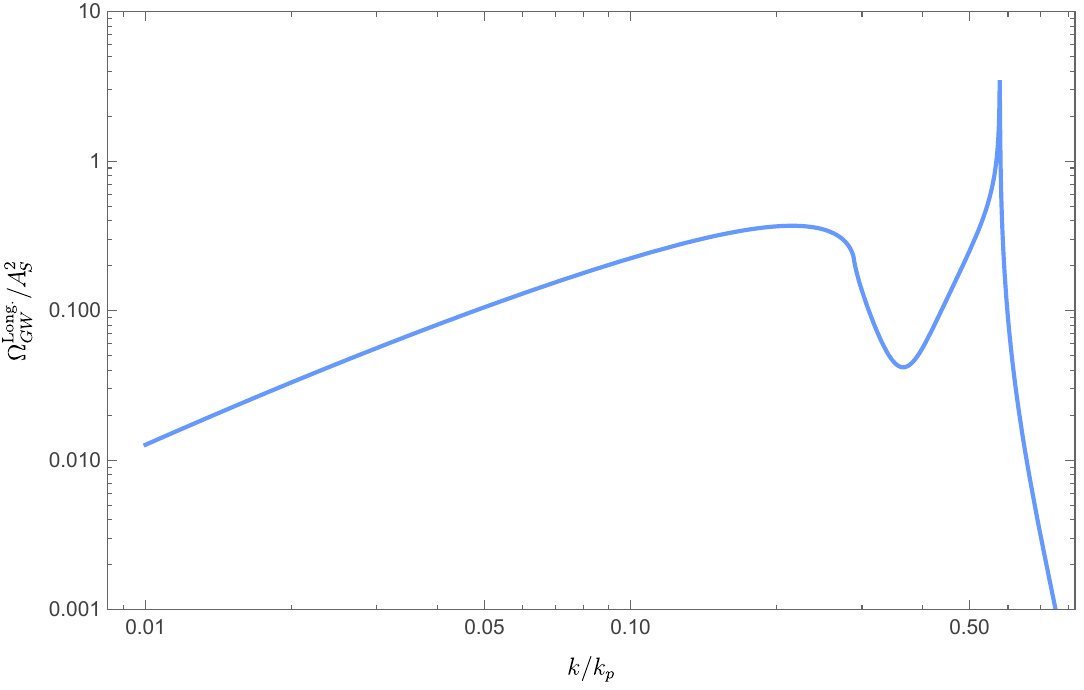}
\caption{Scalar–induced spectrum $\Omega_{\rm GW}(k)/A_s^2$ in the
longitudinal gauge for a Dirac–delta isocurvature peak at $k_p$. 
The evaluation follows the delta line in $(d,s)$: $d=0$ and
$s=\tfrac{2}{\sqrt{3}}(k_p/k)$. 
The spectrum has the standard low-$k$ tail 
$\propto k^{2}\ln^{2}\!k$ for $k\ll k_p$,
peaks at $k=2c_s k_p$ with $c_s=1/\sqrt{3}$, 
and exhibits a sharp cutoff at $k=2k_p$ from momentum conservation. 
At late times ($x\gg1$) the evolution is constant: the longitudinal
baseline is \emph{convergent}.
The GW spectrum is normalized by the scalar power spectrum amplitude $A_s^2$.}
\label{fig:L_spectrum} 
\end{minipage}
\end{figure*}
Employing the transfer functions of this gauge, the source function can be identified as: 
\begin{widetext} 
\begin{align}
\label{trealgf}
f^{\text{Long.}}(d,s,x)&=\frac{16}{3 \kappa ^2 x^6 (s-d)^3 (d+s)^3}\Bigg[\frac{9}{16} x^4 \
(s-d)^2 (d+s)^2+6 x^2 \left(\frac{3}{4} (s-d)^2+\frac{3}{4} (d+s)^2\right)-3 x (s-d) \left(\frac{3}{4} x^2 (d+s)^2\right.\nonumber\\ 
&\quad \left. +18\right) \sin \left(\frac{1}{2} x (s-d)\right)-\sqrt{3} x (d+s) \left(\sqrt{3} \left(\frac{3}{4} x^2 \
(s-d)^2+18\right)-9 \sqrt{3} x (s-d) \sin \left(\frac{1}{2} x (s-d)\right)\right)  \nonumber\\
&\quad \times\sin \left(\frac{1}{2} x (d+s)\right)+2 \left(-3 x^2 \
\left(\frac{3}{4} (d+s)^2-\frac{3}{2} (s-d)^2\right)-3 x (d+s) \
\left(\frac{3}{4} x^2 (s-d)^2-9\right) \right. \nonumber\\
&\quad \left.\times\sin \left(\frac{1}{2} x (d+s)\
\right)-54\right) \cos \left(\frac{1}{2} x (s-d)\right)+2 \cos \left(\frac{1}{2} x (d+s)\right) \
\left(-\frac{9}{4} x^2 (s-d)^2+\frac{9}{2} \right.\nonumber\\ 
&\quad \left.\times x^2 (d+s)^2-3 x (s-d) 
\left(\frac{3}{4} x^2 (d+s)^2-9\right) \sin \left(\frac{1}{2} x (s-d)\right)+\left(\frac{9}{16} x^4 (s-d)^2 (d+s)^2\right.\right. \nonumber\\
&\quad \left.\left.-6 x^2 \left(\frac{3}{4} \
(s-d)^2+\frac{3}{4} (d+s)^2\right)+54\right) \cos \
\left(\frac{1}{2} x (s-d)\right)-54\right)+108 \Bigg].
\end{align}
Incorporating \eqref{trealgf} into Eq. \eqref{ICS}, we derive the explicit expression for ${I}^{\text{Long.}}_{\mathrm{c}}$:
\begin{align}
\label{Iclong}
{I}^{\text{Long.}}_{\mathrm{c}}(d,s,x)=&\frac{3}{4\kappa^2}\,\Biggl\{\,
\frac{8}{x^4\,(d^2-s^2)^3}\Bigl[\,%
2x\cos x\Bigl(\,-2\bigl[x^2(d^2+s^2-1)+d\,x\sin(d x)+2\bigr]+\cos\bigl(\tfrac{s x}{2}\bigr)\Bigl[\bigl(x^2(3d^2+3s^2-4)+8\bigr)\nonumber\\
&\times\cos\,\!\bigl(\tfrac{d x}{2}\bigr)
+4\,d\,x\sin\bigl(\tfrac{d x}{2}\bigr)\Bigr]+(-((d^2-1)x^2)-2)\cos(s x)
+(-((s^2-1)x^2)-2)\cos(d x)+2\,s\,x\sin\bigl(\tfrac{s x}{2}\bigr)\nonumber\\
&\times\bigl[d\,x\sin\tfrac{d x}{2}+2\cos\tfrac{d x}{2}\bigr]
-2\,s\,x\sin(s x)\Bigr)+\sin x\Bigl(\,-4x^2(d^2+s^2-1)\;+2\bigl(x^2(2d^2-s^2+1)-6\bigr)\cos(d x)
\nonumber\\
&+2\bigl(x^2(-d^2+2s^2+1)-6\bigr)\cos(s x)+s\,x\sin\bigl(\tfrac{s x}{2}\bigr)\Bigl[(x^2(d^2+3s^2-4)+24)\cos\tfrac{d x}{2}
+12\,d\,x\sin\tfrac{d x}{2}\Bigr]\nonumber\\
&+\cos\bigl(\tfrac{s x}{2}\bigr)\Bigl[d\,x\,(x^2(3d^2+s^2-4)+24)\sin\tfrac{d x}{2}
+2(x^2(d^2+s^2-4)+24)\cos\tfrac{d x}{2}\Bigr]+2\,s\,x\bigl(-((d^2-1)x^2)\nonumber\\
&-6\bigr)\sin(s x)
+2\,d\,x\bigl(-((s^2-1)x^2)-6\bigr)\sin(d x)-24\Bigr)\Bigr]\Biggr\}
+\;\frac{1}{(d-s)^3(d+s)^3}\Bigl\{
2\,(d^2+s^2-2)^2\nonumber\\&\times \bigl[\mathrm{Si}((d+1)x)+\mathrm{Si}((s+1)x)
+\mathrm{Si}(x-dx)+\mathrm{Si}(x-sx)\bigr]+\;4\bigl(d^4-2d^2(s^2+4)+s^4-8s^2+8\bigr)\,\nonumber\\
& \times \mathrm{Si}(x)+\;\bigl(-3d^4+4d^3s-2d^2(s^2-8)+4ds^3-3s^4+16s^2-16\bigr)
\,\mathrm{Si}\bigl(\tfrac{d-s+2}{2}x\bigr)\nonumber\\
&-\;\bigl(3d^4+4d^3s+2d^2(s^2-8)+4ds^3+3s^4-16s^2+16\bigr)\, 
\mathrm{Si}\bigl(-\tfrac{d x}{2}-\tfrac{s x}{2}+x\bigr)\nonumber\\ 
&+\;\bigl(-3d^4+4d^3s-2d^2(s^2-8)+4ds^3-3s^4+16s^2-16\bigr)  
\,\mathrm{Si}\bigl(\tfrac{-d+s+2}{2}x\bigr)\nonumber\\
&-\;\bigl(3d^4+4d^3s+2d^2(s^2-8)+4ds^3+3s^4-16s^2+16\bigr)  
\,\mathrm{Si}\bigl(\tfrac{d+s+2}{2}x\bigr)\Bigr\},  
\end{align} 
and using \eqref{trealgf} in Eq. \eqref{ICS} we get ${I}^{\text{Long.}}_{\mathrm{s}}$ as:
\begin{align}
\label{Islong} 
{I}^{\text{Long.}}_{\mathrm{s}}(d,s,x)=\,&
\frac{3}{4\kappa^{2}(d-s)^{3}(d+s)^{3}}\Bigl(
-2\!\left(d^{2}+s^{2}-2\right)^{2}\mathrm{Ci}\left(x\left|1-d\right|\right)
-2\!\left(d^{2}+s^{2}-2\right)^{2}\mathrm{Ci}\left(x\left|d+1\right|\right)
-2\!\left(d^{2}+s^{2}-2\right)^{2}\notag\\
&\times\mathrm{Ci}\left(x\left|1-s\right|\right)
-2\!\left(d^{2}+s^{2}-2\right)^{2}\mathrm{Ci}\left(x\left|s+1\right|\right)+\bigl(3d^{4}+4d^{3}s+2d^{2}(s^{2}-8)+4ds^{3}+3s^{4}-16s^{2}\notag\\ 
&+16\bigr)
\mathrm{Ci}\left(\tfrac{x}{2}\left|-d-s+2\right|\right)+\bigl(3d^{4}-4d^{3}s+2d^{2}(s^{2}-8)-4ds^{3}+3s^{4}-16s^{2}+16\bigr)
\mathrm{Ci}\left(\tfrac{x}{2}\left|d-s +2\right|\right)\notag\\
&+\bigl(3d^{4}-4d^{3}s+2d^{2}(s^{2}-8)-4ds^{3}+3s^{4}-16s^{2}+16\bigr)
\mathrm{Ci}\left(\tfrac{x}{2}\left|-d+s+2\right|\right)
+\bigl(3d^{4}+4d^{3}s+2d^{2}\notag\\
&\times(s^{2}-8)+4ds^{3}+3s^{4}-16s^{2}+16\bigr)
\mathrm{Ci}\left(\tfrac{x}{2}\left|d+s+2\right|\right)-4\!\left(d^{4}-2d^{2}(s^{2}+4)+s^{4}-8s^{2}+8\right)\mathrm{Ci}(x)
\Bigr)  
\notag\\
&-\frac{6}{\kappa^{2}x^{4}(d^{2}-s^{2})^{3}}\Bigl(
2x\sin x\Bigl[
2\bigl(x^{2}(d^{2}+s^{2}-1)+dx\sin(dx)+2\bigr)+\cos\left(\tfrac{s x}{2}\right)\Bigl(
-(x^{2}(3d^{2}+3s^{2}-4)\notag\\
&+8)\cos\left(\tfrac{d x}{2}\right)
-4dx\sin\left(\tfrac{d x}{2}\right)
\Bigr)
+\bigl((d^{2}-1)x^{2}+2\bigr)\cos(sx)
+\bigl((s^{2}-1)x^{2}+2\bigr)\cos(dx)
-2sx\sin\left(\tfrac{s x}{2}\right)
\notag\\
&\times\bigl(dx\sin\left(\tfrac{d x}{2}\right)+2\cos\left(\tfrac{d x}{2}\right)\bigr)
+2sx\sin(sx)
\Bigr] 
+\cos x\Bigl[
-4x^{2}(d^{2}+s^{2}-1)
+2\bigl(x^{2}(2d^{2}-s^{2}+1)-6\bigr)\notag\\
&\times\cos(dx)
+2\bigl(x^{2}(-d^{2}+2s^{2}+1)-6\bigr)\cos(sx)
+sx\sin\left(\tfrac{s x}{2}\right)
\bigl((x^{2}(d^{2}+3s^{2}-4)+24)\cos\left(\tfrac{d x}{2}\right)
\notag\\
&+12dx\sin\left(\tfrac{d x}{2}\right)\bigr)
+\cos\left(\tfrac{s x}{2}\right)
\bigl(dx(x^{2}(3d^{2}+s^{2}-4)+24)\sin\left(\tfrac{d x}{2}\right) 
+2(x^{2}(d^{2}+s^{2}-4)+24)\cos\left(\tfrac{d x}{2}\right)\bigr)
\notag\\
&+2sx(-d^{2}x^{2}+x^{2}-6)\sin(sx)
+2dx(-s^{2}x^{2}+x^{2}-6)\sin(dx)
-24 
\Bigr] 
\Bigr). 
\end{align}
\end{widetext}
Regarding isocurvature–induced GWs, we are mostly interested in the small–scale power spectrum, i.e. fluctuations that enter the horizon well before matter–radiation equality. It is therefore an excellent approximation to use the RD solution \eqref{trealg}. Plugging the source function \eqref{trealgf} into the integrals \eqref{ICS} we obtain lengthy analytic expressions for
$ I_c^{\text{Long.}} $ and $ I_s^{\text{Long.}} $, presented in Eqs.~\eqref{Iclong} and \eqref{Islong}.

For analytical simplicity, in Eqs.~\eqref{Iclong} and \eqref{Islong} we take the limit $x\to\infty$. The resulting expressions reduce to
\begin{widetext}

\begin{align}\label{eq:Ic-ds2}
I_{c,\infty}(d,s)
&=\frac{9}{32\,\kappa^2\,(d^{2}-s^{2})^{6}}
\Bigg[
3(d^{2}-s^{2})^{2}
+2(d^{2}+s^{2}-2)^{2}
\ln\!\big|(1-d^{2})(1-s^{2})\big| \notag\\[1ex]
&\quad -2\big(3 d^{4}-4 d^{3}s+2 d^{2}(s^{2}-8)-4 d s^{3}+3 s^{4}-16 s^{2}+16\big)
\ln\!\frac{|d-s+2|}{2} \notag\\[1ex]
&\quad -\big(3 d^{4}+4 d^{3}s+2 d^{2}(s^{2}-8)+4 d s^{3}+3 s^{4}-16 s^{2}+16\big)
\left(\ln\!\frac{|d+s-2|}{2}\notag\right.\\
&\quad\left.+\ln\!\frac{|d+s+2|}{2}\right)
\Bigg],  
\end{align}  
and
\begin{align} 
\label{eq:Is-ds2} 
I_{s,\infty}(d,s)
&= \frac{9\pi^{2}}{32\,\kappa^2\,(d^{2}-s^{2})^{6}}
\Bigg[
- \big(3 d^{4}+4 d^{3}s+2 d^{2}s^{2}-16 d^{2}+4 d s^{3}+3 s^{4}-16 s^{2}+16\big)\,
\notag\\[0.6ex]
&\quad\times\Theta\!\left(1-\Big|\tfrac{d+s}{2}\Big|\right) 
- \big(3 d^{4}-4 d^{3}s+2 d^{2}s^{2}-16 d^{2}-4 d s^{3}+3 s^{4}-16 s^{2}+16\big)\,
\notag\\[0.6ex]
&\quad \times \Theta\!\left(1-\Big|\tfrac{d-s}{2}\Big|\right) 
+\,2(d^{2}+s^{2}-2)^{2}\Big(\Theta(1-|s|)+\Theta(1-| d|)\Big)
+2(d^{2}-s^{2})^{2}\notag\\[0.6ex]
&\quad-16\big(d^{2}+s^{2}-1\big) 
\Bigg]. 
\end{align}
\end{widetext}
With the kernels \eqref{eq:Ic-ds2}–\eqref{eq:Is-ds2} we compute the isocurvature–induced spectrum using Eq.~\eqref{eq:OmegaGW-dsg2}. The same procedure will be applied in the remaining gauges in the next subsections. Let us emphasize that Eqs.~\eqref{eq:Ic-ds2} and \eqref{eq:Is-ds2} are valid for any primordial isocurvature spectrum,
provided the relevant modes enter the horizon well before equality. For illustration we also show the source \eqref{trealgf}, the kernels $I_{c/s}(d,s,x)$, and the spectrum in Figs.~\ref{fig:source-kernel}, \ref{fig:L_kernels}, and \ref{fig:L_spectrum}.

\subsubsection{Dirac--delta isocurvature peak in $(d,s)$}
\label{subsec:long-delta-ds}

In the longitudinal gauge, the source scales as $f_\text{Long.}(d,s,x)\sim x^{-2}$ for $x\gg1$
(up to bounded trigonometric factors). Consequently, with finite values of kernels given in  Eqs.~\eqref{eq:Ic-ds2} and \eqref{eq:Is-ds2} set by the $(d,s)$ geometry and the luminal pieces of the source, the late–time average obeys  
\begin{equation}
\overline{I^2}\;\propto\; I_c^2+I_s^2 \;\sim\; x^{-2}.
\end{equation}
Hence the longitudinal kernel is convergent in radiation domination and the late–time spectrum is finite. 

For a Dirac--delta primordial isocurvature peak,  
$\mathcal{P}_{S} (k)=\mathcal{A}_{S}\,\delta\!\big(\ln(k/k_{p})\big)$,
the $(d,s)$ evaluation reduces to
\begin{equation}
d=0,\qquad s=\frac{2}{\sqrt{3}}\frac{k_{p}}{k},
\end{equation}
and the induced spectrum is
\begin{align}
\label{eq:Omega-delta-ds}  
\Omega_\text{GW,c}(&k)=
\mathcal{A}_{S}^{2}\,
\frac{(3s^{2}-1)^{2}}{36\,s^{2}}\,
\Bigl[I_{c,\infty}(k,0,s)^{2}+I_{s,\infty}(k,0,s)^{2}\Bigr]\,\notag\\&\qquad\times
\Theta\!\left(2 k_{p}-k\right),\notag\\&  
\,\,s=\frac{2}{\sqrt{3}}\frac{k_{p}}{k}.  
\end{align}
\begin{figure}[t]
\centering
\includegraphics[width=\linewidth]{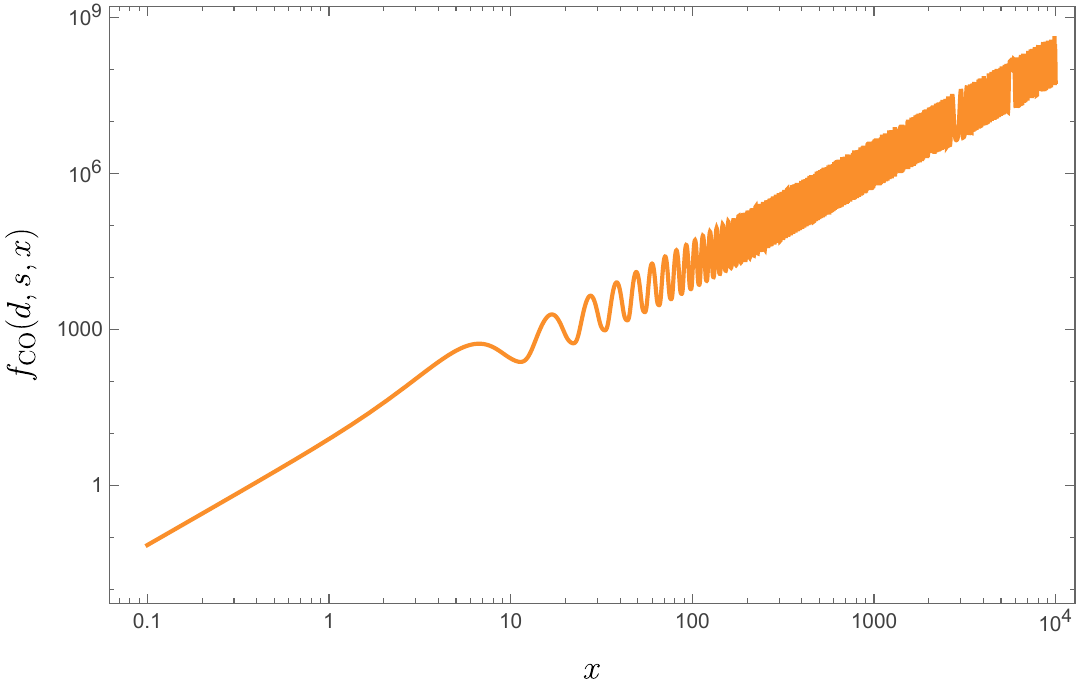}
\caption{The source $f_{\rm CO}(d,s,x)$ evaluated at $(d,s)=(0,1/\sqrt{3})$ versus $x=k\eta$. 
For $x\gg1$ the evolution grows steadily, indicating a strong late–time growth in this slicing for isocurvature initial conditions.} 
\label{fig:CO_source}  
\end{figure} 
It peaks at $k=2c_{s}k_{p}$ ($c_{s}=1/\sqrt{3}$, i.e.\ $s=1$), exhibits a sharp UV cutoff at
$k=2k_{p}$ ($s=1/\sqrt{3}$), and has a $k^{2}\ln^{2}\!k$ tail for $k\ll k_{p}$ ($s\gg1$).
Time–asymptotically ($x\gg1$) the evolution is constant, consistent with the $x^{-2}$ decay of
the kernels. This behavior is illustrated in Fig.~\ref{fig:L_spectrum}.\footnote{
For all figures of the GW spectra we set the scalar spectrum amplitude to $A_s=1$ \cite{Domenech:2023jve}, and the same
normalization is used in all gauges.
}

\subsection{Secondary GWs in Comoving Orthogonal gauge} 
We now turn our attention to the comoving orthogonal gauge, where $\delta V=B=0$. The transfer  function is given by 
\begin{align}
T_{\beta}(x)
&= \frac{3}{\kappa\,x}\Biggl[
\cos\Bigl(\tfrac{x}{\sqrt{3}}\Bigr) - 1
+ \frac{2\sqrt{3}\,x}{3}\,\mathrm{Si}\Bigl(\tfrac{x}{\sqrt{3}}\Bigr)
+ \frac{x^{2}}{6}\notag
\\&\quad- \frac{\sqrt{3}\,x}{3}\,\sin\Bigl(\tfrac{x}{\sqrt{3}}\Bigr)
\Biggr].\label{transcom}
\end{align}

In the comoving orthogonal gauge, the source function can be evaluated as:
\begin{widetext}
\begin{align}
f^{\mathrm{CO}}(d,s,x)
&= \frac{4}{27\,\kappa^{2}\,x^{6}\,(s-d)^{3}(d+s)^{3}}
\Biggl(
4\Biggl(
-\tfrac{9}{2}\,x^{3}(s-d)^{3}\,\mathrm{Si}\Bigl(\tfrac{1}{2}(s-d)x\Bigr)
-\tfrac{9}{8}\,x^{4}(s-d)^{4}
+\Bigl(\tfrac{9}{8}\,x^{4}(s-d)^{4}
+\tfrac{9\,x^{2}(s-d)^{2}}{2\sqrt{2}}
\notag\\&-18\sqrt{2}\Bigr)\cos\Bigl(\tfrac{1}{2}x(s-d)\Bigr)
-9\sqrt{2}\,x(s-d)\sin\Bigl(\tfrac{1}{2}x(s-d)\Bigr)
+18\sqrt{2}
\Biggr)
\Biggl(
-\tfrac{9}{2}\,x^{3}(d+s)^{3}\,\mathrm{Si}\Bigl(\tfrac{1}{2}(d+s)x\Bigr)
-\tfrac{9}{8}\,\notag\\& \times x^{4}(d+s)^{4}
+\Bigl(\tfrac{9}{8}\,x^{4}(d+s)^{4}
+\tfrac{9\,x^{2}(d+s)^{2}}{2\sqrt{2}}
-18\sqrt{2}\Bigr)\cos\Bigl(\tfrac{1}{2}x(d+s)\Bigr)
-9\sqrt{2}\,x(d+s)\sin\Bigl(\tfrac{1}{2}x(d+s)\Bigr)
+18\sqrt{2}
\Biggr) 
\notag\\&+2\Biggl(
-\tfrac{9}{2}\,x^{3}(s-d)^{3}\Bigl(
\sin\Bigl(\tfrac{1}{2}x(s-d)\Bigr)
-2\,\mathrm{Si}\Bigl(\tfrac{1}{2}(s-d)x\Bigr)\Bigr)
+\tfrac{9}{8}\,x^{4}(s-d)^{4}
+\tfrac{9}{4}\,x^{2}(s-d)^{2}\Bigl(4\cos\Bigl(\tfrac{1}{2}x(s-d)\Bigr)\notag\\&+\sqrt{2}-4\Bigr)
-9\sqrt{2}\,x(s-d)\sin\Bigl(\tfrac{1}{2}x(s-d)\Bigr)
-18\sqrt{2}\Bigl(\cos\Bigl(\tfrac{1}{2}x(s-d)\Bigr)-1\Bigr)
\Biggr)
\Biggl(
-\tfrac{9}{2}\,x^{3}(d+s)^{3}\Bigl(
\sin\Bigl(\tfrac{1}{2}x(d+s)\Bigr)
\notag\\&-2\,\mathrm{Si}\Bigl(\tfrac{1}{2}(d+s)x\Bigr)\Bigr)
+\tfrac{9}{8}\,x^{4}(d+s)^{4}
+\tfrac{9}{4}\,x^{2}(d+s)^{2}\Bigl(4\cos\Bigl(\tfrac{1}{2}x(d+s)\Bigr)+\sqrt{2}-4\Bigr)d
-9\sqrt{2}\,x(d+s)\notag\\&\times\sin\Bigl(\tfrac{1}{2}x(d+s)\Bigr)
-18\sqrt{2}\Bigl(\cos\Bigl(\tfrac{1}{2}x(d+s)\Bigr)-1\Bigr)
\Biggr)
\Biggr). 
\end{align} 
\begin{figure}[th]
\centering
\begin{minipage}[t]{0.48\linewidth}
\centering 
\includegraphics[width=\linewidth]{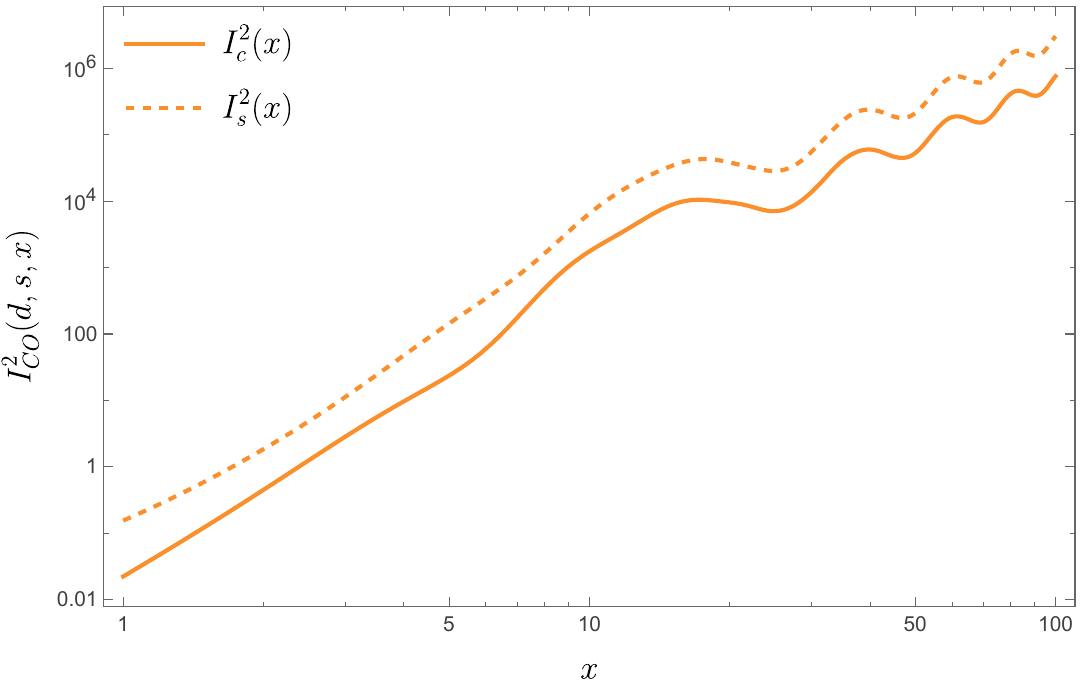}
\caption{$I_c^2(x)$ (solid) and $I_s^2(x)$ (dashed) in the CO gauge at $(d,s)=(0,1/\sqrt{3})$.
The kernels inherit the late-time growth of the source and increase toward large $x$ (here roughly $I^2\!\propto x^{4}$), showing a non–convergent late–time behaviour in this gauge.}
\label{fig:CO_kernels}
\end{minipage}
\hfill
\begin{minipage}[t]{0.48\linewidth}
\centering
\includegraphics[width=\linewidth]{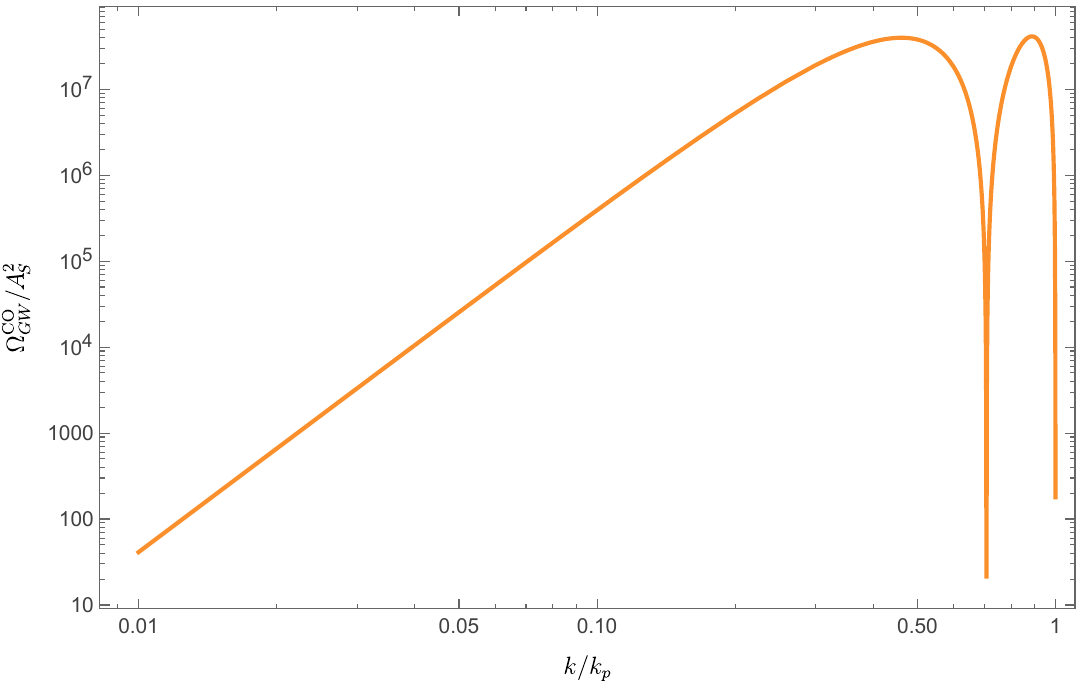}
\caption{$\Omega_{\rm GW}(k)/A_s^2$ in the CO gauge for a Dirac–delta isocurvature peak at $k_p$.
The evaluation follows the delta line in $(d,s)$: $d=0$ and $s=\tfrac{2}{\sqrt{3}}(k_p/k)$.
The late–time evolution grows with $x$, i.e.\ it is \emph{divergent} relative to longitudinal.
The curve exhibits the usual rise at $k\ll k_p$, a peak near $k=2c_s k_p$ with $c_s=1/\sqrt{3}$,
and a sharp cutoff at $k=2k_p$.
The GW spectrum is normalized by the scalar power spectrum amplitude $A_s^2$.
Values exceeding unity arise from gauge-dependent, non-radiative contributions,
signaling the breakdown of linear perturbation theory prior to the radiative projection.}
\label{fig:CO_spectrum}
\end{minipage} 
\end{figure}
By substituting \eqref{transcom} into \eqref{xitransfer}, we have
\begin{align}
I^\text{CO}_{\chi}(d,s,x)&= \frac{16}{243\,x^{4}\,(s-d)^{3}(d+s)^{3}}
\Biggl(x^{2}\Bigl(\tfrac{3}{4}(s-d)^{2}+\tfrac{3}{4}(d+s)^{2}-1\Bigr)
\Bigl(6x(s-d)\,\mathrm{Si}\Bigl(\tfrac{1}{2}(s-d)x\Bigr)
+\tfrac{3}{4}x^{2}(s-d)^{2}-3x(s-d)\notag\\ 
& \quad \times \sin\Bigl(\tfrac{1}{2}x(s-d)\Bigr)
+6\cos\Bigl(\tfrac{1}{2}x(s-d)\Bigr)-6\Bigr)\Bigl(
6x(d+s)\,\mathrm{Si}\Bigl(\tfrac{1}{2}(d+s)x\Bigr)
+\tfrac{3}{4}x^{2}(d+s)^{2} -3x(d+s)\sin\Bigl(\tfrac{1}{2}x(d+s)\Bigr)\notag\\ 
& \quad 
+6\cos\Bigl(\tfrac{1}{2}x(d+s)\Bigr)-6\Bigr)+6\sqrt{2}\,\Bigl(\tfrac{3}{4}x^{2}(d+s)^{2}
-3x(d+s) \sin\Bigl(\tfrac{1}{2}x(d+s)\Bigr)
-6\cos\Bigl(\tfrac{1}{2}x(d+s)\Bigr)+6
\Bigr)\Bigl(6x(s-d)\,\notag\\ 
& \quad \times\mathrm{Si}\Bigl(\tfrac{1}{2}(s-d)x\Bigr) +\tfrac{3}{4}x^{2}(s-d)^{2}
-3x(s-d)\sin\Bigl(\tfrac{1}{2}x(s-d)\Bigr)
+6\cos\Bigl(\tfrac{1}{2}x(s-d)\Bigr)-6\Bigr)+6\sqrt{2}\,
\Bigl(\tfrac{3}{4}x^{2}(s-d)^{2}
-3x\notag\\ 
& \quad \times (s-d)\sin\Bigl(\tfrac{1}{2}x(s-d)\Bigr)
-6\cos\Bigl(\tfrac{1}{2}x(s-d)\Bigr)+6\Bigr)\Bigl(
6x(d+s)\,\mathrm{Si}\Bigl(\tfrac{1}{2}(d+s)x\Bigr)+\tfrac{3}{4}x^{2}(d+s)^{2}
-3x(d+s)\notag\\ 
& \quad \times \sin\Bigl(\tfrac{1}{2}x(d+s)\Bigr)
+6\cos\Bigl(\tfrac{1}{2}x(d+s)\Bigr)-6\Bigr)\Biggr),
\end{align} 
and the analytic expression for the kernels $I^{\mathrm{CO}}_c$ and $I^{\mathrm{CO}}_s$  in the comoving orthogonal gauge are  
\begin{align}
I_c^{\mathrm{CO}}(d,s,x)
&= \frac{3}{4\,\kappa^{2}}
\Biggl\{
\frac{8}{\bigl(d^{2}-s^{2}\bigr)^{3}\,x^{4}}
\Bigl[\sin x\,\Bigl(
-4\,(d^{2}+s^{2}-1)\,x^{2}
+s\Bigl(\bigl((d^{2}+3s^{2}-4)\,x^{2}+24\bigr)\cos\tfrac{d x}{2}\Bigl)\notag\\
&\quad
+12\,d\,x\,\sin\tfrac{d x}{2}\Bigr)\,
\sin\tfrac{s x}{2}\,x
+2d\bigl(-(s^{2}-1)\,x^{2}-6\bigr)\sin(d x)\,x
+2s\bigl(-(d^{2}-1)\,x^{2}-6\bigr)\notag\\
&\quad\times\sin(s x)\,x
+2\bigl((2d^{2}-s^{2}+1)\,x^{2}-6\bigr)\cos(d x)
+2\bigl((-d^{2}+2s^{2}+1)\,x^{2}-6\bigr)\cos(s x)\notag\\
&\quad
+\cos\tfrac{s x}{2}\Bigl(
2\bigl((d^{2}+s^{2}-4)\,x^{2}+24\bigr)\cos\tfrac{d x}{2}
+d\,x\,\bigl((3d^{2}+s^{2}-4)\,x^{2}+24\bigr)\sin\tfrac{d x}{2}
\Bigr)
-24\Bigr)\notag\\ 
&\quad
+2x\cos x\,\Bigl(
-\bigl((s^{2}-1)\,x^{2}+2\bigr)\cos(d x)
-\bigl((d^{2}-1)\,x^{2}+2\bigr)\cos(s x)\notag\\
&\quad
+\cos\tfrac{s x}{2}\Bigl(
\bigl((3d^{2}+3s^{2}-4)\,x^{2}+8\bigr)\cos\tfrac{d x}{2}
+4d x \sin\tfrac{d x}{2}
\Bigr)
+2s x\Bigl(2\cos\tfrac{d x}{2}+d x \sin\tfrac{d x}{2}\Bigr)\notag\\
&\quad\times\sin\tfrac{s x}{2}
-2\bigl((d^{2}+s^{2}-1)\,x^{2}+d\sin(d x)\,x+2\bigr)
-2s x \sin(s x) 
\Bigr)
\Bigr]\notag\\[2pt]
&\quad
+\frac{2}{(d-s)^{3}(d+s)^{3}}
\Bigl[
\mathrm{Si}\bigl((d+1)x\bigr)\,(d^{2}+s^{2}-2)^{2}
+\mathrm{Si}\bigl((s+1)x\bigr)\,(d^{2}+s^{2}-2)^{2}\notag\\
&\quad
+\mathrm{Si}\bigl((1-d)x\bigr)\,(d^{2}+s^{2}-2)^{2}
+\mathrm{Si}\bigl((1-s)x\bigr)\,(d^{2}+s^{2}-2)^{2}
\Bigr]\notag\\
&\quad
-\bigl(3d^{4}+4sd^{3}+2(s^{2}-8)d^{2}+4s^{3}d+3s^{4}-16s^{2}+16\bigr)
\mathrm{Si}\Bigl(\tfrac{-d-s+2}{2}\,x\Bigr)\notag\\
&\quad
+4\bigl(d^{4}-2(s^{2}+4)d^{2}+s^{4}-8s^{2}+8\bigr)\mathrm{Si}(x)\notag\\
&\quad
+\bigl(-3d^{4}+4sd^{3}-2(s^{2}-8)d^{2}+4s^{3}d-3s^{4}+16s^{2}-16\bigr)
\mathrm{Si}\Bigl(\tfrac{d-s+2}{2}\,x\Bigr)\notag\\
&\quad
+\bigl(-3d^{4}+4sd^{3}-2(s^{2}-8)d^{2}+4s^{3}d-3s^{4}+16s^{2}-16\bigr)
\mathrm{Si}\Bigl(\tfrac{-d+s+2}{2}\,x\Bigr)\notag\\
&\quad
-\bigl(3d^{4}+4sd^{3}+2(s^{2}-8)d^{2}+4s^{3}d+3s^{4}-16s^{2}+16\bigr)
\mathrm{Si}\Bigl(\tfrac{d+s+2}{2}\,x\Bigr)\notag\\[2pt]
&\quad \times
\frac{16}{243\,x^{4}\,(s-d)^{3}(d+s)^{3}}
\Bigl\{\;x^{2}\Bigl[\tfrac{3}{4}(s-d)^{2}+\tfrac{3}{4}(d+s)^{2}-1\Bigr]
\Bigl[
6x(s-d)\,\mathrm{Si}\Bigl(\tfrac{1}{2}(s-d)x\Bigr)
\notag\\
&\quad+\tfrac{3}{4}x^{2}(s-d)^{2} 
-3x(s-d)\sin\Bigl(\tfrac{1}{2}x(s-d)\Bigr)
+6\cos\Bigl(\tfrac{1}{2}x(s-d)\Bigr)-6
\Bigr]
\Bigl[
6x(d+s)\, \notag\\
&\quad\times\mathrm{Si}\Bigl(\tfrac{1}{2}(d+s)x\Bigr)
+\tfrac{3}{4}x^{2}(d+s)^{2}
x(d+s)\sin\Bigl(\tfrac{1}{2}x(d+s)\Bigr)+6\cos\Bigl(\tfrac{1}{2}x(d+s)\Bigr)-6
\Bigr] \notag\\[2pt]
&\quad\;+\;6\sqrt{2}\, 
\Bigl[
\tfrac{3}{4}x^{2}(d+s)^{2}
-3x(d+s)\sin\Bigl(\tfrac{1}{2}x(d+s)\Bigr)
			-6\cos\Bigl(\tfrac{1}{2}x(d+s)\Bigr)+6
			\Bigr] \notag\\
			&\quad\times
			\Bigl[
			6x(s-d)\,\mathrm{Si}\Bigl(\tfrac{1}{2}(s-d)x\Bigr)
			+\tfrac{3}{4}x^{2}(s-d)^{2}
			-3x(s-d)\sin\Bigl(\tfrac{1}{2}x(s-d)\Bigr) \notag\\
			&\quad
			+6\cos\Bigl(\tfrac{1}{2}x(s-d)\Bigr)-6
			\Bigr]+\;6\sqrt{2}\,
			\Bigl[ 
			\tfrac{3}{4}x^{2}(s-d)^{2}
			-3x(s-d)\sin\Bigl(\tfrac{1}{2}x(s-d)\Bigr)
			\notag\\[2pt]
			&\quad\;-6\cos\Bigl(\tfrac{1}{2}x(s-d)\Bigr)+6
			\Bigr] \times  
			\Bigl[ 
			6x(d+s)\,\mathrm{Si}\Bigl(\tfrac{1}{2}(d+s)x\Bigr)
			+\tfrac{3}{4}x^{2}(d+s)^{2} 
			-3x(d+s)\notag\\ 
			&\quad\times\sin\Bigl(\tfrac{1}{2}x(d+s)\Bigr) 
			+6\cos\Bigl(\tfrac{1}{2}x(d+s)\Bigr)-6
			\Bigr]  
			\Bigr\}, 
\label{IcCO}
\end{align} 
and  
\begin{align}
I_s^{\mathrm{CO}}(d,s,x)
&= \frac{3}{4\,\kappa^{2}\,(d-s)^{3}(d+s)^{3}}
\Biggl\{
-2\,\mathrm{Ci}\bigl(x\lvert 1-d\rvert\bigr)\,(d^{2}+s^{2}-2)^{2}
-2\,\mathrm{Ci}\bigl(x\lvert d+1\rvert\bigr)\,(d^{2}+s^{2}-2)^{2}
\notag\\
&\quad
-2\,\mathrm{Ci}\bigl(x\lvert 1-s\rvert\bigr)\,(d^{2}+s^{2}-2)^{2}
-2\,\mathrm{Ci}\bigl(x\lvert s+1\rvert\bigr)\,(d^{2}+s^{2}-2)^{2}
-4\bigl(d^{4}-2(s^{2}+4)d^{2}\notag\\
&\quad+s^{4}-8s^{2}+8\bigr)\mathrm{Ci}(x)
+\bigl(3d^{4}+4sd^{3}+2(s^{2}-8)d^{2}+4s^{3}d+3s^{4}-16s^{2}+16\bigr)\notag\\
&\quad
\times\mathrm{Ci}\Bigl(\tfrac{x}{2}\,\lvert{-}d{-}s{+}2\rvert\Bigr)
+\bigl(3d^{4}-4sd^{3}+2(s^{2}-8)d^{2}-4s^{3}d+3s^{4}-16s^{2}+16\bigr)\notag\\
&\quad
\times\mathrm{Ci}\Bigl(\tfrac{x}{2}\,\lvert d{-}s{+}2\rvert\Bigr)
+\bigl(3d^{4}-4sd^{3}+2(s^{2}-8)d^{2}-4s^{3}d+3s^{4}-16s^{2}+16\bigr)\notag\\
&\quad
\times\mathrm{Ci}\Bigl(\tfrac{x}{2}\,\lvert{-}d{+}s{+}2\rvert\Bigr)
+\bigl(3d^{4}+4sd^{3}+2(s^{2}-8)d^{2}+4s^{3}d+3s^{4}-16s^{2}+16\bigr)\notag\\
&\quad
\times\mathrm{Ci}\Bigl(\tfrac{x}{2}\,\lvert d{+}s{+}2\rvert\Bigr)
\Biggr\}
-\frac{6}{(d^{2}-s^{2})^{3}\,x^{4}\,\kappa^{2}}
\Biggl\{
2x\sin x\Bigl(
((s^{2}-1)x^{2}+2)\cos(d x)
+((d^{2}-1)x^{2}\notag\\
&\quad+2)\cos(s x)
			+\cos\tfrac{s x}{2}\Bigl(-((3d^{2}+3s^{2}-4)x^{2}+8)\cos\tfrac{d x}{2}-4d x\sin\tfrac{d x}{2}\Bigr)
			-2s x\bigl(2\cos\tfrac{d x}{2}\notag\\
			&\quad+d x\sin\tfrac{d x}{2}\bigr)\sin\tfrac{s x}{2}
			+2\bigl((d^{2}+s^{2}-1)x^{2}+d\sin(d x)\,x+2\bigr)
			+2s x\sin(s x)
			\Bigr)
			\notag\\
			&\quad
			+\cos x\Bigl(
			-4(d^{2}+s^{2}-1)x^{2}
			+s\bigl((d^{2}+3s^{2}-4)x^{2}+24\bigr)\cos\tfrac{d x}{2}\,\sin\tfrac{s x}{2}\,x
			\notag\\
			&\quad
			+12d x\,\sin\tfrac{d x}{2}\,\sin\tfrac{s x}{2}\,x
			+2d(-s^{2}x^{2}+x^{2}-6)\sin(d x)\,x
			+2s(-d^{2}x^{2}+x^{2}-6)\sin(s x)\,x
			\notag\\
			&\quad
			+2\bigl((2d^{2}-s^{2}+1)x^{2}-6\bigr)\cos(d x)
			+2\bigl((-d^{2}+2s^{2}+1)x^{2}-6\bigr)\cos(s x)
			\notag\\
			&\quad
			+\cos\tfrac{s x}{2}\Bigl(
			2\bigl((d^{2}+s^{2}-4)x^{2}+24\bigr)\cos\tfrac{d x}{2}
			+d x\bigl((3d^{2}+s^{2}-4)x^{2}+24\bigr)\sin\tfrac{d x}{2}
			\Bigr)
			-24
			\Bigr)
			\Biggr\}
			\notag\\[2pt]
			&\quad
			+ \frac{16}{243\,x^{4}\,(s-d)^{3}(d+s)^{3}}
			\Bigl\{
			x^{2}\Bigl[\tfrac{3}{4}(s-d)^{2}+\tfrac{3}{4}(d+s)^{2}-1\Bigr]
			\Bigl[
			6x(s-d)\,\times\mathrm{Si}\Bigl(\tfrac{1}{2}(s-d)x\Bigr)\notag\\
			&\quad
			+\tfrac{3}{4}x^{2}(s-d)^{2}  
			-3x(s-d)\sin\Bigl(\tfrac{1}{2}x(s-d)\Bigr)
			+6\cos\Bigl(\tfrac{1}{2}x(s-d)\Bigr)-6
			\Bigr]
			\Bigl[
			6x(d+s)\,\notag\\
			&\quad\times\mathrm{Si}\Bigl(\tfrac{1}{2}(d+s)x\Bigr)
			+\tfrac{3}{4}x^{2}(d+s)^{2} 
			-3x(d+s)\sin\Bigl(\tfrac{1}{2}x(d+s)\Bigr)
			+6\cos\Bigl(\tfrac{1}{2}x(d+s)\Bigr)-6
			\Bigr]
			\notag\\[2pt] 
			&\quad
			+6\sqrt{2}\,
			\Bigl[
			\tfrac{3}{4}x^{2}(d+s)^{2}
			-3x(d+s)\sin\Bigl(\tfrac{1}{2}x(d+s)\Bigr)
			-6\cos\Bigl(\tfrac{1}{2}x(d+s)\Bigr)+6
			\Bigr] \notag\\
			&\quad\times
			\Bigl[
			6x(s-d)\,\mathrm{Si}\Bigl(\tfrac{1}{2}(s-d)x\Bigr)
			+\tfrac{3}{4}x^{2}(s-d)^{2}
			-3x(s-d)\sin\Bigl(\tfrac{1}{2}x(s-d)\Bigr) \notag\\
			&\quad
			+6\cos\Bigl(\tfrac{1}{2}x(s-d)\Bigr)-6
			\Bigr]
			+6\sqrt{2}\,
			\Bigl[
			\tfrac{3}{4}x^{2}(s-d)^{2}
			-3x(s-d)\sin\Bigl(\tfrac{1}{2}x(s-d)\Bigr)
			\notag\\[2pt]
			&\quad-6\cos\Bigl(\tfrac{1}{2}x(s-d)\Bigr)+6
			\Bigr] \times 
			\Bigl[
			6x(d+s)\,\mathrm{Si}\Bigl(\tfrac{1}{2}(d+s)x\Bigr)
+\tfrac{3}{4}x^{2}(d+s)^{2}
-3x(d+s)\notag\\
&\quad\times\sin\Bigl(\tfrac{1}{2}x(d+s)\Bigr) 
+6\cos\Bigl(\tfrac{1}{2}x(d+s)\Bigr)-6
\Bigr] \Bigr\}. 
\label{IsCO}
\end{align}
\end{widetext}

In radiation domination (RD),  we insert the CO source into the kernel integrals \eqref{ICS}.
The exact analytic expressions for $I_c^{(\mathrm{CO})}(d,s,x)$ and $I_s^{(\mathrm{CO})}(d,s,x)$ are  and given Eqs. \eqref{IcCO} and  \eqref{IsCO}; the figures are obtained by direct evaluation of the exact formulas, and we also use their late–time limits $I_{c,\infty}^{(\mathrm{CO})}(d,s)$ and $I_{s,\infty}^{(\mathrm{CO})}(d,s)$ in \eqref{eq:Ibar2}.
At $(d,s)=(0,1/\sqrt{3})$ the kernels grow as $I\sim x^{2}$ (hence $I^{2}\sim x^{4})$, so the late–time evolution of $\Omega_{\rm GW}$ increases roughly as $x^{6}$.
This growth originates from non–luminal pieces that do not decay in this slicing and represents a clear divergence compared with the longitudinal case. The corresponding behavior of the source function is illustrated in Fig.~\ref{fig:CO_source}, and a comparison of the resulting kernels is shown in Fig.~\ref{fig:CO_kernels}. 
For a Dirac–delta primordial peak, $\mathcal{P}_{S}(k)=\mathcal{A}_{S}\,\delta(\ln(k/k_p))$, we evaluate \eqref{eq:OmegaGW-dsg2} along the delta line $d=0$, $s=\tfrac{2}{\sqrt{3}}(k_p/k)$; see Fig.~\ref{fig:CO_spectrum}.
A gauge–independent treatment is presented later in Sec.~\ref{sec:gauge-independent}.

\subsection{Secondary GWs in Synchronous (TT) gauge} \label{s-gauge}Within the framework of the TT gauge, the metric perturbations conform to $\phi=B=0$, resulting in the retention of solely the $ij$ components in the perturbed metric. During the RD phase, the solutions to the transfer functions are articulated by   
\begin{align}
T_{E}(x)
&= -\frac{3}{\sqrt{2}\,\kappa}\Biggl(
\frac{x^{2}}{6}
- 3\,\mathrm{Ci}\Bigl(\tfrac{x}{\sqrt{3}}\Bigr)
+ \frac{3\,\sin\bigl(\tfrac{x}{\sqrt{3}}\bigr)}{x}\notag
\\
&\quad
- \frac{3}{\sqrt{3}}\,\mathrm{Si}\Bigl(\tfrac{x}{\sqrt{3}}\Bigr)
\Biggr),\notag
\\[4pt]
T_{\sigma}(x)
&= -\frac{3}{\sqrt{2}\,\kappa}\Bigl(
\tfrac{1}{2}-\tfrac{3}{x^{2}}
+\tfrac{3}{x^{2}}\cos\tfrac{x}{\sqrt{3}}
\Bigr),\notag
\\[4pt]
T_{\psi}(x)
&= \frac{3}{\sqrt{2}\,\kappa}\Bigl(
1-\tfrac{\sqrt{3}}{x}\sin\tfrac{x}{\sqrt{3}}
\Bigr).  
\label{transfertt}\end{align}   
\begin{figure}[th]
\centering
\includegraphics[width=\linewidth]{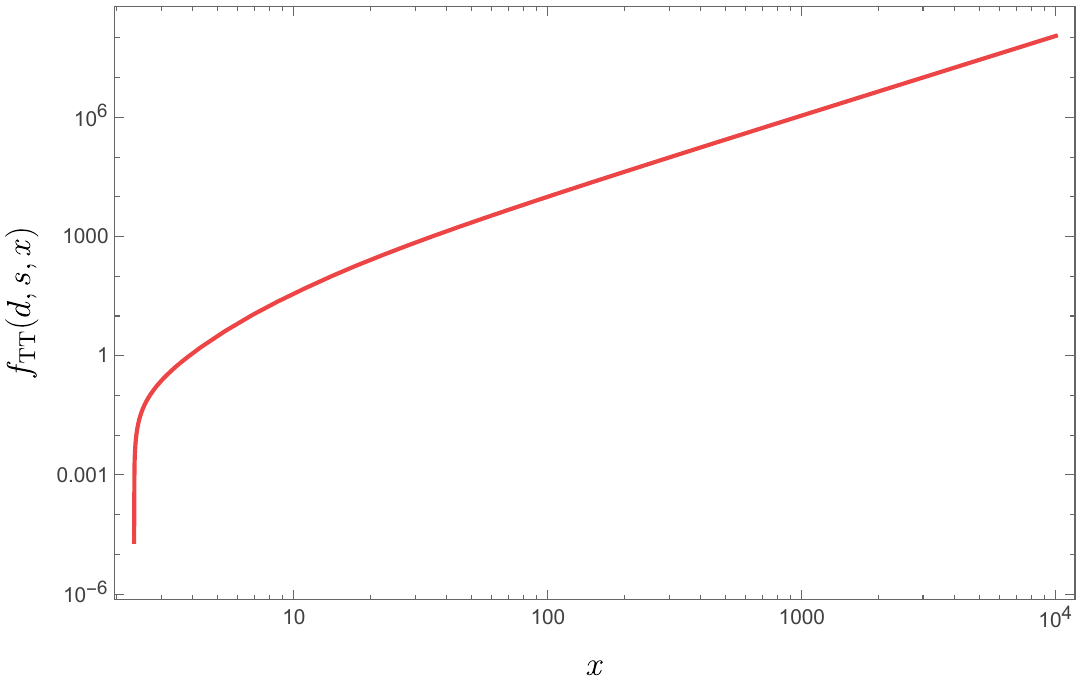}
\caption{Source term $f_{\rm TT}(d,s,x)$ at $(d,s)=(0,1/\sqrt{3})$ versus the dimensionless time variable $x\equiv k\eta$.
After horizon entry the profile remains oscillatory but its evolution grows steadily over several decades in $x$,
indicating strong late–time growth sourcing in this slicing.}
\label{fig:TT_source}
\end{figure}
\begin{figure*}[th]
\centering
\begin{minipage}[t]{0.48\linewidth}
\centering
\includegraphics[width=\linewidth]{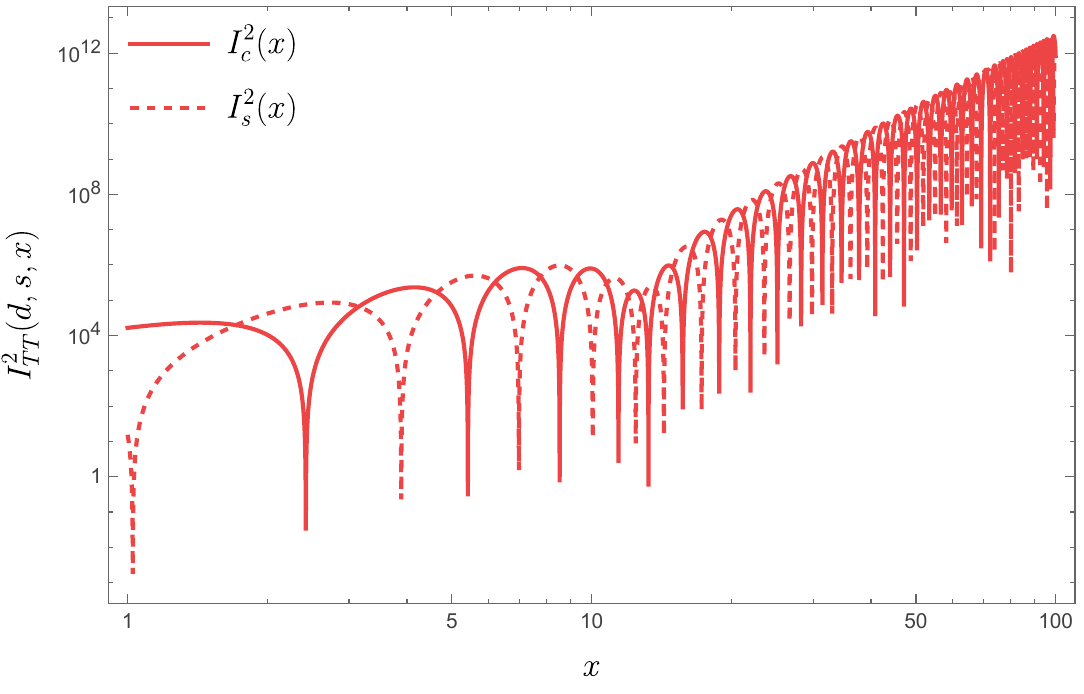}
\caption{Squared kernels in the TT gauge, $I_c^{2}(x)$ (solid) and $I_s^{2}(x)$ (dashed), at $(d,s)=(0,1/\sqrt{3})$.
A pronounced large–$x$ background develops instead of a bounded evolution, tracking the late-time growth of the source
and signalling a non–convergent late–time readout in this gauge.}
\label{fig:TT_kernels} 
\end{minipage} 
\hfill
\begin{minipage}[t]{0.48\linewidth}
\centering
\includegraphics[width=\linewidth]{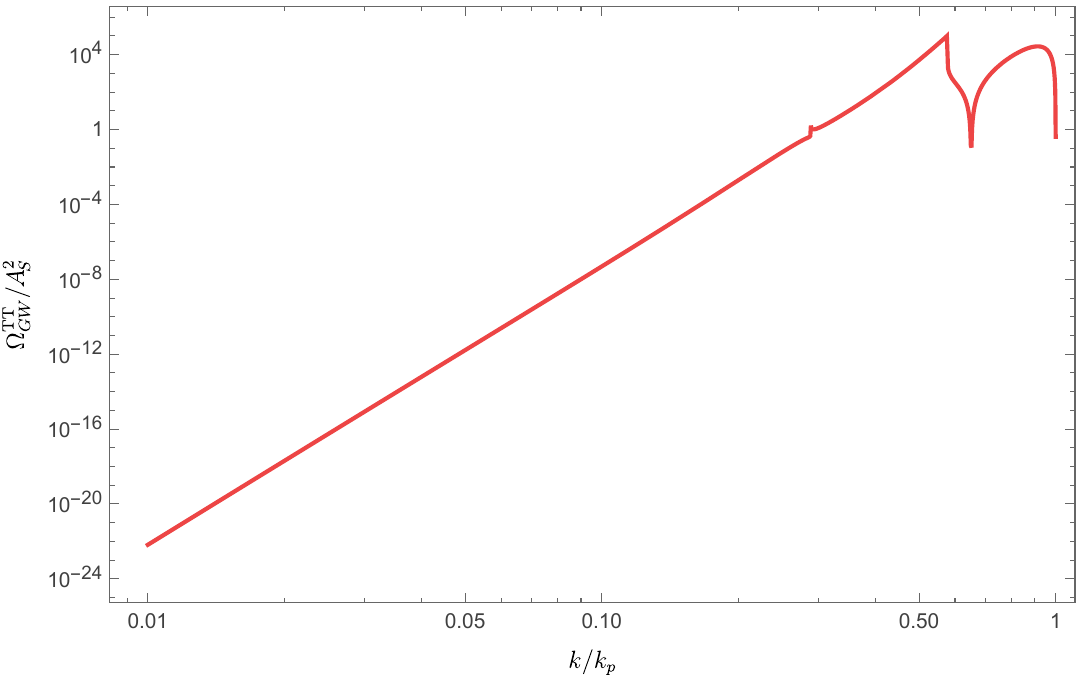} 
\caption{$\Omega_{\rm GW}(k)/A_s^2$ from the TT–gauge kernels for a Dirac–delta isocurvature peak at $k_p$.
The IR rise and the familiar peak near $k=2c_s k_p$ with the UV cutoff at $k=2k_p$ are visible,
but the pre–projection late–time evolution increases with $x$ and is \emph{divergent} relative
to the longitudinal baseline.
The GW spectrum is normalized by the scalar power spectrum amplitude $A_s^2$.
The apparent $\Omega_{\rm GW}>1$ behavior reflects gauge-dependent,
non-radiative terms and indicates the breakdown of linear perturbation theory before projection.} 
\label{fig:TT_spectrum}
\end{minipage}
\end{figure*}      
Furthermore, within the context of the TT gauge, by employing the gauge conditions in perturbation theory, the transfer function corresponding to the coordinate transformation can be explicitly evaluated. Consequently, with the use of the initial conditions, the non-zero transfer functions are determined as follows:   
\begin{align}
T_{\alpha}(x)
&= -\frac{3}{\sqrt{2}\,\kappa}\Bigl(
\frac{x}{3}
- \frac{3}{x}\cos\tfrac{x}{\sqrt{3}}
+ \frac{3}{x^{2}}\sin\tfrac{x}{\sqrt{3}}
\Bigr),
\notag\\[4pt]
T_{\beta}(x)
&= -\frac{3}{\sqrt{2}\,\kappa}\Bigl(
\frac{x^{2}}{6}
- 3\,\mathrm{Ci}\Bigl(\tfrac{x}{\sqrt{3}}\Bigr)
+ \frac{3}{x}\sin\tfrac{x}{\sqrt{3}}\notag
\\&-\frac{3}{\sqrt{3}}\,\mathrm{Si}\Bigl(\tfrac{x}{\sqrt{3}}\Bigr)
\Bigr).\label{aplabetatt} 
\end{align}			

In the transverse-traceless (TT) gauge, gauge modes also contribute to the transfer function $T_E(x)$, due to residual gauge freedom during the radiation-dominated era. By accounting for these residual gauge transformations, the source function inherits contributions from the transfer function structure. As a result, the source term in the TT gauge can be analytically evaluated, and the expression takes the following form:
\begin{widetext}
\begin{align}
f^{\mathrm{TT}}(d, s, x) &= 
\frac{32}{27 (s-d)^4 (d+s)^4 x^5 \kappa ^2}
\Bigl(
\tfrac92 x
\Bigl(
\tfrac92 (s-d)^2 \Bigl(\tfrac34 (s-d)^2+\tfrac34 (d+s)^2-1\Bigr)\,
\text{Si}\Bigl(\tfrac12(s-d)x\Bigr)\,x^4
\notag\\
&\quad+\tfrac32(s-d)
\Bigl(
-\tfrac34 (s-d)^2\Bigl(\tfrac34 (s-d)^2+\tfrac34 (d+s)^2-1\Bigr)x^4
			-2\Bigl(\tfrac{27}{64}(s-d)^6-\tfrac{9}{16}(s-d)^4\notag\\
			&\quad+\tfrac34(s-d)^2
			+\tfrac98(d+s)^4
			+\tfrac34(d+s)^2\bigl(\tfrac{9}{16}(s-d)^4+3(s-d)^2+1\bigr)-3\Bigr)x^2
			+4\bigl(\tfrac34 (s-d)^2\notag\\
			&\quad-1\bigr)\bigl(\tfrac34 (s-d)^2+\tfrac94 (d+s)^2-3\bigr)
			\Bigr)x
			+3(s-d)\Bigl(
			-\tfrac{45}{8}(s-d)^4+15(s-d)^2
			+3\bigl(\tfrac34 (s-d)^2\notag\\
			&\quad+\tfrac34 (d+s)^2-1\bigr)x^2
			-\tfrac92(d+s)^2\bigl(\tfrac34 (s-d)^2-1\bigr)-6
			\Bigr)\cos\Bigl(\tfrac12(s-d)x\Bigr)x
			+2\Bigl(
			\tfrac34\Bigl(
			18(s-d)^2\notag\\
			&\quad+\bigl(\tfrac{15}{4}(s-d)^2+\tfrac{45}{4}(d+s)^2-9\bigr)x^2-36
			\Bigr)(s-d)^2
			+\tfrac92(d+s)^2\bigl(\tfrac34 (d+s)^2-1\bigr)x^2
			\Bigr)\notag\\
			&\quad \times\sin\Bigl(\tfrac12(s-d)x\Bigr)
			\Bigr)
			\text{Si}\Bigl(\tfrac12(d+s)x\Bigr)(d+s)^2
			+3\cos\Bigl(\tfrac12(d+s)x\Bigr)
			\Bigl(
			2\Bigl(
			\tfrac{9}{16}(5x^2+24)(s-d)^4
			\notag\\
			&\quad+\tfrac34\bigl(\tfrac34(d+s)^2(13x^2+12)-9(x^2+6)\bigr)(s-d)^2
			+\tfrac92(d+s)^2\bigl(\tfrac34(d+s)^2-1\bigr)x^2
			\Bigr)\notag\\
			&\quad\times \sin\Bigl(\tfrac12(s-d)x\Bigr)\sqrt{3}
			+\tfrac12(s-d)x
			\Bigl(
			\tfrac98\bigl(\bigl(\tfrac{15}{4}(s-d)^2-6\bigr)x^2+30\bigr)(d+s)^4
			-\tfrac34\bigl(\tfrac34\bigl(3x^4\notag\\
			&\quad+35x^2-48\bigr)(s-d)^2+6(x^2+33)\bigr)(d+s)^2
			+18(x^2+8)
			+\tfrac94(s-d)^2\Bigl(
			-\bigl(\bigl(\tfrac34(s-d)^2\notag\\
			&\quad-1\bigr)x^4\bigr)
			-2\bigl(\tfrac{9}{16}(s-d)^4-\tfrac34(s-d)^2+2\bigr)x^2
			+3(s-d)^2-30
			\Bigr)
			+3(s-d)x
			\Bigl(
			-\tfrac{45}{8}(d+s)^4\notag\\
			&\quad
			+\tfrac34\bigl(-\tfrac92(s-d)^2+3x^2+20\bigr)(d+s)^2
			+3\bigl(\tfrac34(s-d)^2-1\bigr)(x^2+2)
			\Bigr)
			\text{Si}\Bigl(\tfrac12(s-d)x\Bigr)
			\Bigr)\sqrt{3}
			\notag\\
			&\quad+(s-d)x
			\Bigl(
			\tfrac{135}{8}(s-d)^4-\tfrac{27}{4}(x^2+11)(s-d)^2
			+\tfrac{135}{8}(d+s)^4+9(x^2+8)
			\notag\\
			&\quad-\tfrac34(d+s)^2\bigl(\tfrac34(x^2-36)(s-d)^2+9(x^2+11)\bigr)
			\Bigr)\cos\Bigl(\tfrac12(s-d)x\Bigr)(-\sqrt{3})
			\Bigr)\sqrt{3}(d+s)
			\notag\\
			&\quad+\tfrac12\Bigl(
			3(s-d)x
			\Bigl(
			\tfrac{1}{32}(-81)x^2(d+s)^6
			-\tfrac{27}{16}(x^4-2x^2-4)(d+s)^4
			+\tfrac34\Bigl(
			-3\bigl(\tfrac34(s-d)^2-1\bigr)x^4
			\notag\\
			&\quad+\bigl(\tfrac{45}{8}(s-d)^4-\tfrac{105}{4}(s-d)^2-12\bigr)x^2
			+36(s-d)^2-90
			\Bigr)(d+s)^2
			+18(x^2+8)
			\notag\\
			&\quad-\tfrac92(s-d)^2\bigl(\tfrac32(x^2-5)(s-d)^2+x^2+33\bigr)
			\Bigr)\cos\Bigl(\tfrac12(s-d)x\Bigr)\sqrt{3}
			+2\Bigl(
			-\tfrac{9}{16}(5x^2+24)\notag\\
			&\quad\times\bigl(\tfrac34(d+s)^2x^2+6\bigr)(s-d)^4
			+\tfrac94\Bigl(
			\tfrac34\bigl((3-\tfrac{15}{4}(d+s)^2)x^4
			+\bigl(3(d+s)^2-22\bigr)x^2-24\bigr)(d+s)^2
			\notag\\
			&\quad+18(x^2+6)
			\Bigr)(s-d)^2
			-\tfrac92(d+s)^2\bigl(\tfrac34(d+s)^2-1\bigr)x^2
			\bigl(\tfrac34(d+s)^2x^2+6\bigr)
			\Bigr)\sin\Bigl(\tfrac12(s-d)x\Bigr)\sqrt{3}
			\notag\\
			&\quad+\tfrac32(s-d)x
			\Bigl(
			\tfrac{9}{16}(s-d)^2(d+s)^2\bigl(\tfrac34(s-d)^2+\tfrac34(d+s)^2-1\bigr)x^6
			+2\Bigl(\tfrac{27}{32}(s-d)^6+\tfrac{9}{16}(s-d)^4\notag\\
			&\quad-\tfrac94(s-d)^2
			+\tfrac{27}{64}(d+s)^6\bigl(\tfrac34(s-d)^2+2\bigr)
			+\tfrac{9}{16}(d+s)^4\bigl(\tfrac98(s-d)^4+3(s-d)^2+1\bigr)
			+\tfrac34(d+s)^2\notag\\
			&\quad\times\bigl(\tfrac{27}{64}(s-d)^6+\tfrac94(s-d)^4-3\bigr)\Bigr)x^4
			+4\Bigl(
			\tfrac{81}{64}(d+s)^6
			-\tfrac{27}{16}\bigl(\tfrac34(s-d)^2-1\bigr)(d+s)^4
			+\tfrac{9}{16}(s-d)^2\notag\\
			&\quad\times\bigl(29-\tfrac94(s-d)^2\bigr)(d+s)^2
			+3\bigl(\tfrac{27}{64}(s-d)^6+\tfrac{9}{16}(s-d)^4-3\bigr)\Bigr)x^2
			-3(s-d)
			\Bigl(
			\tfrac34(d+s)^2\bigl(\tfrac34\notag\\
			&\quad\times(s-d)^2+\tfrac34(d+s)^2-1\bigr)x^4
			+2\Bigl(
			\tfrac{27}{64}(d+s)^6-\tfrac{9}{16}(d+s)^4+\tfrac34(d+s)^2
			+\tfrac98(s-d)^4
			+\tfrac34\notag\\
			&\quad\times(s-d)^2\bigl(\tfrac{9}{16}(d+s)^4+3(d+s)^2+1\bigr)-3\Bigr)x^2
			-4\bigl(\tfrac34(d+s)^2-1\bigr)\bigl(\tfrac94(s-d)^2+\tfrac34(d+s)^2\notag\\
			&\quad-3\bigr)
			\Bigr)\text{Si}\Bigl(\tfrac12(s-d)x\Bigr)x
			-12\Bigl(
			\tfrac{9}{8}(s-d)^4-\tfrac{45}{4}(s-d)^2+\tfrac{9}{8}(d+s)^4
			+\tfrac34(d+s)^2\bigl(3(s-d)^2\notag\\
			&\quad-15\bigr)+24\Bigr)
			\Bigr)\sqrt{3}
			\Bigr)\sqrt{3}(d+s)
			+(s-d)\sin\Bigl(\tfrac12(d+s)x\Bigr)
			\Bigl[
			\tfrac{27}{4}(s-d)x\sin\Bigl(\tfrac12(s-d)x\Bigr)\sqrt{3}(d+s)^2
			\notag\\
			&\quad+\Bigl(
			-\tfrac{9}{16}(5x^2+24)\bigl(\tfrac34(s-d)^2x^2+6\bigr)(d+s)^4
			+\tfrac94\Bigl(
			\tfrac34\bigl(x^2\bigl(\tfrac34(4-5x^2)(s-d)^2+3x^2-22\bigr)\notag\\
			&\quad-24\bigr)(s-d)^2
			+18(x^2+6)\Bigr)(d+s)^2
			-\tfrac92(s-d)^2\bigl(\tfrac34(s-d)^2-1\bigr)x^2\bigl(\tfrac34(s-d)^2x^2+6\bigr)
			\Bigr)\sqrt{3}
			\notag\\
			&\quad+6\Bigl(
			\tfrac{9}{16}(5x^2+24)(d+s)^4
			+\tfrac34\bigl(\tfrac34(s-d)^2(13x^2+12)-9(x^2+6)\bigr)(d+s)^2
			+\tfrac92(s-d)^2\bigl(\tfrac34\notag\\
			&\quad\times(s-d)^2-1\bigr)x^2
			\Bigr)\cos\Bigl(\tfrac12(s-d)x\Bigr)\sqrt{3}
			+3(s-d)x
\Bigl(
\tfrac{9}{16}(5x^2+24)(d+s)^4
+\tfrac94\bigl(\bigl(\tfrac{15}{4}(s-d)^2\notag\\
&\quad-3\bigr)x^2-12\bigr)(d+s)^2
+\tfrac92(s-d)^2\bigl(\tfrac34(s-d)^2-1\bigr)x^2
\Bigr)\text{Si}\Bigl(\tfrac12(s-d)x\Bigr)\sqrt{3}
\Bigr]\sqrt{3}
\Bigr). 
\label{FuvS}
\end{align}
\end{widetext}

Substituting the TT source, Eq.~\eqref{FuvS},  into the kernel definitions \eqref{ICS} 
yields exact analytic expressions for  $I_c(d,s,x)$ and $I_s(d,s,x)$, but the resulting expressions    
are several pages lengthy \footnote{The analytic expressions of $I_{c/s}(d,s,x)$ are available but are too long to print; all figures are obtained by direct evaluation of the exact expressions. An ancillary repository in Mathematica files used to evaluate them accompanies this work.}.     
We therefore do not reproduce them in the main text. At large $x$, the source exhibits clear late-time growth (Fig.~\ref{fig:TT_source}), which is mirrored by a corresponding increase in the squared kernels (Fig.~\ref{fig:TT_kernels}), in sharp contrast with the bounded evolution in the longitudinal gauge.    

For a Dirac–delta isocurvature peak,
$\mathcal{P}_{S}(k)=\mathcal{A}_{S}\,\delta\!\big(\ln(k/k_{p})\big)$,
we evaluate the spectrum along the delta line in $(d,s)$,
$d=0$ and $s=\tfrac{2}{\sqrt{3}}(k_{p}/k)$, using
Eqs.~\eqref{eq:OmegaGW-dsg2} and \eqref{eq:Ibar2} with the TT kernels
$I_{c,\infty}(d,s)$ and $I_{s,\infty}(d,s)$. The resulting $\Omega_{\rm GW}(k)$ is shown in
Fig.~\ref{fig:TT_spectrum}. The IR rise, peak near $k=2c_s k_p$, and UV cutoff at $k=2k_p$
are present, while the late–time evolution grows with $x$, i.e.\ the raw (pre–projection) readout
is divergent in this gauge.

As in the longitudinal case, we focus on modes reentering during RD and insert the gauge–specific
source into \eqref{ICS}. When convenient we use the $x\to\infty$ limits $I_{c,\infty}^{({\rm TT})}(d,s)$
and $I_{s,\infty}^{({\rm TT})}(d,s)$ to build Eq.~\eqref{eq:OmegaGW-dsg2}.
The procedure applies to any ${\cal P}_S(k)$ provided the relevant scales reenter well before equality.

From the large-$x$ behaviour of the functions (Figs.~\ref{fig:TT_kernels}, \ref{fig:TT_spectrum}) and as summarized in Sec.~\ref{summary}, the late-time growth is approximately \emph{$I\sim x^{3}$}, implying a pre-projection scaling $\Omega_{\rm GW}\propto x^{8}$ in this slicing.

\subsection{Secondary GWs in Total Matter gauge} In the framework of the total matter gauge, where $\delta V=E=0$, the transfer functions become 
\begin{figure}[th] 
\centering
\includegraphics[width=\linewidth]{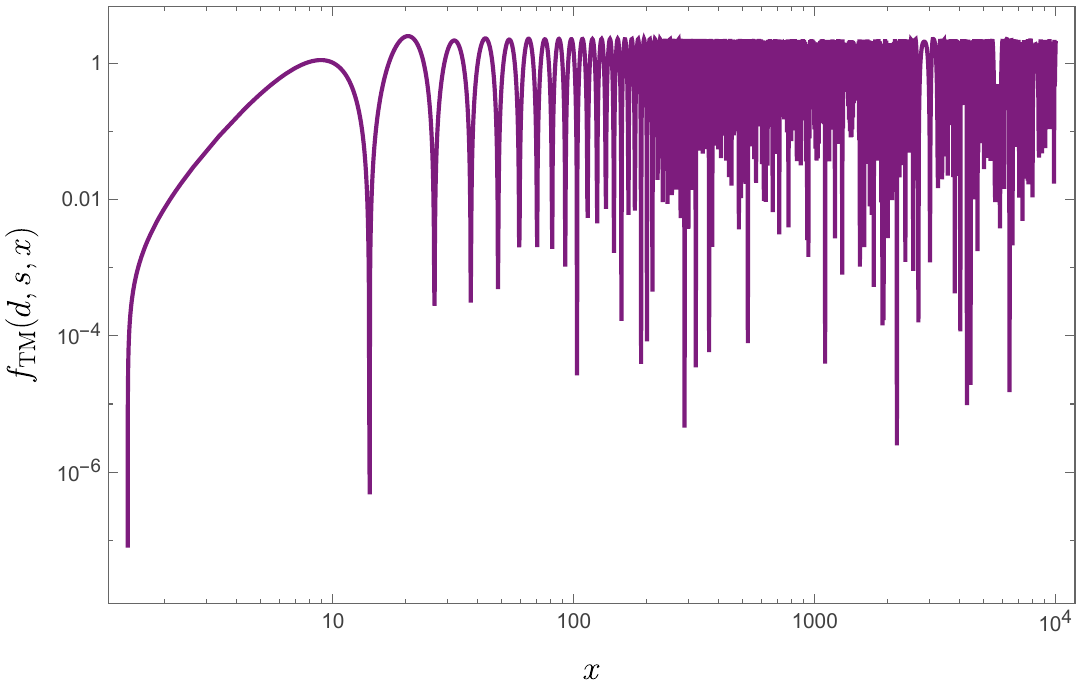}
\caption{Source term in the total–matter gauge at $(d,s)=(0,1/\sqrt{3})$.
After horizon entry the time dependence remains oscillatory with a slowly increasing mean level; the source grows with $x$, unlike the bounded longitudinal case.}\label{fig:TM_source_single} 
\end{figure}
\begin{align}
T_{\phi}(x)
&= \frac{1}{6}\,\frac{3x-\kappa}{x^{3}\kappa^{2}}
\Bigl(6 + x^{2} - 2\sqrt{3}\,x \sin\tfrac{x}{\sqrt{3}}
\notag
\\&- 6\cos\tfrac{x}{\sqrt{3}}\Bigr)
\Bigl(\sin\tfrac{x}{\sqrt{3}} + \cos\tfrac{x}{\sqrt{3}}\Bigr), 
\label{ttransferr}
\\
T_{B}(x)
&= \frac{1}{4\sqrt{2}\,\kappa^{2} x^{3}}\,(3x-\kappa)
\Bigl(6 + x^{2}
- 2\sqrt{3}\,x \sin\tfrac{x}{\sqrt{3}}
\notag \\ 
&\quad- 6\cos\tfrac{x}{\sqrt{3}}\Bigr),\label{tttransfer2}
\end{align}
\begin{align}
T_{\psi}(x)
&= \frac{\sqrt{2}}{6\,\kappa^{3}}(3x-\kappa)
\Bigl[(6+x^{2})\bigl(\sin\tfrac{x}{\sqrt{3}}+\cos\tfrac{x}{\sqrt{3}}\bigr)
\notag\\
&\quad-2\sqrt{3}\,x\,\sin\tfrac{x}{\sqrt{3}}\bigl(\sin\tfrac{x}{\sqrt{3}}+\cos\tfrac{x}{\sqrt{3}}\bigr)
\notag\\
&\quad-6\,\cos\tfrac{x}{\sqrt{3}}\bigl(\sin\tfrac{x}{\sqrt{3}}+\cos\tfrac{x} {\sqrt{3}}\bigr)\Bigr]\notag\\
&\quad+\frac{(3x-\kappa)}{4\,\kappa^{3}}
\Bigl[(6+x^{2})-2\sqrt{3}\,x\,\sin\tfrac{x}{\sqrt{3}}\notag\\
&\quad-6\,\cos\tfrac{x}{\sqrt{3}}\Bigr]
\Bigl(\mathrm{Si}\tfrac{x}{\sqrt{3}}-\mathrm{Ci}\tfrac{x}{\sqrt{3}}\Bigr).
\label{tttransfer3}
\end{align}  

Since $S$ is a gauge invariant quantity, the analytical expression for the transfer function $T_S(x)$ will be the same as presented in longitudinal gauge.    
\begin{figure*}[th]
\centering
\begin{minipage}[t]{0.48\linewidth}
\centering
\includegraphics[width=\linewidth]{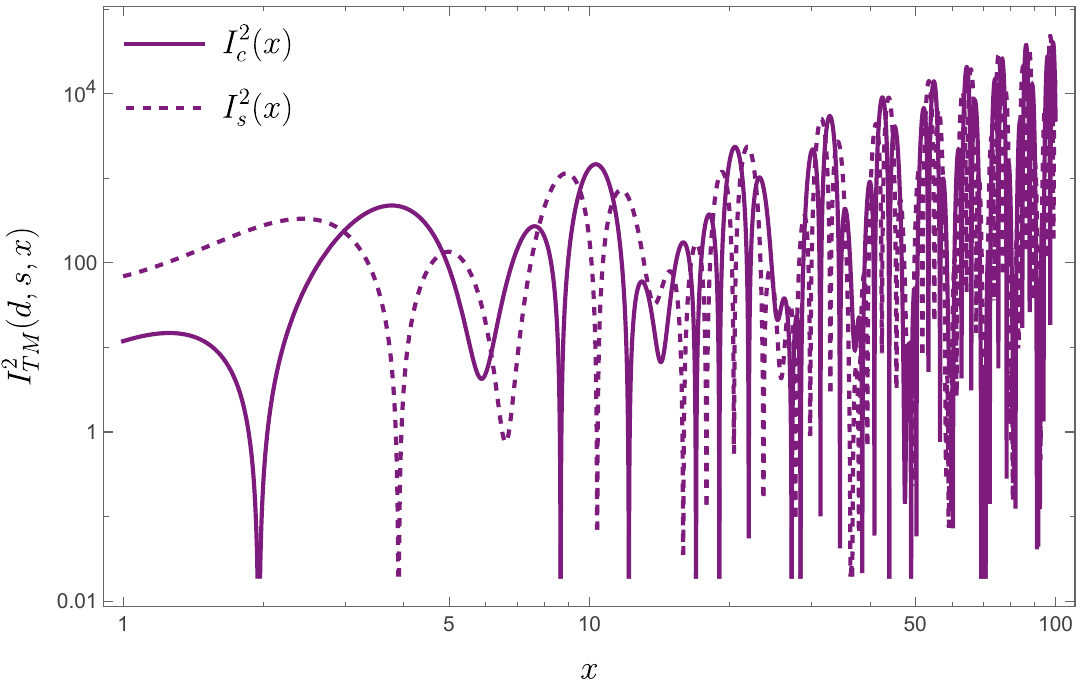}
\caption{Squared kernels $I_c^2(x)$ (solid) and $I_s^2(x)$ (dashed) in the total–matter gauge at $(d,s)=(0,1/\sqrt{3})$.
 A mild late–time uplift is visible and the kernels do not saturate, indicating a non-convergent raw readout.}
\label{fig:TM_kernels}
\end{minipage}
\hfill
\begin{minipage}[t]{0.48\linewidth}
\centering\includegraphics[width=\linewidth]{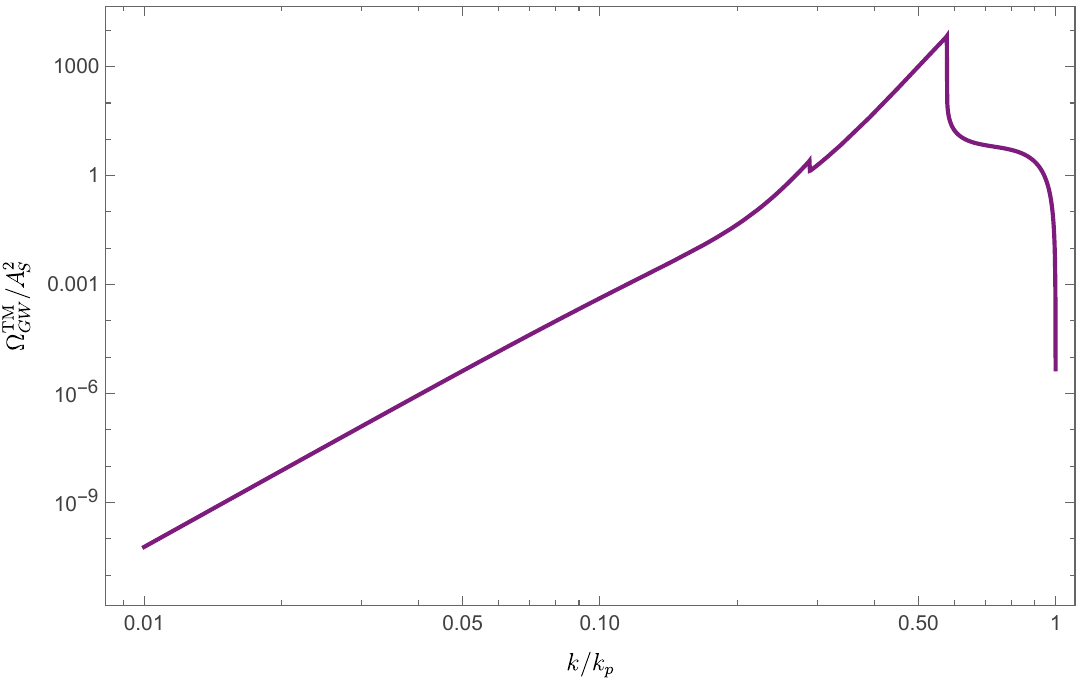} 
\caption{$\Omega_{\rm GW}(k)/A_s^2$ in the total–matter gauge for a Dirac–delta
isocurvature peak at $k_p$. The spectrum shows the standard IR rise
$\propto k^{2}\ln^{2}k$, a resonant feature near $k=2c_s k_p$, and a sharp
cutoff at $k=2k_p$. The pre-projection late–time evolution grows with $x$
in this slicing.
The GW spectrum is normalized by the scalar power spectrum amplitude $A_s^2$.} 
\label{fig:TM_spectrum}
\end{minipage} 
\end{figure*}
In total matter gauge, substituting the results of equations \eqref{ttransferr}, \eqref{tttransfer2}, and \eqref{tttransfer3} into Eq. \eqref{sourcesymmetric} and after derivation in $(d, s)$ domain instead, we get the final expression as 
\begin{widetext}
\begin{align}
f^{\mathrm{TM}}(d,s,x)
&= \frac{4}{3\,\kappa^{2}\,x^{6}\,(s-d)^{3}(d+s)^{3}}
\Bigg(
-6x^{2}\Bigl[\tfrac{3}{4}(s-d)^{2}+\tfrac{3}{4}(d+s)^{2}-6\Bigr]
+3x(d+s)\Bigl[x^{2}\Bigl(\tfrac{3}{4}(s-d)^{2}-6\Bigr)+36\Bigr]\notag\\
&\quad\times
\sin\Bigl(\tfrac{1}{2}x(d+s)\Bigr)
+\sqrt{3}\,x(s-d)\,\sin\Bigl(\tfrac{1}{2}x(s-d)\Bigr)
\Bigl\{ 
\sqrt{3}\Bigl[x^{2}\Bigl(\tfrac{3}{4}(d+s)^{2}-6\Bigr)+36\Bigr]
+3\sqrt{3}\,x(x^{2}-6)\notag\\
&\quad \times(d+s)\sin\Bigl(\tfrac{1}{2}x(d+s)\Bigr)
\Bigr\}
+\Bigl[
-\tfrac{3}{4}x^{4}(s-d)^{2}\Bigl(\tfrac{3}{4}(d+s)^{2}-6\Bigr)
+6x^{2}\Bigl(-\tfrac{15}{4}(s-d)^{2}+\tfrac{3}{4}(d+s)^{2}-6\Bigr)
\notag\\
&\quad
-3x(d+s)\Bigl(x^{2}\Bigl(\tfrac{3}{4}(x^{2}-5)(s-d)^{2}-6\Bigr)+36\Bigr)
\sin\Bigl(\tfrac{1}{2}x(d+s)\Bigr)
+216 
\Bigr]\cos\Bigl(\tfrac{1}{2}x(s-d)\Bigr)
+\cos\Bigl(\tfrac{1}{2}x\notag\\
&\quad\times(d+s)\Bigr)
\Bigl\{
-\tfrac{3}{4}x^{4}(d+s)^{2}\Bigl(\tfrac{3}{4}(s-d)^{2}-6\Bigr)
+6x^{2}\Bigl(\tfrac{3}{4}(s-d)^{2}-\tfrac{15}{4}(d+s)^{2}-6\Bigr)
-3x(s-d)\Bigl(x^{2}\notag\\
&\quad \times\Bigl(\tfrac{3}{4}(x^{2}-5)(d+s)^{2}-6\Bigr)+36\Bigr)
\sin\Bigl(\tfrac{1}{2}x(s-d)\Bigr)
+\Bigl[
\tfrac{9}{16}x^{6}(s-d)^{2}(d+s)^{2}
-2x^{4}\Bigl(\tfrac{3}{4}\bigl(\tfrac{3}{2}(d+s)^{2}\notag\\
&\quad +3\bigr)(s-d)^{2}
+\tfrac{9}{4}(d+s)^{2}\Bigr)
+6x^{2}\Bigl(\tfrac{15}{4}(s-d)^{2}+\tfrac{15}{4}(d+s)^{2}+6\Bigr)-216
\Bigr]\cos\Bigl(\tfrac{1}{2}x(s-d)\Bigr)
\notag\\
&\quad+216
\Bigr\}
-216
\Bigg).
\label{sourceTM} 
\end{align}
\end{widetext}
    
Incorporating \eqref{sourceTM} into Eq. \eqref{ICS}, we derive the explicit expression for ${I}^{\text{TM}}_{\mathrm{c}}$ as follows: \begin{widetext}
\begin{align}
I^{\text{TM}}_c(d,s,x)
&= \frac{3}{8 (d-s)^3 (d+s)^3 \kappa^2} \Bigg[
\frac{\sin x}{x^4} \Big(
d^4 x^5 \cos(sx) + s^4 x^5 \cos(sx) - 2 d^2 s^2 x^5 \cos(sx) - 8 d^3 x^4 \sin(dx) + 8 d s^2 x^4 \notag\\
			&\quad \times\sin(dx) - 8 s^3 x^4 \sin(sx) + 8 d^2 s x^4 \sin(sx)  - 24 d^3 x^3 \cos\left(\tfrac{s x}{2}\right) \sin\left(\tfrac{d x}{2}\right)
			- 40 d s^2 x^3 \cos\left(\tfrac{s x}{2}\right) \sin\left(\tfrac{d x}{2}\right) \notag\\
			&\quad + 96 d x^3 \cos\left(\tfrac{s x}{2}\right) \sin\left(\tfrac{d x}{2}\right)
			+ 32 d s^2 x^3 \sin(dx) - 48 d x^3 \sin(dx) + 32 d^2 s x^3 \sin(sx) - 48 s x^3 \sin(sx)\notag\\
			&\quad  + 16 d^2 x^2 + 16 s^2 x^2 + 8 d^2 x^2 \cos(sx) - 40 s^2 x^2 \cos(sx) - 48 x^2 \cos(sx) - 96 d s x^2 \sin\left(\tfrac{d x}{2}\right) \sin\left(\tfrac{s x}{2}\right)\notag\\
			&\quad  - 96 x^2 - 192 d x \cos\left(\tfrac{s x}{2}\right) \sin\left(\tfrac{d x}{2}\right)
			+ 96 d x \sin(dx) + 96 s x \sin(sx)  + \big( (d^2 - s^2)^2 x^5 - 8 (5 d^2 - s^2 \notag\\
			&\quad+ 6) x^2 + 96 \big) \cos(dx)
			+ 96 \cos(sx) + 8 \cos\left(\tfrac{d x}{2}\right) \Big(
			2 \big( (d^2 + s^2 + 12) x^2 - 24 \big) \cos\left(\tfrac{s x}{2}\right) - s x \big( (5 d^2 \notag\\
			&\quad + 3 s^2 - 12) x^2 + 24 \big) \sin\left(\tfrac{s x}{2}\right)
			\Big) + 192 \Big)  - 8 x \cos x \Big(
			- 2 d^2 x^2 - 2 s^2 x^2 - d^2 x^2 \cos(sx) + s^2 x^2 \notag\\
			&\quad\times \cos(sx) + 6 x^2 \cos(sx)
			+ 4 d s x^2 \sin\left(\tfrac{d x}{2}\right) \sin\left(\tfrac{s x}{2}\right)
			+ 12 x^2  + 8 d x \cos\left(\tfrac{s x}{2}\right) \sin\left(\tfrac{d x}{2}\right)
			- 4 d x \notag\\
			&\quad \times \sin(dx) - 4 s x \sin(sx) + \big( (d^2 - s^2 + 6) x^2 - 4 \big) \cos(dx) - 4 \cos(sx) + 2 \cos\left(\tfrac{d x}{2}\right) \Big(
			\big( (d^2 \notag\\
			&\quad  + s^2 - 12) x^2 + 8 \big) \cos\left(\tfrac{s x}{2}\right)
			+ 4 s x \sin\left(\tfrac{s x}{2}\right)
			\Big) - 8 \Big)  + 2 \Big(
			(d^4 + (4 s - 2) d^3 + (6 s^2 - 6 s - 4)\notag\\
			&\quad \times d^2 + 2 s^2 (2 s - 3) d
			+ s^4 - 2 s^3 - 4 s^2 + 24) \times \text{Si}\left(-\tfrac{d x}{2} - \tfrac{s x}{2} + x\right)
			+ 8 (d^2 + s^2 - 6) \text{Si}(x) \notag\\
			&\quad- (d^4 + 2 d^3 + (6 s^2 - 2) d^2 + 6 s^2 d + s^4 - 2 s^2 + 12)\,\text{Si}((d+1)x) + (d^4 + (2 - 4 s) d^3\notag\\
			&\quad + (6 s^2 - 6 s - 4) d^2 + 2 s^2 (3 - 2 s) d
			+ s^4 - 2 s^3 - 4 s^2 + 24)\text{Si}\left(\tfrac{1}{2} (d - s + 2) x\right) 
			- (d^4 \notag\\ 
			&\quad +(6 s^2 + 6 s - 2) d^2 + s^4 + 2 s^3 - 2 s^2 + 12)\,\text{Si}((s+1)x) + (d^4 - 2 (2 s + 1) d^3 + (6 s^2 \notag\\
			&\quad + 6 s - 4) d^2 - 2 s^2 (2 s + 3) d 
			+ s^4 + 2 s^3 - 4 s^2 + 24)  \text{Si}\left(\tfrac{1}{2} (-d + s + 2) x\right) 
			+ (d^4 + (4 s\notag\\
			&\quad  + 2) d^3 + (6 s^2 + 6 s - 4) d^2 \phantom{\times} + 2 s^2 (2 s + 3) d + s^4 + 2 s^3 - 4 s^2 + 24)\, 
			\text{Si}\left(\tfrac{1}{2} (d + s + 2) x\right)\notag\\
			&\quad - (d^4 - 2 d^3 + (6 s^2 - 2) d^2 - 6 s^2 d + s^4 - 2 s^2 + 12)\,\text{Si}(x - d x)  - (d^4 + (6 s^2 - 6 s-2) d^2\notag\\
	&\quad  + s^4 - 2 s^3 - 2 s^2 + 12)\,\text{Si}(x - s x)
\Big) \Bigg],
\end{align}
and $I_s$ can be written as 
\begin{align}
I^{\text{TM}}_s(d,s,x)
&= \frac{3}{4 \kappa^2 (d - s)^3 (d + s)^3} \Big[
\big(d^4 + d^2 (6 s^2 - 6 s - 2) + s^4 - 2 s^3 - 2 s^2 + 12\big)\,\text{Ci}(x |1 - s|)
			+ \big(d^4 + d^2 (6 s^2 + 6 s - 2) \big)\,\notag\\
			&\quad \times\text{Ci}(x |s + 1|) + \big(s^4 + 2 s^3 - 2 s^2 + 12\big)\,\text{Ci}(x |s + 1|)
			+ \big(d^4 - 2 d^3 + d^2 (6 s^2 - 2) - 6 d s^2 + s^4 - 2 s^2 + 12\big)\,\notag\\
			&\quad\times  \text{Ci}(x |1 - d|) + \big(d^4 + 2 d^3 + d^2 (6 s^2 - 2) + 6 d s^2 + s^4 - 2 s^2 + 12\big)\,\text{Ci}(x |d + 1|)
			- \big(d^4 + d^3 (4 s - 2) + d^2   \notag\\
			&\quad\times (6 s^2 - 6 s - 4) + 2 d s^2 (2 s - 3)
			+ s^4 - 2 s^3 - 4 s^2 + 24\big)\,\text{Ci}\left(\tfrac{1}{2} x |-d - s + 2|\right) - \big(d^4 + d^3 (2 - 4 s) \notag\\
			&\quad + d^2 (6 s^2 - 6 s - 4) + 2 d s^2 (3 - 2 s)
			+ s^4 - 2 s^3 - 4 s^2 + 24\big)\,\text{Ci}\left(\tfrac{1}{2} x |d - s + 2|\right)  - \big(d^4 - 2 d^3 (2 s + 1)\notag\\
			&\quad + d^2 (6 s^2 + 6 s - 4) - 2 d s^2 (2 s + 3)
			+ s^4 + 2 s^3 - 4 s^2 + 24\big)\,\text{Ci}\left(\tfrac{1}{2} x |-d + s + 2|\right) - \big(d^4 + d^3 (4 s + 2)\notag\\
			&\quad  + d^2 (6 s^2 + 6 s - 4) + 2 d s^2 (2 s + 3)
			+ s^4 + 2 s^3 - 4 s^2 + 24\big)\,\text{Ci}\left(\tfrac{1}{2} x |d + s + 2|\right) - 8 (d^2 + s^2 - 6)\,\text{Ci}(x)
			\Big] \notag\\ 
			&\quad - \frac{3}{8 \kappa^2 x^4 (d - s)^3 (d + s)^3} \Big( 
			8 x \sin(x) \Big(
			\big(x^2 (d^2 - s^2 + 6) - 4\big)\cos(d x)
			+ 2 \cos\left(\tfrac{d x}{2}\right)\!(\big(x^2 (d^2 + s^2 - 12)  \notag\\  
			&\quad+ 8\big)\cos\left(\tfrac{s x}{2}\right)
			+ 4 s x \sin\left(\tfrac{s x}{2}\right) )  
			- d^2 x^2 \cos(s x) - 2 d^2 x^2
			+ 4 d s x^2 \sin\left(\tfrac{d x}{2}\right) \sin\left(\tfrac{s x}{2}\right)
			+ 8 d x \sin\left(\tfrac{d x}{2}\right) \notag\\
			&\quad \times \cos\left(\tfrac{s x}{2}\right) 
			- 4 d x \sin(d x) - 2 s^2 x^2
			+ s^2 x^2 \cos(s x) + 6 x^2 \cos(s x)
			- 4 s x \sin(s x) - 4 \cos(s x) + 12 x^2\notag\\
			&\quad  - 8
			\Big) 
			+ \cos(x) \Big(
			d^4 x^5 \cos(s x)
			- 24 d^3 x^3 \sin\left(\tfrac{d x}{2}\right) \cos\left(\tfrac{s x}{2}\right)
			- 8 d^3 x^4 \sin(d x)
			- 2 d^2 s^2 x^5 \cos(s x) \notag\\
			&\quad 
			+ 8 \cos\left(\tfrac{d x}{2}\right) \Big(
			2 \big(x^2 (d^2 + s^2 + 12) - 24\big) \cos\left(\tfrac{s x}{2}\right)
			- s x \big(x^2 (5 d^2 + 3 s^2 - 12) + 24\big) \sin\left(\tfrac{s x}{2}\right)
			\Big) \notag\\
			&\quad 
			+ \big(x^5 (d^2 - s^2)^2 - 8 x^2 (5 d^2 - s^2 + 6) + 96\big) \cos(d x)
			+ 8 d^2 s x^4 \sin(s x) + 32 d^2 s x^3 \sin(s x) \notag\\
			&\quad 
			+ 8 d^2 x^2 \cos(s x) + 16 d^2 x^2
			+ 8 d s^2 x^4 \sin(d x) + 32 d s^2 x^3 \sin(d x) 
			- 40 d s^2 x^3 \sin\left(\tfrac{d x}{2}\right) \cos\left(\tfrac{s x}{2}\right)
			\notag\\
			&\quad + 96 d x^3 \sin\left(\tfrac{d x}{2}\right) \cos\left(\tfrac{s x}{2}\right) 
			- 96 d s x^2 \sin\left(\tfrac{d x}{2}\right) \sin\left(\tfrac{s x}{2}\right)
			- 192 d x \sin\left(\tfrac{d x}{2}\right) \cos\left(\tfrac{s x}{2}\right) 
			- 48 d x^3 \notag\\
			&\quad \times\sin(d x) + 96 d x \sin(d x) 
			+ s^4 x^5 \cos(s x) - 8 s^3 x^4 \sin(s x)
+ 16 s^2 x^2 - 40 s^2 x^2 \cos(s x)
\notag\\
&\quad - 48 s x^3 \sin(s x) - 48 x^2 \cos(s x) 
+ 96 s x \sin(s x) + 96 \cos(s x) - 96 x^2 + 192
\Big)
\Big).
\end{align}
\end{widetext}

Substituting the TM source into the kernel integrals \eqref{ICS} yields the explicit analytical form
$\,I_c^{\rm TM}(d,s,x)$ and $I_s^{\rm TM}(d,s,x)$, but the exact analytical expressions in $x\to\infty$ are also long, so we do not reproduce them here.  After horizon entry in RD, the time dependence remains oscillatory with a slowly increasing mean level; the source (Fig. \ref{fig:TM_source_single}) grows with $x$, unlike the bounded longitudinal case. 
Also, for subhorizon evolution, the kernels grow $\propto x$ at late times, hence  $\,I_{c/s}^{2}\propto x^2$  (Fig.~\ref{fig:TM_kernels}).   
From the late–time evolution (and consistent with the summary in Sec.~\ref{summary}),  
the growth is approximately  
\emph{$I\sim x$} as $x\to\infty$, implying a pre-projection scaling
$\Omega_{\rm GW}\propto x^{4}$ in this slicing. 

For a Dirac–delta isocurvature peak,
$\mathcal{P}_{S}(k)=\mathcal{A}_{S}\,\delta\!\big(\ln(k/k_{p})\big)$,
we evaluate along the delta line $d=0$, $s=\tfrac{2}{\sqrt{3}}(k_p/k)$ using
Eqs.~\eqref{eq:OmegaGW-dsg2} and \eqref{eq:Ibar2} with the TM kernels
$I_{c,\infty}^{\rm TM}(d,s)$ and $I_{s,\infty}^{\rm TM}(d,s)$.
The resulting $\Omega_{\rm GW}(k)$ (Fig.~\ref{fig:TM_spectrum}) shows the familiar
$k^{2}\ln^{2}k$ rise at low $k$, a resonant feature near $k=2c_s k_p$, and a sharp
cutoff at $k=2k_p$, together with the slow late-time growth of the evolution noted above.
The same procedure applies to general ${\cal P}_S(k)$, provided the relevant modes
reenter well before equality.

\subsection{Secondary GWs in Uniform Curvature gauge} \label{ugauge} 

In the uniform curvature gauge, the metric perturbations satisfy $\psi=E=0$, and the transfer functions of the remaining perturbations read  
\begin{align} 
T_B(x) &= -\frac{3}{2\sqrt{2}\,\kappa x^2}\left[6 + x^2 - 2\sqrt{3}\,x \sin\left(\frac{x}{\sqrt{3}}\right)\right.\notag\\
&\quad\left.- 6\cos\left(\frac{x}{\sqrt{3}}\right)\right],\label{uc1transfer}\\
T_\phi(x) &= -\frac{3}{\sqrt{2}\,x\,\kappa}\left[1 - \cos\left(\frac{x}{\sqrt{3}}\right)\right].\label{uc02transfer}
\end{align}     
\begin{figure}[th]
\centering
\includegraphics[width=\linewidth]{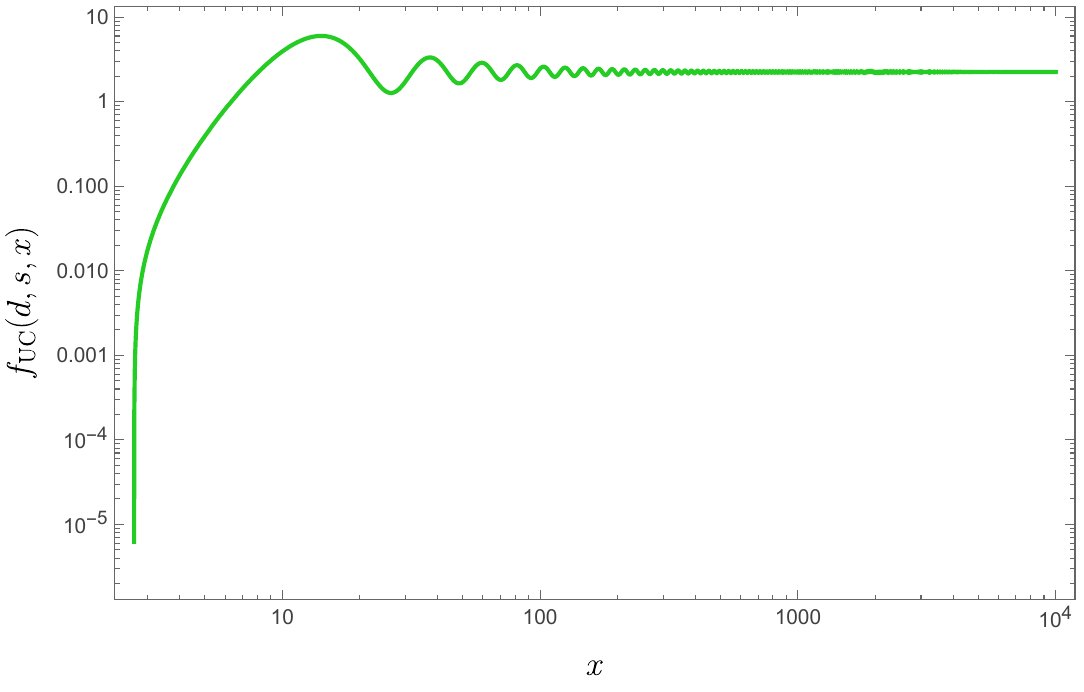}
\caption{Source term $f_{\rm UC}(d,s,x)$ at $(d,s)=(0,1/\sqrt{3})$ versus $x\equiv k\eta$.
After horizon entry the profile remains oscillatory but its evolution shows a non-oscillatory growth, i.e.\ the sourcing is not bounded in this slicing.} 
\label{fig:UC_source}
\end{figure}

As $S$ is a gauge-invariant quantity, the expression for $T_S(x)$ remains unchanged as in the TT gauge. We can find the source function in this gauge as follows:
\begin{widetext}
\begin{align}
f_{UC}(d,s,x)
&= \frac{4}{9 \kappa^2 x^4 (s-d)^3 (d+s)^3} \Bigg(
\frac{9}{4}(d+s)^2 \Big( x^2 \big( \tfrac{3}{4}(x^2+4)(s-d)^2 + 6 \big) + 12 \Big)
			+ 18 \Big( \tfrac{3}{4}(x^2+2)(s-d)^2 + 6 \Big) 
			\notag\\
			&\quad+ \sqrt{3}\,x(d+s)\,\sin\!\left(\tfrac{1}{2}x(d+s)\right)
			\Bigg(
			6 \Big( \tfrac{3}{4}(s-d)^2 + \tfrac{3}{4}(d+s)^2 + 3 \Big)
			\Big( \tfrac{\sqrt{3}}{2}\,x(s-d)\,\sin\!\left(\tfrac{1}{2}x(s-d)\right)
			+ \sqrt{3}\,\notag\\
			&\quad\times \cos\!\left(\tfrac{1}{2}x(s-d)\right) \Big) 
			- \sqrt{3} \Big(\tfrac{3}{4}\,x^2 (s-d)^2 \Big( \tfrac{3}{4}(d+s)^2 + 3 \Big)
			+ 6 \Big( \tfrac{3}{4}(s-d)^2 + \tfrac{3}{4}(d+s)^2 + 3 \Big) \Big) 
			\Bigg) 
			- 2 \Big( \tfrac{3}{4}\notag\\
			&\quad\times(d+s)^2 \big( x^2 \big( \tfrac{3}{4}(s-d)^2 + 3 \big) + 6 \big)
			+ 6 \big( \tfrac{3}{4}(s-d)^2 + 3 \big) \Big)
			\Big( \tfrac{3}{2}\,x(s-d)\,\sin\!\left(\tfrac{1}{2}x(s-d)\right)
			+ 3 \cos\!\left(\tfrac{1}{2}x(s-d)\right) \Big) \notag\\
			&\quad
			+ 6 \cos\!\left(\tfrac{1}{2}x(d+s)\right)
			\Bigg(
			- \tfrac{3}{4}\,x^2 (s-d)^2 \Big( \tfrac{3}{4}(d+s)^2 + 3 \Big)
			+ 2 \Big( \tfrac{3}{4}(s-d)^2 + \tfrac{3}{4}(d+s)^2 + 3 \Big)
\Big( \tfrac{3}{2}\,x(s-d)\,\notag\\
&\quad\times \sin\!\left(\tfrac{1}{2}x(s-d)\right)
+ 3 \cos\!\left(\tfrac{1}{2}x(s-d)\right) \Big) 
- 6 \Big( \tfrac{3}{4}(s-d)^2 + \tfrac{3}{4}(d+s)^2 + 3 \Big)
\Bigg)
\Bigg).
\end{align}

Within this framework, the kernel functions can be determined using equations \eqref{uc1transfer} and \eqref{uc02transfer}  substituted into Eq. \eqref{iuv}, resulting in  
\begin{align}
I^{UC}_{c}(d,s,x&) = 
\frac{3}{4 (d - s)^3 (d + s)^3} \Bigg( 
4 (d^4 - 2 d^2 s^2 + s^4 - 8) \, \text{Si}(x) 
			+ 2 (d^4 + 2 d^2 s^2 + s^4 - 4)  \times \big( \text{Si}((d+1)x) + \text{Si}((s+1)x) \big) 
			\notag\\ 
			&   + 2 (d^4+ 2 d^2 s^2 + s^4 - 4)  
			\big( \text{Si}(x - dx) + \text{Si}(x - sx) \big) - (3 d^4 + 4 d^3 s + 2 d^2 s^2 + 4 d s^3 + 3 s^4 - 16) 
			\,\text{Si}\left(-\tfrac{dx}{2}- \tfrac{sx}{2} + x\right) \notag\\ 
			&  + (-3 d^4 + 4 d^3 s - 2 d^2 s^2 + 4 d s^3 - 3 s^4 + 16) 
			\big(  
			\text{Si}\left( \tfrac{1}{2}(d - s + 2)x \right)
			+ \text{Si}\left( \tfrac{1}{2}(-d + s + 2)x \right)
			\big)  - (3 d^4 + 4 d^3 s + 2 d^2 s^2 \notag\\ 
			&  + 4 d s^3 + 3 s^4 - 16) 
			\, \text{Si}\left( \tfrac{1}{2}(d + s + 2)x \right)
			\Big)  - \frac{3}{4 x^2 (d^2 - s^2)^3} \Big(
			\sin(x) \bigg(
			- x^3 (d^2 - s^2)^2 
			+ 8 d x (d^2 + s^2 + 2) \sin(dx) 
			\notag\\ 
			&  + 8 (d^2 + s^2 + 2) \times\cos(dx)  + 2 d x \Big( 
			x (d - s)(d + s)(d^2 - s^2 + 4)  
			- 8(d^2 + s^2 + 2)  
			\Big) \sin\left( \tfrac{dx}{2} \right) \cos\left( \tfrac{sx}{2} \right) + 2 \cos\left( \tfrac{dx}{2} \right) \notag\\ 
			&  \times\Big(
			sx \Big( 
			x(d - s)(d + s)(d^2 - s^2 - 4) 
			- 8(d^2 + s^2 + 2) 
			\Big) \sin\left( \tfrac{sx}{2} \right)  - 16(d^2 + s^2 + 2) \cos\left( \tfrac{sx}{2} \right)
			\Big) 
			+ 8 (d^2 + s^2 + 2)\notag\\ 
			&  \times (sx \sin(sx) + \cos(sx))
			+ 16(d^2 + s^2 + 2)
			\bigg)  + 16 x (d^2 + s^2 + 2) \cos(x) 
			\left( \cos\left( \tfrac{dx}{2} \right) - \cos\left( \tfrac{sx}{2} \right) \right)^2\Bigg), 
\end{align}  
and   
\begin{figure*}[th]
\centering
\begin{minipage}[t]{0.48\linewidth}
\centering
\includegraphics[width=\linewidth]{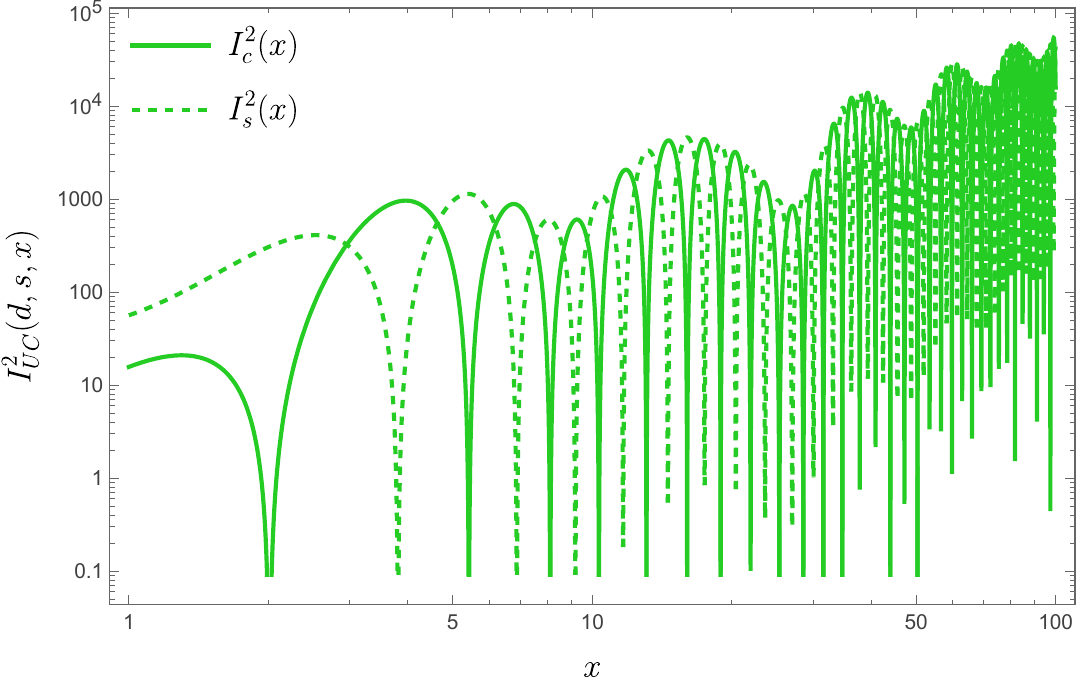}
\caption{Squared kernels in the UC gauge, $I_c^2(x)$ (solid) and $I_s^2(x)$ (dashed),
evaluated at $(d,s)=(0,1/\sqrt{3})$.  Both components grow at large $x$, indicating a
non-convergent (pre-projection) late-time readout compared with the longitudinal baseline.} 
\label{fig:UC_kernels}
\end{minipage}
\hfill
\begin{minipage}[t]{0.48\linewidth}
\centering
\includegraphics[width=\linewidth]{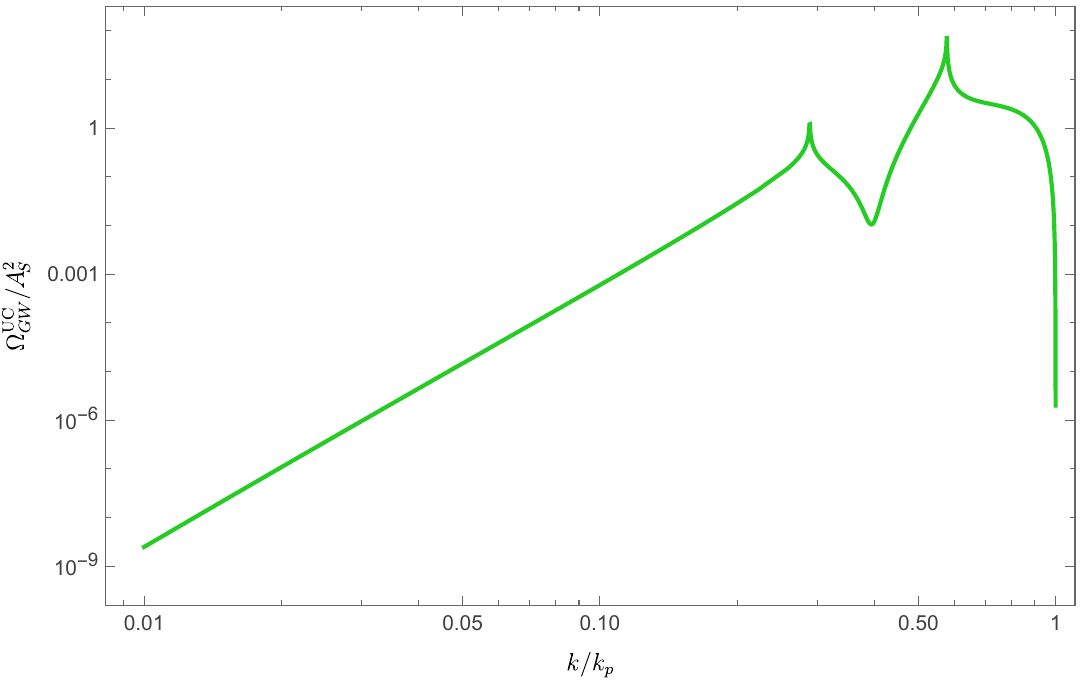}
\caption{$\Omega_{\rm GW}(k)/A_s^2$ in the UC gauge for a Dirac–delta isocurvature peak at $k_p$.
The spectrum displays the usual IR rise and kinematic features (peak near $k=2c_sk_p$,
cutoff at $k=2k_p$ with $c_s=1/\sqrt{3}$), but its late-time evolution grows with $x$ and
is \emph{divergent} in this gauge.
The GW spectrum is normalized by the scalar power spectrum amplitude $A_s^2$.}
\label{fig:UC_spectrum}
\end{minipage}
\end{figure*}
\begin{align}
I^{\mathrm{UC}}_{s}(d,s,x)
&= \frac{3}{4\,(d-s)^{3}(d+s)^{3}}
			\Biggl\{
			-2\bigl(d^{4}+2d^{2}s^{2}+s^{4}-4\bigr)
			\Bigl[
			\mathrm{Ci}\bigl(x\lvert 1-d\rvert\bigr)
			+\mathrm{Ci}\bigl(x\lvert d+1\rvert\bigr)
			\notag\\
			&\quad+\mathrm{Ci}\bigl(x\lvert 1-s\rvert\bigr)
			+\mathrm{Ci}\bigl(x\lvert s+1\rvert\bigr)
			\Bigr]
			+\bigl(3d^{4}+4d^{3}s+2d^{2}s^{2}+4ds^{3}+3s^{4}-16\bigr)
			\notag\\
			&\quad\times\mathrm{Ci}\Bigl(\tfrac{x}{2}\,\lvert{-}d{-}s{+}2\rvert\Bigr)
+\bigl(3d^{4}-4d^{3}s+2d^{2}s^{2}-4ds^{3}+3s^{4}-16\bigr) 
\mathrm{Ci}\Bigl(\tfrac{x}{2}\,\lvert d{-}s{+}2\rvert\Bigr) 
\notag\\ 
&\quad 
+\bigl(3d^{4}-4d^{3}s+2d^{2}s^{2}-4ds^{3}+3s^{4}-16\bigr)
\mathrm{Ci}\Bigl(\tfrac{x}{2}\,\lvert{-}d{+}s{+}2\rvert\Bigr)
\notag\\
&\quad
+\bigl(3d^{4}+4d^{3}s+2d^{2}s^{2}+4ds^{3}+3s^{4}-16\bigr)
\mathrm{Ci}\Bigl(\tfrac{x}{2}\,\lvert d{+}s{+}2\rvert\Bigr)
			-4\,\mathrm{Ci}(x)\,\bigl(d^{4}-2d^{2}s^{2}\notag\\
			&\quad+s^{4}-8\bigr)
			\Biggr\}
			+\frac{3\,\cos x}{4\,x^{2}\,(d^{2}-s^{2})^{3}}
			\Biggl(
			-\,x^{3}(d^{2}-s^{2})^{2}
			+8d x\,(d^{2}+s^{2}+2)\sin(d x)
			\notag\\
			&\quad+8(d^{2}+s^{2}+2)\cos(d x)
			+2d x\Bigl[x(d-s)(d+s)(d^{2}-s^{2}+4)-8(d^{2}+s^{2}+2)\Bigr]
			\notag\\
			&\quad\times\sin\Bigl(\tfrac{d x}{2}\Bigr)\cos\Bigl(\tfrac{s x}{2}\Bigr) 
			+2\cos\Bigl(\tfrac{d x}{2}\Bigr)\Bigl\{
			s x\Bigl[x(d-s)(d+s)(d^{2}-s^{2}-4)-8(d^{2}+s^{2}+2)\Bigr]\notag\\ 
			&\quad \times 
			\sin\Bigl(\tfrac{s x}{2}\Bigr)
			-16(d^{2}+s^{2}+2)\cos\Bigl(\tfrac{s x}{2}\Bigr)
			\Bigr\}
			+8(d^{2}+s^{2}+2)\bigl[s x\sin(s x)+\cos(s x)\bigr]
			\notag\\ 
			&\quad+16(d^{2}+s^{2}+2)
			\Biggr)
			-48\,x\,(d^{2}+s^{2}+2)\,\sin x\,
\Bigl(\cos\tfrac{d x}{2}-\cos\tfrac{s x}{2}\Bigr)^{2}.
\label{IsUC}
\end{align}
\end{widetext} 

Inserting the UC source into the kernel integrals \eqref{ICS} yields analytical form of 
expressions for $I_c(d,s,x)$ and $I_s(d,s,x)$. From the large-$x$ limit of those expressions we find, at fixed $(d,s)$, where 
$
I_{c}^{(\mathrm{UC})}(d,s,x)\propto x,\quad\text{and}\quad
I_{s}^{(\mathrm{UC})}(d,s,x)\propto x \quad (x\gg1), 
$
so that $\overline{I^2}\sim x^{2}$ and the pre-projection energy density grows as
$\Omega_{\rm GW}\propto x^{4}$. The late-time behavior of the source function is shown in Fig. \ref{fig:UC_source}.  
This late-time dependence is visible in
Figs.~\ref{fig:UC_kernels}–\ref{fig:UC_spectrum} and contrasts with the bounded
longitudinal baseline.

For a Dirac–delta isocurvature peak,
$\mathcal{P}_{S}(k)=\mathcal{A}_{S}\,\delta\!\big(\ln(k/k_{p})\big)$,
we evaluate along the delta line $d=0$, $s=\tfrac{2}{\sqrt{3}}(k_p/k)$ using
Eqs.~\eqref{eq:OmegaGW-dsg2} and \eqref{eq:Ibar2}; the resulting $\Omega_{\rm GW}(k)$
is shown in Fig.~\ref{fig:UC_spectrum}.
        
\subsection{Secondary GWs in Uniform Density gauge} 

The uniform density gauge is described by the condition $\delta\rho = E = 0$. We find the only nonzero transfer function as follows:  
\begin{align}
T_{B}(x)
&= \frac{6}{x}  + \frac{3}{2\sqrt{2}\,\kappa\,x^{3}}
\Bigl[
6 + x^{2}
- 2\sqrt{3}\,x \sin\Bigl(\tfrac{x}{\sqrt{3}}\Bigr)
\notag\\
&\quad- 6\cos\Bigl(\tfrac{x}{\sqrt{3}}\Bigr)
\Bigr].
\label{transferud}
\end{align} 
\begin{figure}[t]
\centering
\includegraphics[width=\linewidth]{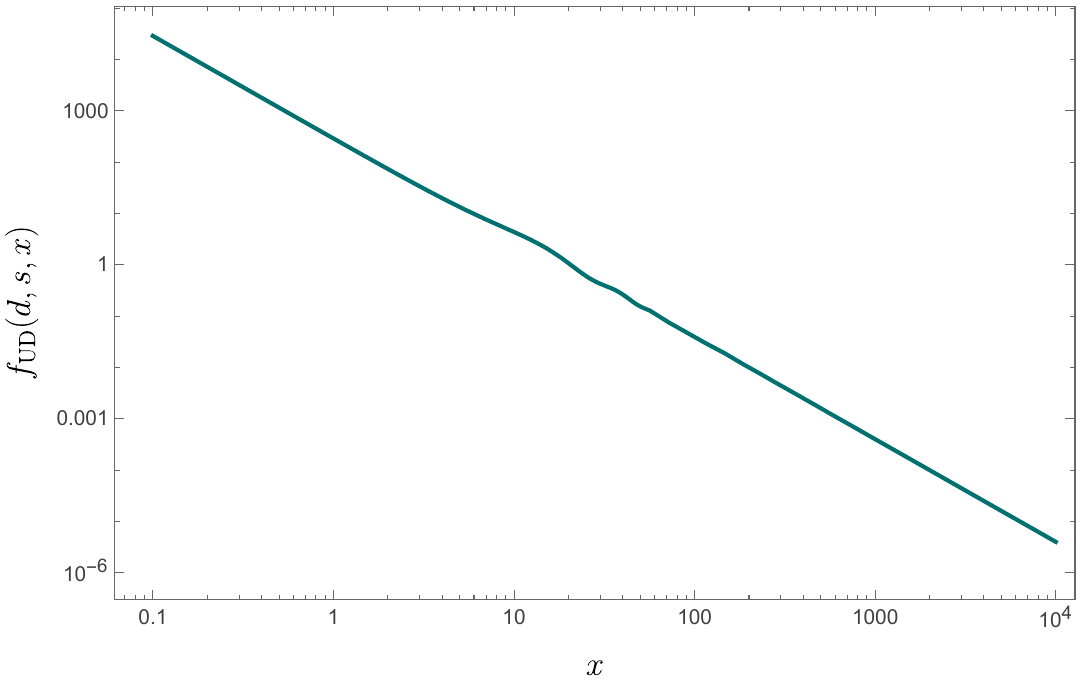}
\caption{Source term  $f_{\rm UD}(d,s,x)$ at $(d,s)=(0,1/\sqrt{3})$ versus $x\equiv k\eta$.
After horizon entry the profile remains oscillatory but its time evolution shows a pronounced late-time dependence over many decades in $x$, characteristic of the UD slicing for isocurvature. } 
\label{fig:UD_source}
\end{figure}
Although the transfer function of $S$ is gauge invariant,  here in this gauge it remains as in the TT gauge. One can use the Eq.  \eqref{transferud}  into Eq.  \eqref{sourcesymmetric}, we derive the source function as:  
\begin{widetext}
\begin{align}
f^\text{UD}(d,s,x)=&\,
\frac{8}{9 \kappa ^2 x^6 (s-d)^4 (d+s)^4}
\Bigg(
\left(\frac{3}{4} \left(\sqrt{2}+8\right) x^2 (s-d)^2
-3 \sqrt{2} x (s-d) \sin \left(\tfrac{1}{2} x (s-d)\right)
\right.
\notag\\
&\left.
-6 \sqrt{2} \left(\cos \left(\tfrac{1}{2} x (s-d)\right)-1\right)\right)
\left(\frac{3}{4} \left(\sqrt{2}+8\right) x^2 (d+s)^2
-3 \sqrt{2} x (d+s) \sin \left(\tfrac{1}{2} x (d+s)\right)
\right.
\notag\\
&\left.
-6 \sqrt{2} \left(\cos \left(\tfrac{1}{2} x (d+s)\right)-1\right)
\right)
\Bigg).\label{sourceUD}
\end{align}
Using \eqref{sourceUD}, we can find the analytical expression of $I_c^\text{UD}$ as follows: 
\begin{align}
I_c^\text{UD}(d,s,x)
&= -\frac{2}{3 \kappa ^2 x^4 (d-s)^4 (d+s)^4}
\Bigg(
2 x \cos(x)\,\Big[
\big( x^2 (-2 d^2 - 3 s^2 + 1) - 2 \big)\cos(d x)
+ \cos\!\left(\tfrac{d x}{2}\right)\Big(
\big( x^2 \big( (24 \sqrt{2}+11) d^2 
\notag\\
&\quad+ (24 \sqrt{2}+11) s^2 - 4 \big) + 8 \big)\cos\!\left(\tfrac{s x}{2}\right)
+ 4 s x \sin\!\left(\tfrac{s x}{2}\right) 
\Big) 
- 3 d^2 x^2 \cos(s x)
- 24 \sqrt{2}\, d^2 x^2
- 6 d^2 x^2
+ 2 d s x^2 \notag\\
&\quad\times \sin\!\left(\tfrac{d x}{2}\right)\sin\!\left(\tfrac{s x}{2}\right) 
+ 48 \sqrt{2}\, d s x^2 \sin\!\left(\tfrac{d x}{2}\right)\sin\!\left(\tfrac{s x}{2}\right)
+ 4 d x \sin\!\left(\tfrac{d x}{2}\right)\cos\!\left(\tfrac{s x}{2}\right)
- 2 d x \sin(d x) 
- 24 \sqrt{2}\, s^2 x^2 \notag\\
&\quad
- 6 s^2 x^2
- 2 s^2 x^2 \cos(s x)
+ x^2 \cos(s x)
- 2 s x \sin(s x)
- 2 \cos(s x) + 2 x^2 - 4
\Big] 
+ \sin(x)\,
\Big(3 d^3 x^3 \sin\!\left(\tfrac{d x}{2}\right)\notag\\
&\quad\times\cos\!\left(\tfrac{s x}{2}\right)
+ 24 \sqrt{2}\, d^3 x^3 \sin\!\left(\tfrac{d x}{2}\right)\cos\!\left(\tfrac{s x}{2}\right) 
+ \cos\!\left(\tfrac{d x}{2}\right)
\Big(s x \big( x^2 ( -3 (8 \sqrt{2}+5) d^2 + 3 (8 \sqrt{2}+1) s^2 - 4 ) + 24 \big)\notag\\
&\quad\times 
\sin\!\left(\tfrac{s x}{2}\right) 
+ 2 \big( x^2 ( 3 (8 \sqrt{2}+3) d^2 + 3 (8 \sqrt{2}+3) s^2 - 4 ) + 24 \big)
\cos\!\left(\tfrac{s x}{2}\right)
\Big) 
+ 6 d^2 s x^3 \sin(s x)
- 6 d^2 x^2 \cos(s x)
\hfill\notag\\
&\quad - 48 \sqrt{2}\, d^2 x^2
- 12 d^2 x^2 
+ 6 d s^2 x^3 \sin(d x)
- 24 \sqrt{2}\, d s^2 x^3 \sin\!\left(\tfrac{d x}{2}\right)\cos\!\left(\tfrac{s x}{2}\right)
- 15 d s^2 x^3 \sin\!\left(\tfrac{d x}{2}\right)\cos\!\left(\tfrac{s x}{2}\right) 
\hfill\notag\\
&\quad - 2 (3 s^2 - 1) x^2 \cos(d x)
- 4 d x^3 \sin\!\left(\tfrac{d x}{2}\right)\cos\!\left(\tfrac{s x}{2}\right)
+ 12 d s x^2 \sin\!\left(\tfrac{d x}{2}\right)\sin\!\left(\tfrac{s x}{2}\right) 
+ 96 \sqrt{2}\, d s x^2 \sin\!\left(\tfrac{d x}{2}\right)\sin\!\left(\tfrac{s x}{2}\right)
\hfill\notag\\
&\quad+ 24 d x \sin\!\left(\tfrac{d x}{2}\right)\cos\!\left(\tfrac{s x}{2}\right)
+ 2 d x^3 \sin(d x) - 12 d x \sin(d x) 
- 48 \sqrt{2}\, s^2 x^2 - 12 s^2 x^2
+ 2 s x^3 \sin(s x) + 2 x^2 \cos(s x)\notag\\
&\quad- 12 s x \sin(s x) - 12 \cos(s x) + 4 x^2 - 24
\Big) 
\Bigg) 
+ \frac{8}{12 \kappa^2 (d-s)^4 (d+s)^4}
\Bigg( 
(s+1)^2 (3 d^2 + 2 s - 1)\,\text{Si}((s+1)x)
\notag\\
&\quad+ (s-1)^2 (3 d^2 - 2 s - 1) 
+ (s-1)^2 (3 d^2 - 2 s - 1)\,\text{Si}(x - s x)
- \Big( 3 (8\sqrt{2}+33) d^4
- 6 d^2 \big( (8\sqrt{2}+33) s^2 \notag\\
&\quad+ 16\sqrt{2} + 4 \big)
+ 3 (8\sqrt{2}+33) s^4 
- 24 (4\sqrt{2} + 1) s^2 + 8 \Big)\,\text{Si}(x)
+ (d+s-2)\Big(3 (8\sqrt{2}+1) d^3
+ d^2 (-3 (8\sqrt{2}\notag\\
&\quad +5) s + 48\sqrt{2} + 22 )  
- d \big( 3 (8\sqrt{2}+5) s^2 + (96\sqrt{2}+4) s + 4 \big)   
+ 3 (8\sqrt{2}+1) s^3
+ (48\sqrt{2}+22) s^2 - 4 s - 8
\Big)\,\notag\\
&\quad \times \text{Si}\left(-\tfrac{d x}{2} - \tfrac{s x}{2} + x\right)  
+ (d-s+2)\Big(  
3 (8\sqrt{2}+1) d^3
+ d^2 ( 3 (8\sqrt{2}+5) s - 48\sqrt{2} - 22 )
- d \big( 3 (8\sqrt{2}+5) s^2 \notag\\ 
&\quad + (96\sqrt{2}+4) s + 4 \big) 
- 3 (8\sqrt{2}+1) s^3  
- 2 (24\sqrt{2}+11) s^2 + 4 s + 8 
\Big)\,\text{Si}\left(\tfrac{1}{2}(d - s + 2)x\right)  
+ (d - s - 2)\notag\\  
&\quad \times \Big( 
3 (8\sqrt{2}+1) d^3
+ d^2 ( 3 (8\sqrt{2}+5) s + 48\sqrt{2} + 22 ) 
+ d ( -3 (8\sqrt{2}+5) s^2 + (96\sqrt{2}+4) s - 4 )  - 3 (8\sqrt{2}+1) s^3 \notag\\  
&\quad
+ (48\sqrt{2}+22) s^2 + 4 s - 8 
\Big)\,\text{Si}\left(\tfrac{1}{2}(-d + s + 2)x\right) 
+ (d + s + 2)\Big( 
3 (8\sqrt{2}+1) d^3  
- d^2 ( 3 (8\sqrt{2}+5) s + 48\sqrt{2}\notag\\  
&\quad + 22 )
+ d ( -3 (8\sqrt{2}+5) s^2 + (96\sqrt{2}+4) s - 4 ) 
+ 3 (8\sqrt{2}+1) s^3
- 2 (24\sqrt{2}+11) s^2 - 4 s + 8
\Big)\,\notag\\  
&\quad\text{Si}\left(\tfrac{1}{2}(d + s + 2)x\right)
- 8 (d - 1)^2 (2 d - 3 s^2 + 1)\,\text{Si}(x - d x) 
+ 8 (d + 1)^2 (2 d + 3 s^2 - 1)\,\text{Si}((d + 1)x)
\Bigg), 
\end{align} 
\begin{figure*}[th]
  \centering
  \begin{minipage}[t]{0.48\linewidth}
    \centering
    \includegraphics[width=\linewidth]{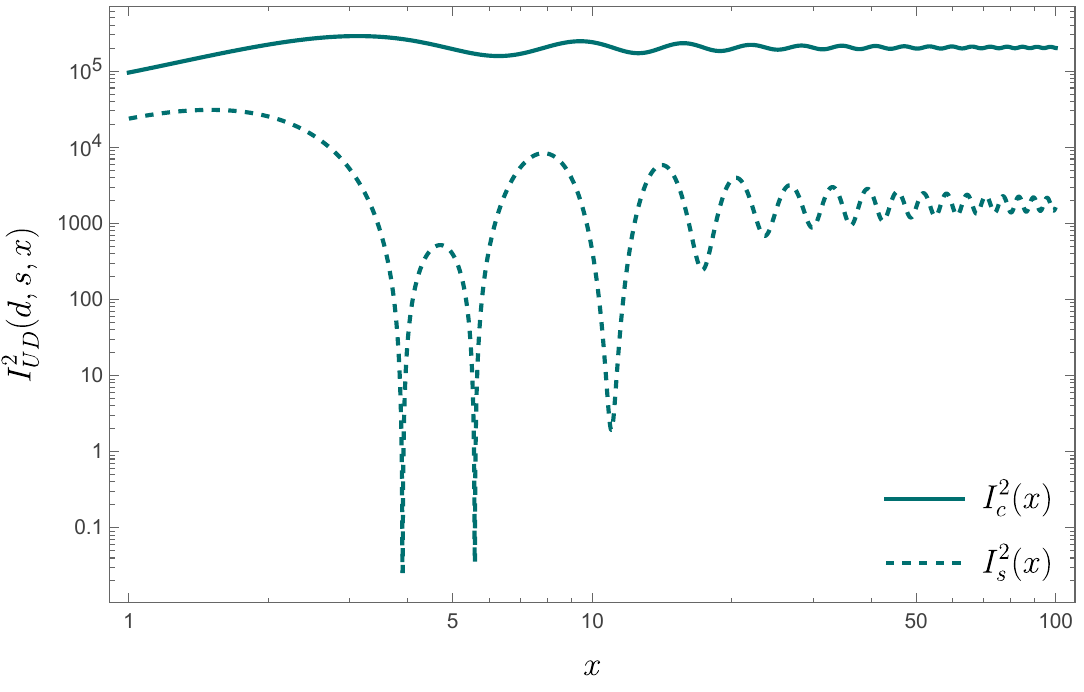}
    \caption{Squared kernels in UD, $I_c^2(x)$ (solid) and $I_s^2(x)$ (dashed), at $(d,s)=(0,1/\sqrt{3})$.
  The large-$x$ behavior does not settle to a bounded evolution: $I_c^2$ approaches a broad plateau while $I_s^2$ continues to rise, indicating a non-decaying late-time readout in this gauge.} 
    \label{fig:UD_kernels}
  \end{minipage}
  \hfill
 \begin{minipage}[t]{0.48\linewidth}
  \centering
\includegraphics[width=\linewidth]{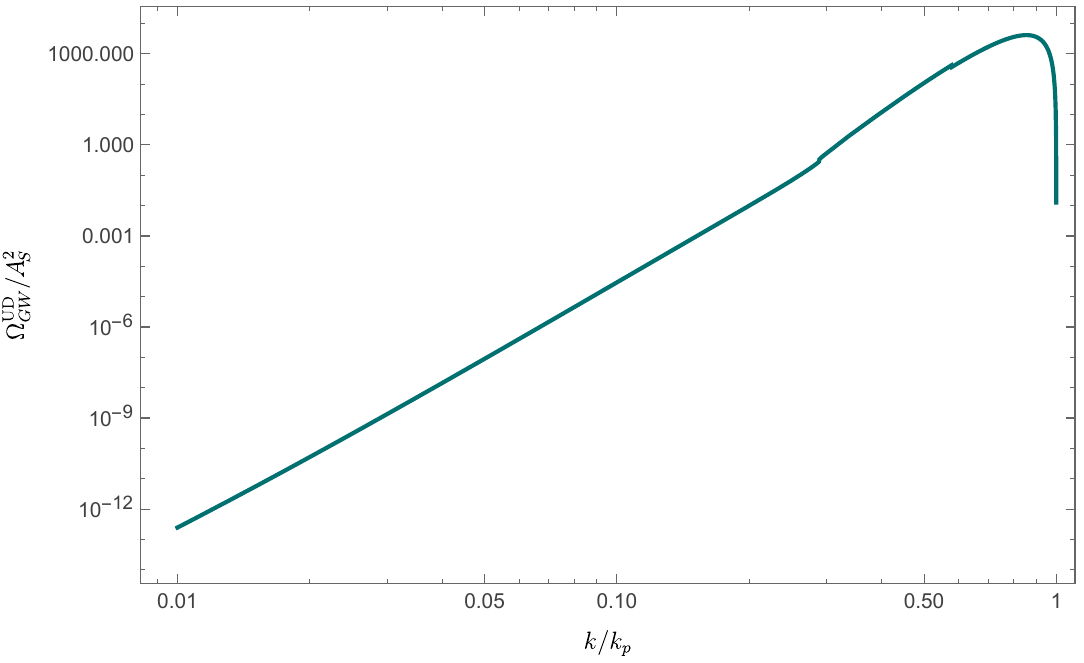}
\caption{$\Omega_{\rm GW}(k)/A_s^2$ in the UD gauge for a Dirac–delta isocurvature peak at $k_p$.
The pre-projection late-time evolution grows with $x$ (divergent relative to longitudinal), while the spectral shape retains the familiar IR rise, peak near $k=2c_s k_p$, and UV cutoff at $k=2k_p$.
The GW spectrum is normalized by the scalar power spectrum amplitude $A_s^2$.}
  \label{fig:UD_spectrum}
\end{minipage}
\end{figure*}
and $I_s^{\text{UD}}$ as follows:  
\begin{align} 
I_s^\text{UD}(d,s,x)=& -\frac{8}{12 \kappa ^2 (d-s)^4 (d+s)^4} (s-1)^2 \left(3 d^2-2 s-1\right) \text{Ci}(x | 1-s| ) 
+ 8 (s+1)^2 \left(3 d^2+2 s-1\right) \text{Ci}(x | s+1| ) 
\notag
\\
&+ (d+s-2) \Big(3 (8 \sqrt{2}+1) d^3 + d^2 (-3 (8 \sqrt{2}+5) s+ 48 \sqrt{2}+22)  - d (3 (8 \sqrt{2}+5) s^2 + (96 \sqrt{2}+4) s \notag
\\
&+ 4) + 3 (8 \sqrt{2}+1) s^3 + (48 \sqrt{2}+22) s^2 - 4 s - 8 \Big) \text{Ci}\left(\tfrac{1}{2} x | -d-s+2| \right) 
+ (d-s+2) \Big(3 (8 \sqrt{2}+1) d^3 \notag
\\
&+ d^2 (3 (8 \sqrt{2}+5) s - 48 \sqrt{2} - 22) - d (3 (8 \sqrt{2}+5) s^2 + (96 \sqrt{2}+4) s + 4) - 3 (8 \sqrt{2}+1) s^3 - 2 (24 \sqrt{2}\notag
\\
&+11) s^2 + 4 s + 8 \Big) 
\text{Ci}\left(\tfrac{1}{2} x | d-s+2| \right) 
+ (d-s-2)\times \Big(3 (8 \sqrt{2}+1) d^3 + d^2 (3 (8 \sqrt{2}+5) s + 48 \sqrt{2} \notag
\\
&+ 22) 
+ d (-3 (8 \sqrt{2}+5) s^2 + (96 \sqrt{2}+4) s - 4) - 3 (8 \sqrt{2}+1) s^3 
+ (48 \sqrt{2}+22) s^2 + 4 s - 8 \Big) \notag\\&\times \text{Ci}\left(\tfrac{1}{2} x | -d+s+2| \right) 
+ (d+s+2) \Big(3 (8 \sqrt{2}+1) d^3 - d^2 (3 (8 \sqrt{2}+5) s + 48 \sqrt{2} + 22) 
+ d (-3 (8 \sqrt{2}+5) s^2 \notag\\& + (96 \sqrt{2}+4) s - 4) + 3 (8 \sqrt{2}+1) s^3  - 2 (24 \sqrt{2}+11) s^2 - 4 s + 8 \Big) \text{Ci}\left(\tfrac{1}{2} x | d+s+2| \right)  
- 8 (d-1)^2 (2 d - 3 s^2 \notag\\&+ 1) \text{Ci}(x | 1-d| )  
+ 8 (d+1)^2 (2 d + 3 s^2 - 1) \text{Ci}(x | d+1| )  
- 4 \text{Ci}(x) \Big(3 (8 \sqrt{2}+33) d^4 - 6 d^2 ((8 \sqrt{2}+33) s^2 \notag\\&+ 16 \sqrt{2} + 4) + 3 (8 \sqrt{2}+33) s^4 
- 24 (4 \sqrt{2}+1) s^2 + 8\Big) + \frac{8}{6 \kappa ^2 x^4 (d-s)^4 (d+s)^4} x \sin (x) \Big( (x^2 (2 d^2 + 3 s^2 - 1) \notag\\&+ 2) \cos (d x) 
- \cos \left(\tfrac{d x}{2}\right) \Big( (x^2 ((24 \sqrt{2}+11) d^2 + (24 \sqrt{2}+11) s^2 - 4) + 8) \cos \left(\tfrac{s x}{2}\right) 
+ 4 s x \sin \left(\tfrac{s x}{2}\right) \Big) + 3 d^2 x^2\notag\\&\times \cos (s x) + 6 d^2 x^2 + 24 \sqrt{2} d^2 x^2 
- 48 \sqrt{2} d s x^2 \sin \left(\tfrac{d x}{2}\right) \sin \left(\tfrac{s x}{2}\right) - 2 d s x^2 \sin \left(\tfrac{d x}{2}\right) \sin \left(\tfrac{s x}{2}\right) 
- 4 d x \sin \left(\tfrac{d x}{2}\right) \notag\\&\cos \left(\tfrac{s x}{2}\right) + 2 d x \sin (d x) + 6 s^2 x^2 + 24 \sqrt{2} s^2 x^2 
+ 2 s^2 x^2 \cos (s x) - x^2 \cos (s x) + 2 s x \sin (s x) + 2 \cos (s x) - 2 x^2\notag\\& + 4 \Big) 
+ 4 \cos (x) \Big( 3 d^3 x^3 \sin \left(\tfrac{d x}{2}\right) \cos \left(\tfrac{s x}{2}\right) + 24 \sqrt{2} d^3 x^3 \sin \left(\tfrac{d x}{2}\right) \cos \left(\tfrac{s x}{2}\right) 
+ \cos \left(\tfrac{d x}{2}\right) \Big( s x (x^2 (-3 (8 \sqrt{2}+5) d^2 \notag\\&+ 3 (8 \sqrt{2}+1) s^2 - 4) + 24) \sin \left(\tfrac{s x}{2}\right)  
+ 2 (x^2 (3 (8 \sqrt{2}+3) d^2 + 3 (8 \sqrt{2}+3) s^2 - 4) + 24) \cos \left(\tfrac{s x}{2}\right) \Big) 
+ 6 d^2 s x^3\notag\\&\times \sin (s x) - 6 d^2 x^2 \cos (s x) - 48 \sqrt{2} d^2 x^2 - 12 d^2 x^2 
+ 6 d s^2 x^3 \sin (d x) - 24 \sqrt{2} d s^2 x^3 \sin \left(\tfrac{d x}{2}\right) \cos \left(\tfrac{s x}{2}\right) 
\notag\\&- 15 d s^2 x^3 \times\sin \left(\tfrac{d x}{2}\right) \cos \left(\tfrac{s x}{2}\right) - 2 ((3 s^2 - 1) x^2 + 6) \cos (d x) 
- 4 d x^3 \sin \left(\tfrac{d x}{2}\right) \cos \left(\tfrac{s x}{2}\right) + 12 d s x^2 \sin \left(\tfrac{d x}{2}\right)\notag\\& \times \sin \left(\tfrac{s x}{2}\right) 
+ 96 \sqrt{2}d s x^2 \sin \left(\tfrac{d x}{2}\right) \sin \left(\tfrac{s x}{2}\right) + 24 d x \sin \left(\tfrac{d x}{2}\right) \cos \left(\tfrac{s x}{2}\right)  + 2 d x^3 \sin (d x) - 12 d x \sin (d x) \notag\\& - 48 \sqrt{2} s^2 x^2 - 12 s^2 x^2 
+ 2 s x^3 \sin (s x) + 2 x^2 \cos (s x) - 12 s x \sin (s x) - 12 \cos (s x) + 4 x^2 - 24 \Big).
\end{align} 
\end{widetext} 

The UD source grows steadily after horizon entry  (Fig.~\ref{fig:UD_source}). 
Accordingly, the kernel functions in Fig.~\ref{fig:UD_kernels} remain of $\mathcal{O}(1)$ at late times rather than showing the $x^{-1}$ decay seen in the longitudinal benchmark. 
Hence, before radiative projection, the GW energy density increases quadratically with conformal time,
\begin{equation}
\Omega_{\rm GW}(x)\;\propto\;x^{2},\qquad x\gg1,
\end{equation} 
i.e.\ it diverges at late times in the UD gauge\footnote{For \emph{isocurvature}-induced GWs in the UD gauge, the late-time growth is the least divergent among the gauges analyzed here: $\Omega_{\rm GW}(x)\propto x^{2}$ for $x\gg1$. By contrast, for adiabatic-induced GWs it is much steeper, $\Omega_{\rm GW}(x)\propto x^{6}$; as we noted in our previous studies  cf.~\cite{Lu:2020diy,Ali:2023moi}, such rapid growth in UD might signal a breakdown of linear perturbation theory.}.     

For a Dirac–delta isocurvature peak, $\mathcal{P}_{S}(k)=\mathcal{A}_{S}\,\delta\!\big(\ln(k/k_{p})\big)$, we evaluate the spectrum along the delta line in $(d,s)$, $d=0$ and $s=\tfrac{2}{\sqrt{3}}(k_{p}/k)$, using Eqs.~\eqref{eq:OmegaGW-dsg2} and \eqref{eq:Ibar2} with the UD kernels. 
The resulting $\Omega_{\rm GW}(k)$ (Fig.~\ref{fig:UD_spectrum}) shows the standard IR rise, a peak at $k=2c_s k_p$, and a sharp cutoff at $k=2k_p$, with a late–time evolution that increases with $x$. 
Exact closed-form expressions for $I_{c/s}^{(\mathrm{UD})}(d,s,x)$ are very long; all figures are obtained by direct evaluation of the exact formulas.
\begin{figure}[th]
\centering
\includegraphics[width=\linewidth]{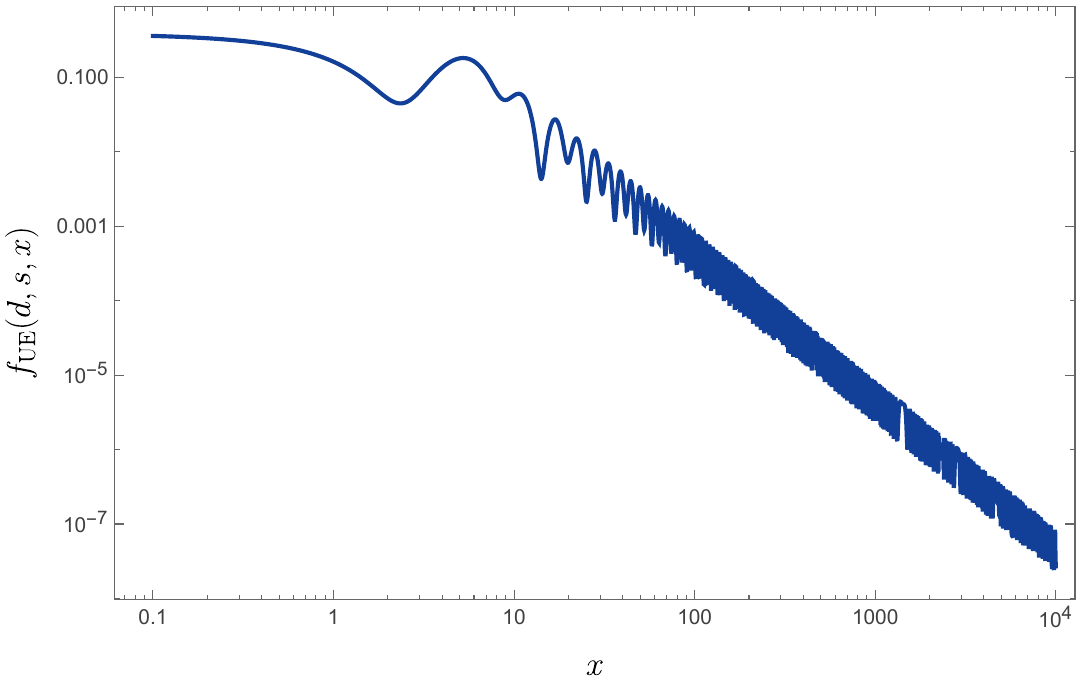}
\caption{Source term $f_{\rm UE}(d,s,x)$ at $(d,s)=(0,1/\sqrt{3})$ versus $x\equiv k\eta$.
After horizon entry the profile remains oscillatory with a gently decaying evolution and only a mild late-time dependence,
indicating that late–time sourcing in this slicing is well behaved.} 
\label{fig:UE_source}
\end{figure}
\subsection{Secondary GWs in Uniform Expansion gauge} Let us begin by considering the uniform expansion gauge with conditions $3(\mathcal{H}\phi+\psi^\prime)-\nabla^2\sigma=0$ and $E=0$.
The transfer function is given below:
\begin{align}
T_{\alpha}(x)
&= \frac{9}{8\,\kappa^{2} x^{4}}
\Bigl(
2\sqrt{2}\,\kappa\Bigl(
- x^{2}
+ 2\sqrt{3}\,x\,\sin\tfrac{x}{\sqrt{3}}
\notag\\
&\quad+ 6\cos\tfrac{x}{\sqrt{3}}
- 6
\Bigr)
+ 2x^{2}\Bigl(\cos\tfrac{x}{\sqrt{3}} - 1\Bigr)
\notag\\
&\quad+ 3x^{2}
- 6\sqrt{3}\,x\,\sin\tfrac{x}{\sqrt{3}}
- 18\cos\tfrac{x}{\sqrt{3}}
+ 18
\Bigr).  
\label{transferexpa}
\end{align}
\begin{figure*}[th]  
\centering
\begin{minipage}[t]{0.48\linewidth}
\centering
\includegraphics[width=\linewidth]{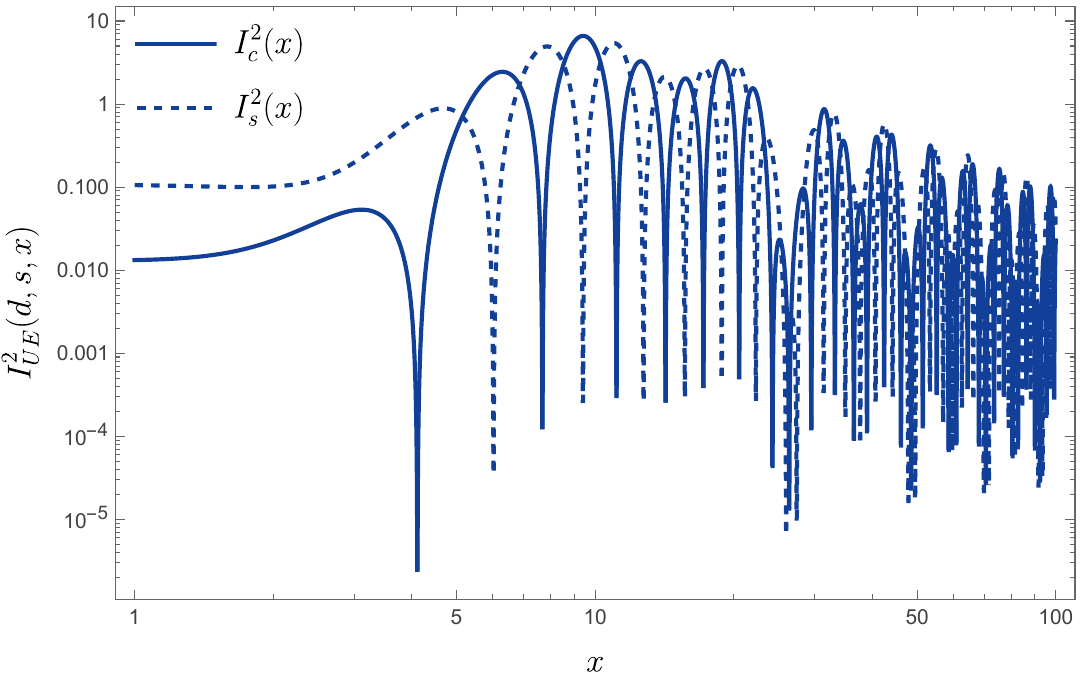}
\caption{Evolution of the source term $f_{\rm UE}(d,s,x)$ as a function of $x\equiv k\eta$ at $(d,s)=(0,1/\sqrt{3})$. For $x\gtrsim1$ the source remains bounded with weak time dependence, indicating well-behaved late-time behavior in the uniform-expansion gauge.}
\label{fig:UE_kernels}
\end{minipage}
\hfill
\begin{minipage}[t]{0.48\linewidth}
\centering
\includegraphics[width=\linewidth]{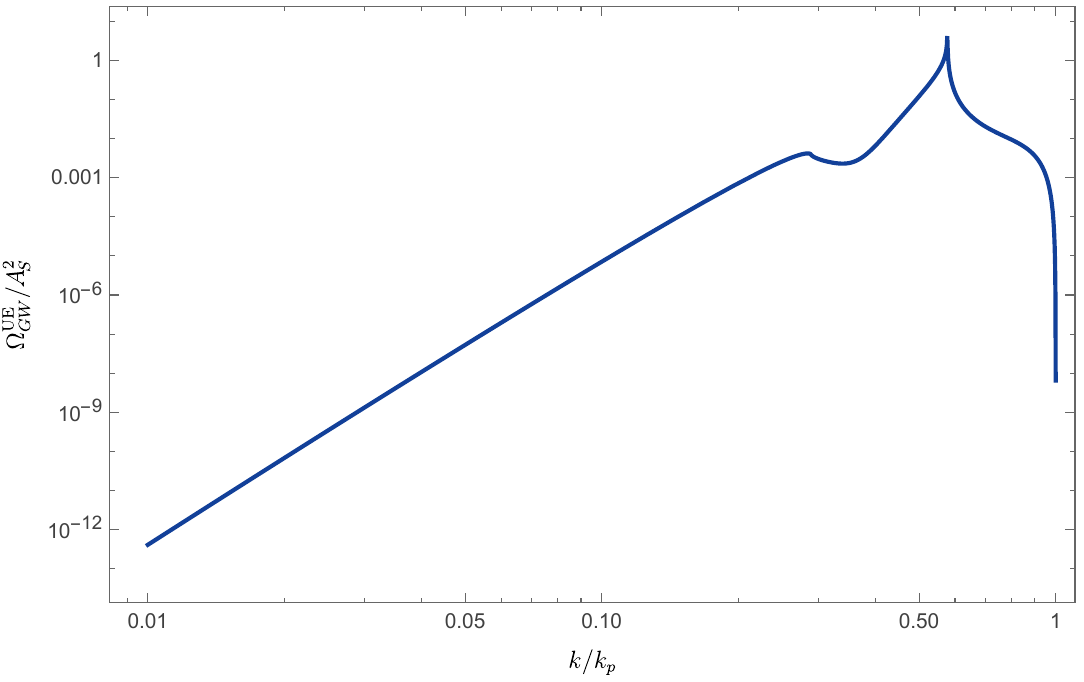}
\caption{$\Omega_{\rm GW}(k)/A_s^2$ in the UE gauge for a Dirac--delta isocurvature peak at $k_p$.
The pre–projection spectrum is nearly time independent at late times (baseline–like), with the standard IR tail
$\propto k^{2}\ln^{2}k$, a peak at $k=2c_s k_p$ ($c_s=1/\sqrt{3}$), and a sharp cutoff at $k=2k_p$.
The GW spectrum is normalized by the scalar power spectrum amplitude $A_s^2$.}
\label{fig:UE_spectrum} 
\end{minipage}
\end{figure*}
We find the source function in the UE gauge as follows:  
\begin{widetext}
\begin{align}
f^{\mathrm{UE}}(d,s,x)
&= \frac{4}{9\,\kappa^{4}\,x^{8}\,(s-d)^{4}(d+s)^{4}}
\Bigg\{
\Big[
3\!\biggl(
\tfrac{3}{4}(1-2\sqrt{2}\kappa)\,x^{2}(s-d)^{2}
+ 4\sqrt{6}\kappa\,x(s-d)
- 36\sqrt{2}\kappa + 54
\biggr)
+ 3x(s-d)\!\notag\\
&\quad\times\!\biggl(
\tfrac{3}{4}x^{2}(s-d)^{2}
- 2\sqrt{6}\kappa\,x(s-d)
+ 18\sqrt{2}\kappa - 27
\biggr)
\sin\!\left(\tfrac{x}{2}(s-d)\right) 
+ 2\!\biggl(
\tfrac{3}{2}\sqrt{\tfrac{3}{2}}\,\kappa\,x^{3}(s-d)^{3}
+ \tfrac{9}{2}(\sqrt{2}\kappa\notag\\
&\quad -2)\,x^{2}(s-d)^{2}
- 6\sqrt{6}\kappa\,x(s-d)
+ 54\sqrt{2}\kappa - 81
\biggr)
\cos\!\left(\tfrac{x}{2}(s-d)\right)
\Big] \times
\Big[
3\!\biggl(
\tfrac{3}{4}(1-2\sqrt{2}\kappa)\,x^{2}(d+s)^{2}
\notag\\[2pt]
&\quad+ 4\sqrt{6}\kappa\,x(d+s)
- 36\sqrt{2}\kappa + 54
\biggr)
+ 3x(d+s)\!\times\!\biggl(
\tfrac{3}{4}x^{2}(d+s)^{2}
- 2\sqrt{6}\kappa\,x(d+s)
+ 18\sqrt{2}\kappa - 27
\biggr) 
\notag\\
&\quad \times  \sin\!\left(\tfrac{x}{2}(d+s)\right) 
+ 2\!\biggl(
\tfrac{3}{2}\sqrt{\tfrac{3}{2}}\,\kappa\,x^{3}(d+s)^{3}
- \tfrac{9}{2}(\sqrt{2}\kappa-2)\,x^{2}(d+s)^{2}
- 6\sqrt{6}\kappa\,x(d+s)
+ 54\sqrt{2}\kappa \notag\\[2pt]
&\quad- 81
\biggr)
\cos\!\left(\tfrac{x}{2}(d+s)\right)
\Big] 
+ 2\Big[
\tfrac{3}{2}\sqrt{\tfrac{3}{2}}\,\kappa\,x^{3}(s-d)^{3}
- \frac{9\kappa}{\sqrt{2}}\,x^{2}(s-d)^{2}
+ \tfrac{9}{4}x^{2}(s-d)^{2}
+ 6\sqrt{6}\kappa\,x(s-d)
\notag\\
&\quad- 36\sqrt{2}\kappa + 54 
- 3x(s-d)\!\biggl(
2\sqrt{2}\kappa\!\left(\tfrac{\sqrt{3}}{2}x(s-d)-3\right)
+ 9
\biggr)
\sin\!\left(\tfrac{x}{2}(s-d)\right) 
+ 6\!\biggl(
\tfrac{\sqrt{3}}{2}x(s-d)-3
\biggr)
\!\biggl(
\tfrac{\sqrt{3}}{2}x(s-d)\notag\\
&\quad-2\sqrt{2}\kappa+3
\biggr)
\cos\!\left(\tfrac{x}{2}(s-d)\right)
\Big]\times
\Big[
\tfrac{3}{2}\sqrt{\tfrac{3}{2}}\,\kappa\,x^{3}(d+s)^{3}
- \frac{9\kappa}{\sqrt{2}}\,x^{2}(d+s)^{2}
+ \tfrac{9}{4}x^{2}(d+s)^{2}
+ 6\sqrt{6}\kappa\,x(d+s)
\notag\\
&\quad- 36\sqrt{2}\kappa + 54 
- 3x(d+s)\!\biggl(
2\sqrt{2}\kappa\!\left(\tfrac{\sqrt{3}}{2}x(d+s)-3\right)
+ 9
\biggr)
\sin\!\left(\tfrac{x}{2}(d+s)\right) 
+ 6\!\biggl(
\tfrac{\sqrt{3}}{2}x(d+s)-3
\biggr)
\notag\\
&\quad\!\times \biggl(
\tfrac{\sqrt{3}}{2}x(d+s)-2\sqrt{2}\kappa+3
\biggr)
\cos\!\left(\tfrac{x}{2}(d+s)\right)
\Big]
\Bigg\}.
\label{UEex}
\end{align}
\end{widetext}
This expression \eqref{UEex} decays as $x^{-1}$ when $x\rightarrow\infty$, 
indicating that $\Omega_{\mathrm{GW}}^{\mathrm{UE}}$ approaches $\Omega_{\mathrm{GW}}^{\mathrm{Long.}}$ at late time. For secondary GWs, we obtain the same energy density, $\Omega_{\mathrm{GW}}$,  in the uniform expansion gauge as that in the longitudinal gauge.  
	
By substituting \eqref{transferexpa} into \eqref{xitransfer}, we have 
\begin{widetext} 
\begin{align}
I_\chi^{\mathrm{UE}}(d,s,x)&= -\frac{32}{27\,\kappa^{4}\,x^{8}\,(s-d)^{5}(d+s)^{5}}
\Bigl[
-\frac{3\kappa x^{2}(s-d)^{2}}{\sqrt{2}}
+2\!\left(\frac{3}{4}x^{2}(s-d)^{2}+6\sqrt{2}\kappa-9\right)\!
\cos\Bigl(\tfrac{1}{2}x(s-d)\Bigr)\notag\\
&\quad+\frac{3}{4}x^{2}(s-d)^{2}
+3\left(2\sqrt{2}\kappa-3\right)x(s-d)\,
\sin\Bigl(\tfrac{1}{2}x(s-d)\Bigr)
-12\sqrt{2}\kappa+18
\Bigr]
\times
\Bigl[
-\frac{3\kappa x^{2}(d+s)^{2}}{\sqrt{2}}\notag				\\ 
&\quad+2\!\left(\frac{3}{4}x^{2}(d+s)^{2}+6\sqrt{2}\kappa-9\right)\!
\cos\Bigl(\tfrac{1}{2}x(d+s)\Bigr)
+\frac{3}{4}x^{2}(d+s)^{2}
+3\left(2\sqrt{2}\kappa-3\right)x(d+s)\,\notag			\\ 
&\quad\times \sin\Bigl(\tfrac{1}{2}x(d+s)\Bigr)
-12\sqrt{2}\kappa+18 
\Bigr], 
\end{align} 
and the analytic expressions for $Ic$ and $Is$ are given as:  
\begin{align} 
I_c^{\mathrm{UE}}(d,s,x) &= \frac{3}{4\,\kappa^{2}}
\Biggl\{
\frac{8}{\bigl(d^{2}-s^{2}\bigr)^{3}\,x^{4}}
\Bigl[
\sin x\,\Bigl(
-4\,(d^{2}+s^{2}-1)\,x^{2}
+s\Bigl(\bigl((d^{2}+3s^{2}-4)\,x^{2}+24\bigr)\cos\tfrac{d x}{2}\Bigl)
+12\,d\,x\,\sin\tfrac{d x}{2}\Bigr)\,\notag\\
&\quad\times 
\sin\tfrac{s x}{2}\,x
+2d\bigl(-(s^{2}-1)\,x^{2}-6\bigr)\sin(d x)\,x
+2s\bigl(-(d^{2}-1)\,x^{2}-6\bigr)\times\sin(s x)\,x
+2\bigl((2d^{2}-s^{2}+1)\,x^{2}\notag\\
&\quad-6\bigr)\cos(d x)
+2\bigl((-d^{2}+2s^{2}+1)\,x^{2}-6\bigr)\cos(s x)
+\cos\tfrac{s x}{2}\Bigl(
2\bigl((d^{2}+s^{2}-4)\,x^{2}+24\bigr)\cos\tfrac{d x}{2}
+d\,x\,\bigl((3d^{2}\notag\\
&\quad+s^{2}-4)\,x^{2}+24\bigr)\sin\tfrac{d x}{2}
\Bigr)
-24\Bigr)
+2x\cos x\,\Bigl(
-\bigl((s^{2}-1)\,x^{2}+2\bigr)\cos(d x)
-\bigl((d^{2}-1)\,x^{2}+2\bigr)\cos(s x)\notag\\
&\quad
+\cos\tfrac{s x}{2}\Bigl(
\bigl((3d^{2}+3s^{2}-4)\,x^{2}+8\bigr)\cos\tfrac{d x}{2}
+4d x \sin\tfrac{d x}{2}
\Bigr)
+2s x\Bigl(2\cos\tfrac{d x}{2}+d x \sin\tfrac{d x}{2}\Bigr)\sin\tfrac{s x}{2}
-2\bigl((d^{2}+s^{2}\notag\\
&\quad-1)\,x^{2}+d\sin(d x)\,x+2\bigr)
-2s x \sin(s x)
\Bigr)
\Bigr] 
+\frac{2}{(d-s)^{3}(d+s)^{3}}
\Bigl[
\mathrm{Si}\bigl((d+1)x\bigr)\,(d^{2}+s^{2}-2)^{2}
+\mathrm{Si}\bigl((s+1)\notag\\
&\quad \times x\bigr)\,(d^{2}+s^{2}-2)^{2}
+\mathrm{Si}\bigl((1-d)x\bigr)\,(d^{2}+s^{2}-2)^{2}
+\mathrm{Si}\bigl((1-s)x\bigr)\,(d^{2}+s^{2}-2)^{2}
\Bigr]
-\bigl(3d^{4}+4sd^{3}+2(s^{2}\notag\\
&\quad-8)d^{2}+4s^{3}d+3s^{4}-16s^{2}+16\bigr)
\mathrm{Si}\Bigl(\tfrac{-d-s+2}{2}\,x\Bigr)
+4\bigl(d^{4}-2(s^{2}+4)d^{2}+s^{4}-8s^{2}+8\bigr)\mathrm{Si}(x)
+\bigl(-3d^{4}\notag\\
&\quad+4sd^{3}-2(s^{2}-8)d^{2}+4s^{3}d-3s^{4}+16s^{2}-16\bigr)
\mathrm{Si}\Bigl(\tfrac{d-s+2}{2}\,x\Bigr)
+\bigl(-3d^{4}+4sd^{3}-2(s^{2}-8)d^{2}+4s^{3}d\notag\\
&\quad-3s^{4}+16s^{2}-16\bigr)
\mathrm{Si}\Bigl(\tfrac{-d+s+2}{2}\,x\Bigr)
-\bigl(3d^{4}+4sd^{3}+2(s^{2}-8)d^{2}+4s^{3}d+3s^{4}-16s^{2}+16\bigr)
\mathrm{Si}\Bigl(\tfrac{d+s+2}{2}\,x\Bigr)\notag\\
&\quad  
-\frac{32}{27\,\kappa^{4}\,x^{8}\,(s-d)^{5}(d+s)^{5}} \Bigl[
\Bigl(
\tfrac{3}{4}(s-d)^{2}x^{2}
-\tfrac{3(s-d)^{2}\kappa\,x^{2}}{\sqrt{2}}
+3(2\sqrt{2}\kappa-3)\,x(s-d)\times\sin\tfrac{(s-d)x}{2}
-12\sqrt{2}\kappa \notag\\
&\quad 
+2\bigl(\tfrac{3}{4}(s-d)^{2}x^{2}+6\sqrt{2}\kappa-9\bigr)\cos\tfrac{(s-d)x}{2}
+18
\Bigr)\times
\Bigl(
\tfrac{3}{4}(d+s)^{2}x^{2}
-\tfrac{3(d+s)^{2}\kappa\,x^{2}}{\sqrt{2}}
+3(2\sqrt{2}\kappa-3)\,x(d+s)\notag\\
&\quad\sin\tfrac{(d+s)x}{2}
-12\sqrt{2}\kappa
+2\bigl(\tfrac{3}{4}(d+s)^{2}x^{2}+6\sqrt{2}\kappa-9\bigr)\cos\tfrac{(d+s)x}{2}
+18
\Bigr)
\Bigr]
\Biggr\},\label{IcUE}
\end{align}
and
\begin{align}
I_s^{\mathrm{UE}}(d,s,x)
&= \frac{3}{4\,\kappa^{2}\,(d-s)^{3}(d+s)^{3}}
\Bigr(
-2\,\mathrm{Ci}\bigl(x\lvert 1-d\rvert\bigr)\,(d^{2}+s^{2}-2)^{2}
-2\,\mathrm{Ci}\bigl(x\lvert d+1\rvert\bigr)\,(d^{2}+s^{2}-2)^{2}
-2\,\mathrm{Ci}\bigl(x\lvert 1-s\rvert\bigr)\,(d^{2}\notag\\
&\quad +s^{2}-2)^{2}
-2\,\mathrm{Ci}\bigl(x\lvert s+1\rvert\bigr)\,(d^{2}+s^{2}-2)^{2}
-4\bigl(d^{4}-2(s^{2}+4)d^{2}+s^{4}-8s^{2}+8\bigr)\mathrm{Ci}(x)
+\bigl(3d^{4}+4sd^{3}+2(s^{2}\notag\\
&\quad-8)d^{2}+4s^{3}d+3s^{4}-16s^{2}+16\bigr)
\times\mathrm{Ci}\Bigl(\tfrac{x}{2}\,\lvert{-}d{-}s{+}2\rvert\Bigr)
			+\bigl(3d^{4}-4sd^{3}+2(s^{2}-8)d^{2}-4s^{3}d+3s^{4}-16s^{2}\notag\\
			&\quad+16\bigr)
			\times\mathrm{Ci}\Bigl(\tfrac{x}{2}\,\lvert d{-}s{+}2\rvert\Bigr)
			+\bigl(3d^{4}-4sd^{3}+2(s^{2}-8)d^{2}-4s^{3}d+3s^{4}-16s^{2}+16\bigr)
			\times\mathrm{Ci}\Bigl(\tfrac{x}{2}\,\lvert{-}d{+}s{+}2\rvert\Bigr)
			+\bigl(3d^{4}\notag\\
			&\quad+4sd^{3}+2(s^{2}-8)d^{2}+4s^{3}d+3s^{4}-16s^{2}+16\bigr)
			\times\mathrm{Ci}\Bigl(\tfrac{x}{2}\,\lvert d{+}s{+}2\rvert\Bigr)
			\Bigr)
			-\frac{6}{(d^{2}-s^{2})^{3}\,x^{4}\,\kappa^{2}}
			\Bigr(
			2x\sin x\Bigl(
			((s^{2}\notag\\
			&\quad-1)x^{2}+2)\cos(d x)
			+((d^{2}-1)x^{2}+2)\cos(s x)
			+\cos\tfrac{s x}{2}\Bigl(-((3d^{2}+3s^{2}-4)x^{2}+8)\cos\tfrac{d x}{2}-4d x\sin\tfrac{d x}{2}\Bigr)
			\notag\\
			&\quad-2s x\bigl(2\cos\tfrac{d x}{2}
			+d x\sin\tfrac{d x}{2}\bigr)\sin\tfrac{s x}{2}
			+2\bigl((d^{2}+s^{2}-1)x^{2}+d\sin(d x)\,x+2\bigr)
			+2s x\sin(s x) 
			\Bigr) 
			+\cos x\Bigl( 
			-4(d^{2}\notag\\
			&\quad+s^{2}-1)x^{2}
			+s\bigl((d^{2}+3s^{2}-4)x^{2}+24\bigr)\cos\tfrac{d x}{2}\,\sin\tfrac{s x}{2}\,x 
			+12d x\,\sin\tfrac{d x}{2}\,\sin\tfrac{s x}{2}\,x
			+2d(-s^{2}x^{2}+x^{2}-6)\notag\\
			&\quad \times\sin(d x)\,x  
			+2s(-d^{2}x^{2}+x^{2}-6)\sin(s x)\,x 
			+2\bigl((2d^{2}-s^{2}+1)x^{2}-6\bigr)\cos(d x) 
			+2\bigl((-d^{2}+2s^{2}+1)x^{2}\notag\\
			&\quad-6\bigr)\cos(s x)  
			+\cos\tfrac{s x}{2}\Bigl(
			2\bigl((d^{2}+s^{2}-4)x^{2}+24\bigr)\cos\tfrac{d x}{2}
			+d x\bigl((3d^{2}+s^{2}-4)x^{2}+24\bigr)\sin\tfrac{d x}{2}
			\Bigr)
			-24
			\Bigr)
			\Bigr)
			\notag\\
			&\quad-\frac{32}{27\,(s-d)^{5}(d+s)^{5}\,x^{8}}
			\Biggl(
			\Bigl(
			\tfrac{3}{4}(s-d)^{2}x^{2}-\tfrac{3}{\sqrt{2}}(s-d)^{2}x^{2}
			+3(2\sqrt{2}-3)x(s-d)\sin\tfrac{(s-d)x}{2}
			+2\bigl(\tfrac{3}{4}(s-d)^{2}x^{2}\notag\\
&\quad+6\sqrt{2}-9\bigr)\cos\tfrac{(s-d)x}{2}
-12\sqrt{2}+18
\Bigr)\times
\Bigl(\tfrac{3}{4}(d+s)^{2}x^{2}-\tfrac{3}{\sqrt{2}}(d+s)^{2}x^{2}
+3(2\sqrt{2}-3)x(d+s)\sin\tfrac{(d+s)x}{2}
\notag\\
&\quad+2\bigl(\tfrac{3}{4}(d+s)^{2}x^{2}+6\sqrt{2}-9\bigr)\cos\tfrac{(d+s)x}{2}
-12\sqrt{2}+18
\Bigr)
\Biggr)
-\frac{32}{27\,(s-d)^{5}(d+s)^{5}\,x^{8}\,\kappa^{4}}
\Biggl(
\Bigl(
\tfrac{3}{4}(s-d)^{2}x^{2}\notag\\
&\quad-\tfrac{3\kappa}{\sqrt{2}}(s-d)^{2}x^{2}
+3(2\sqrt{2}\kappa-3)x\times(s-d)\sin\tfrac{(s-d)x}{2}
-12\sqrt{2}\kappa
+2\bigl(\tfrac{3}{4}(s-d)^{2}x^{2}+6\sqrt{2}\kappa-9\bigr)\cos\tfrac{(s-d)x}{2}
\notag\\
&\quad+18
\Bigr)
\Bigl(
\tfrac{3}{4}(d+s)^{2}x^{2}-\tfrac{3\kappa}{\sqrt{2}}(d+s)^{2}x^{2}
-12\sqrt{2}\kappa
+2\bigl(\tfrac{3}{4}(d+s)^{2}x^{2}\notag\\
&\quad+6\sqrt{2}\kappa-9\bigr)\cos\tfrac{(d+s)x}{2}
+18
\Bigr)
\Biggr).
\label{IsUE}
\end{align}
\end{widetext}
    
The UE integrals  are obtained by inserting the UE source into the time integrals \eqref{ICS}. The evolution of the second-order source term transfer functions is illustrated in Fig. \ref{fig:UE_source}.  
The resulting time evolution of the kernels and gravitational wave spectrum is displayed in Figs.~\ref{fig:UE_kernels} and  \ref{fig:UE_spectrum} respectively.  
At late times the kernel functions decrease roughly as $1/x$, so $\overline{I^2}\sim x^{-2}$ and the pre–projection energy density approaches a constant, in line with the longitudinal baseline. 

For the Dirac--delta isocurvature peak,
$\mathcal{P}_S(k)=\mathcal{A}_S\,\delta\!\big(\ln(k/k_p)\big)$,
we evaluate along the delta line $d=0$, $s=\tfrac{2}{\sqrt{3}}(k_p/k)$ using
Eqs.~\eqref{eq:OmegaGW-dsg2} and \eqref{eq:Ibar2} with the UE kernels.
The resulting $\Omega_{\rm GW}(k)$ (Fig.~\ref{fig:UE_spectrum}) shows the standard
$k^{2}\ln^{2}\!k$ infrared rise, a peak at $k=2c_s k_p$, and the cutoff at $k=2k_p$.

\subsection{Secondary GWs in Newtonian–motion (Nm) gauge}
\label{subsec:Nm}

In the Newtonian–motion gauge we impose the conditions in Eq.~\eqref{eq:gauge-conditions},
which reduce the scalar sector to the two metric potentials $\alpha$ and $\beta$ obeying, during
radiation domination and to $\mathcal{O}(\kappa^{-1})$,
\begin{align}
T_\alpha'(x)
+\Bigl(\frac{1}{x}+\frac{1}{4\sqrt{2}\,\kappa}\Bigr)T_\alpha(x)&=0,
\label{eq:NmTaEOM}\\
T_\beta''(x)
+\Bigl(\frac{1}{x}+\frac{1}{4\sqrt{2}\,\kappa}\Bigr)T_\beta'(x)&=0.
\label{eq:NmTbEOM}
\end{align} 

Keeping the exact prefactor $1/(x+4\sqrt{2}\,\kappa)$ and expanding a posteriori in $\kappa^{-1}$,
Eqs.~\eqref{eq:NmTaEOM}–\eqref{eq:NmTbEOM} integrate to
\begin{align}
T_\alpha(x)&=\frac{C_\alpha}{x\,(4\sqrt{2}\,\kappa+x)},\label{eq:NmTaGen}\\
T_\beta'(x)&=\frac{C_\beta}{x\,(4\sqrt{2}\,\kappa+x)},\\
T_\beta(x)&=C_0+C_\beta\ln\!\Bigl(\frac{x}{\,4\sqrt{2}\,\kappa+x}\Bigr),
\label{eq:NmTbGen}
\end{align}
with constants $C_\alpha,C_\beta,C_0$ fixed by super-horizon matching to the chosen isocurvature initial data.
The Nm kernel admits the standard decomposition $I_{\rm Nm}=I_\chi+I_{\rm Long}$, where $I_\chi$
contains only non-oscillatory phases.  Thus the radiative part that sources the free tensor mode is 
the longitudinal one.  

\begin{figure}[t] 
\centering
\includegraphics[width=\linewidth]{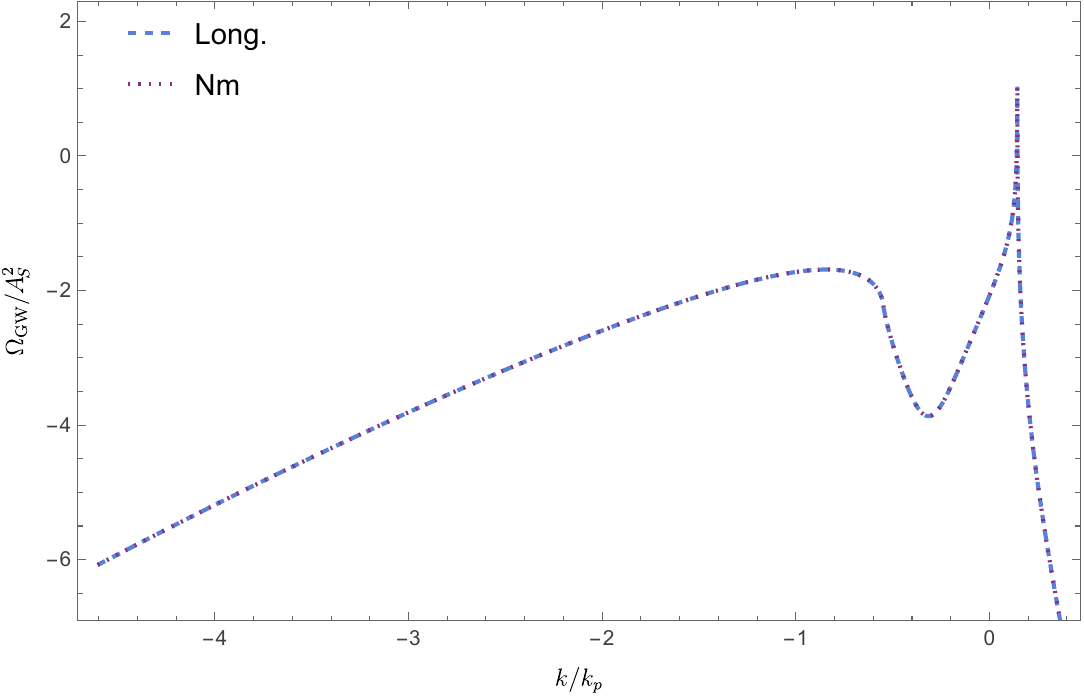}
\caption{Nm vs.\ longitudinal (Dirac–delta peak at $k_p$): identical late–time spectrum after the radiative (luminal) selection. The curves coincide over the full support $k\le 2k_p$, indicating that the Nm choice adds no gauge–specific modulation; retaining the radiative $\sin x/\cos x$ pieces reproduces the longitudinal observable.}  
\label{fig:Nm-vs-Long}
\end{figure}

At late times ($x\gg1$) the time-evolution of Nm gauge kernel follows the same $x^{-1}$ decay as in the case of
longitudinal gauge, so the pre–projection energy density approaches a constant. Consequently, the observable
spectrum $\Omega_{\rm GW}(k)$ coincides with the longitudinal result. Accordingly, we only show the energyspectrum in Fig.~\ref{fig:Nm-vs-Long} and do not duplicate the source or kernel figures.

For a Dirac–delta isocurvature peak,
$\mathcal{P}_{S}(k)=\mathcal{A}_{S}\,\delta\!\big(\ln(k/k_{p})\big)$,
we evaluate along the delta line $d=0$, $s=\tfrac{2}{\sqrt{3}}(k_{p}/k)$ using
Eqs.~\eqref{eq:OmegaGW-dsg2} and \eqref{eq:Ibar2} with the Nm kernels.
The spectrum shown in Fig.~\ref{fig:Nm-vs-Long} follows the baseline shape with
IR $k^{2}\ln^{2}\!k$, a peak at $k=2c_s k_p$ ($c_s=1/\sqrt{3}$), and a cutoff at $k=2k_p$.

\subsection{Secondary GWs in N-body (Nb) gauge} 

In the N-body (Nb) gauge, using conditions \eqref{eq:gauge-conditions}, we find the following equation:  
\begin{align}
\,T_B''(x)
+\Bigl(\tfrac{2}{x}+\tfrac{1}{4\sqrt{2}\,\kappa}\Bigr)T_B'(x)
+\tfrac{1}{3}\,T_B(x)=0.
\end{align}  
We normalize to unit primordial isocurvature amplitude and impose regular super-horizon behavior. This uniquely fixes the overall normalization. The resulting transfer functions are as follows \begin{align}
T_B(x)&=\frac{\sin\bigl(x/\sqrt{3}\bigr)}{x}
\Bigl[1-\frac{x}{8\sqrt{2}\,\kappa}\Bigr], \label{Nbg1}
\\[4pt]
\,T_\phi(x)&= -\frac{3}{\sqrt{2}\,\kappa\,x}
\Bigl[1-\cos\bigl(\tfrac{x}{\sqrt{3}}\bigr)\Bigr], \label{Nbg2}
\\[4pt]
\,T_\psi(x)&= \frac{1}{\sqrt{2}\,\kappa}
\Biggl[\frac{\sin\bigl(x/\sqrt{3}\bigr)}{\sqrt{3}\,x}
-\frac{1-\cos\bigl(x/\sqrt{3}\bigr)}{x^{2}}\Biggr].
\label{Nbg3}
\end{align} 
Utilizing these Eqs. \eqref{Nbg1}-\eqref{Nbg3}, we find the source function as follows:  
   \begin{figure}[th]
\centering
\includegraphics[width=\linewidth]{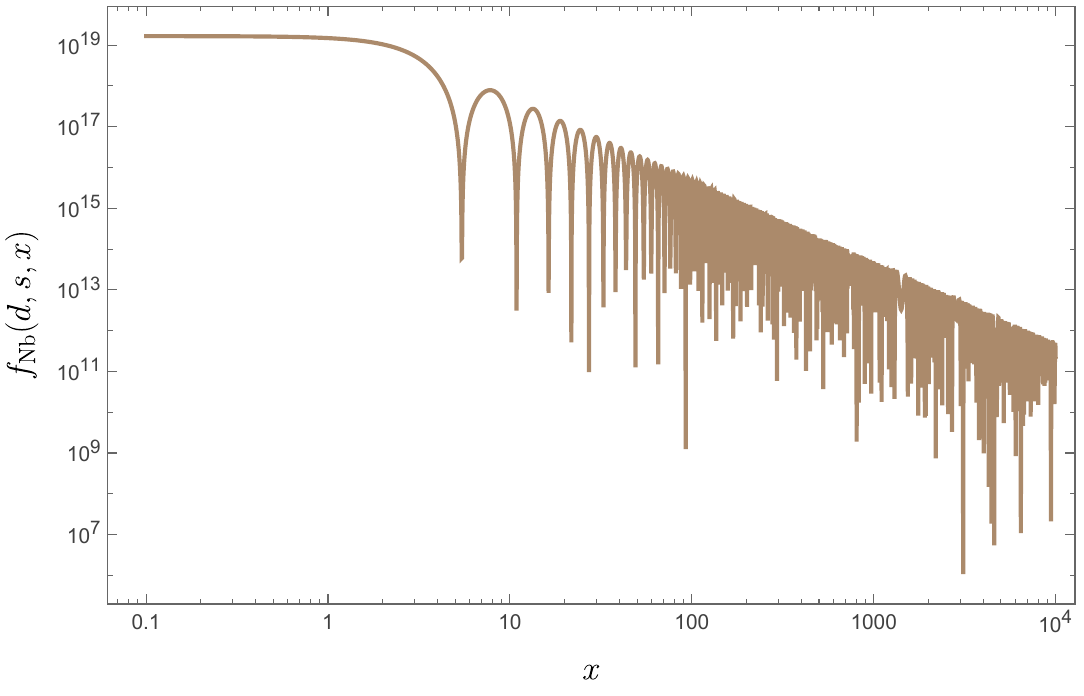}
\caption{Source $f_{\rm Nb}(d,s,x)$ in the N–body gauge at $(d,s)=(0,1/\sqrt{3})$ as a function of $x\equiv k\eta$. It is nearly constant on superhorizon scales, turns over near horizon entry, and for $x\gg1$ shows a decaying time dependence $\propto x^{-2}$. The phase and intermediate–time behaviour differ from longitudinal, whereas the late–time decay is the same.} 
\label{fig:NB-source}
\end{figure}
\begin{widetext}
\begin{align}
f^{\text{Nb}}(d,s,x)
&= \frac{8}{1296 (s-d)^2 (d+s)^2 x^4 \kappa^2}\,
\cos\big(\tfrac12(d+s)x\big)\Big(
48\Big(\tfrac92(s-d)x\sqrt{3}
+ \tfrac12(d+s)x\big(\tfrac12(d+s)x\big(2-\tfrac32\sqrt{3}(s-d)x\big)\notag\\&\times\sqrt{3}+9\big)\sqrt{3}
- 15\Big) + 16\Big(\tfrac{9}{16}(s-d)^2(d+s)^2 x^4
+ \tfrac{27}{4}(s-d)(d+s)\big(\tfrac12(s-d)\sqrt{3}+\tfrac12(d+s)\sqrt{3}\big)x^3
- 6\big(\tfrac34(s\notag\\&-d)^2+\tfrac34(d+s)^2\big)x^2 - 27\big(\tfrac12(s-d)\sqrt{3}+\tfrac12(d+s)\sqrt{3}\big)x + 45\Big)
\cos\big(\tfrac12(s-d)x\big) + \Big(72\big(\tfrac12\sqrt{3}(d+s)x\big(\tfrac12\sqrt{3}(d\notag\\&+s)x-3\big)-6\big)\kappa\sqrt{2}
+ \tfrac12(s-d)x\Big(6(40\sqrt{3}+9) - \tfrac12\sqrt{3}(d+s)x\big(48(d+s)x+\tfrac92(d+s)\sqrt{3}x+144\sqrt{3}-27\big)\Big)\notag\\&\times\sqrt{3}\Big) 
\sin\big(\tfrac12(s-d)x\big)\Big) + 24\Big(48\Big(5-3\big(\tfrac12(s-d)\sqrt{3}+\tfrac12(d+s)\sqrt{3}\big)x\Big)
+ \Big(72\big(\tfrac12(s-d)x\sqrt{3}+2\big)\kappa\sqrt{2}  + \tfrac12(d\notag\\&+s)x\big(-\tfrac12 9\sqrt{3}(s-d)x+72\times(s-d)x-80\sqrt{3}-18\big)\sqrt{3}\Big) 
\sin\big(\tfrac12(d+s)x\big)\Big)  + 8\cos\big(\tfrac12(s-d)x\big)\Big(48\Big(-\frac98 \notag\\&\times\sqrt{3}(s-d)^2(d+s)x^3
+ \tfrac32(s-d)^2 x^2
+ 9\big(\tfrac12(s-d)\sqrt{3}+\tfrac12(d+s)\sqrt{3}\big)x - 15\Big) + \Big(72\big(\tfrac12\sqrt{3}(s-d)x\big(\tfrac12\sqrt{3}(s\notag\\&-d)x-3\big)-6\big)\kappa\sqrt{2}
+ \tfrac12(d+s)x\Big(6(40\sqrt{3}+9) - \tfrac12\sqrt{3}(s-d)x\big(48(s-d)x+\tfrac92(s-d)\sqrt{3}x+144\sqrt{3}-27\big)\Big)\notag\\& \times\sqrt{3}\Big)
\sin\big(\tfrac12(d+s)x\big)\Big)  + 3\sin\big(\tfrac12(s-d)x\big)\Big(
	8\Big(72\times\big(\tfrac12(d+s)x\sqrt{3}+2\big)\kappa\sqrt{2}
+ \tfrac12(s-d)x\big(-\tfrac12 9\sqrt{3}(d+s)x\notag\\&+72(d+s)x-80\sqrt{3}-18\big)\sqrt{3}\Big) + x\Big(\tfrac14(-9)(s-d)(d+s)x^3
+ 24\big(\tfrac12(s-d)\sqrt{3}+\tfrac12(d+s)\sqrt{3}\big)\big(\kappa\sqrt{2}+\tfrac34(s\notag\\&-d)(d+s)\sqrt{3}\big)x^2 - 32\Big(9\kappa\sqrt{\tfrac32}(d+s)^2
- \tfrac34(s-d)(3\sqrt{3}+20)(d+s)
+ 6\kappa\big(\tfrac32\sqrt{\tfrac32}(s-d)^2+2\kappa\big)\Big)x \notag\\&- 384\sqrt{6}\big(\tfrac12(s-d)\sqrt{3}+\tfrac12(d+s)\sqrt{3}\big)\kappa\Big)
\sin\big(\tfrac12(d+s)x\big)\Big). 
\label{eq:fNb}
\end{align}
\end{widetext}
\begin{figure*}[th]
\centering
\begin{minipage}[t]{0.48\linewidth}
\centering
\includegraphics[width=\linewidth]{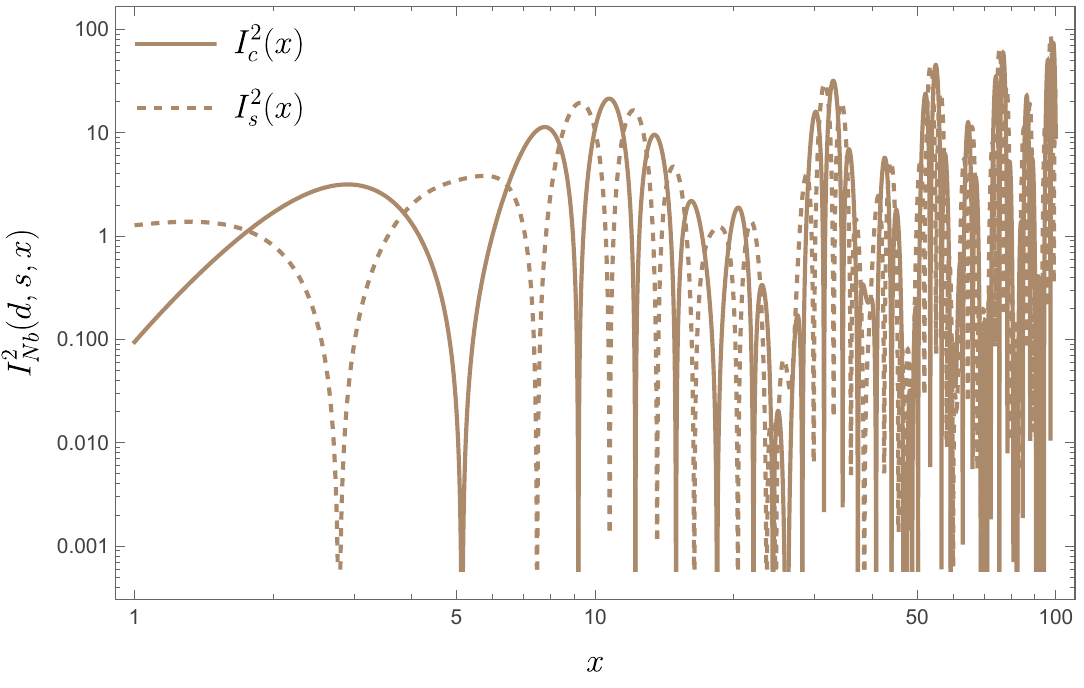}
\caption{Kernel squares $I_c^{2}(x)$ (solid) and $I_s^{2}(x)$ (dashed) in the N--body gauge at $d=0$, $s=1/\sqrt{3}$.
Near horizon crossing the response is oscillatory; for $x\gg1$ the evolution decays as $x^{-2}$ (i.e.\ $I_{c/s}\sim x^{-1}$), and the late--time saturation relevant for the tensor power matches the longitudinal case.}
\label{fig:NB-kernels}
\end{minipage}
\hfill
\begin{minipage}[t]{0.48\linewidth}
\centering
\includegraphics[width=\linewidth]{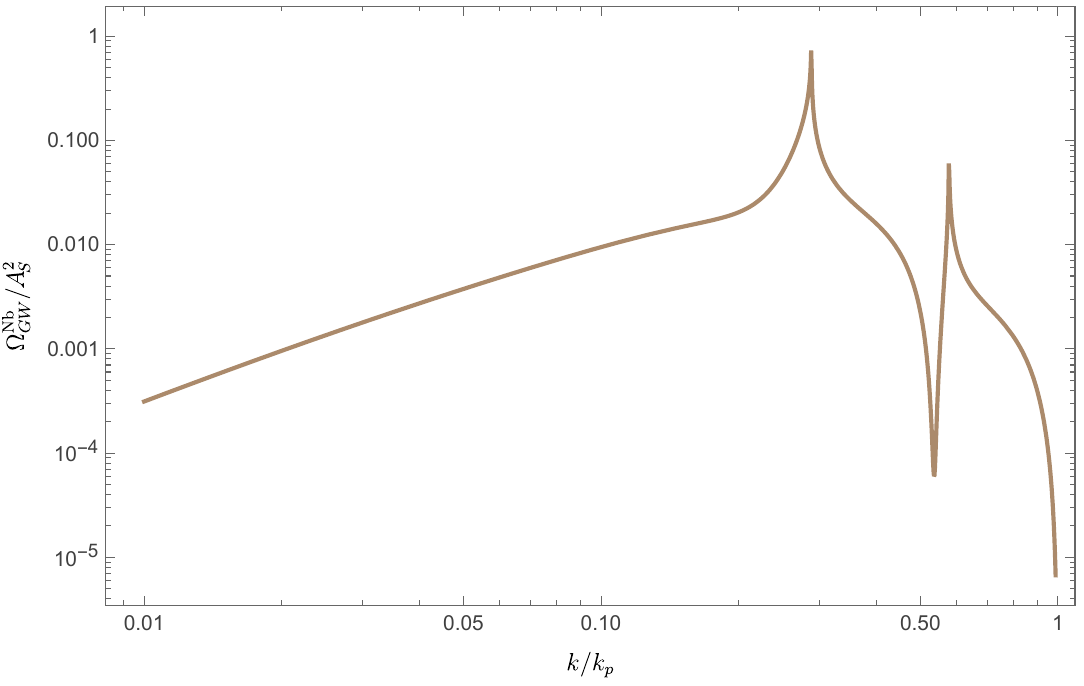}
\caption{
$\Omega_{\rm GW}(k)/A_s^2$ in the N--body gauge for a Dirac--delta isocurvature peak at $k_p$.
The spectrum shows the IR rise $\propto k^{2}\ln^{2}k$ for $k\ll k_p$, a peak at $k=2c_s k_p$ with $c_s=1/\sqrt{3}$, and a sharp cutoff at $k=2k_p$.
At late times the evolution is constant (convergent) and matches the longitudinal result within accuracy.
The GW spectrum is normalized by the scalar power spectrum amplitude $A_s^2$.
}
\label{fig:NB-spectrum}
\end{minipage} 
\end{figure*}  

Substituting the N--body source \eqref{eq:fNb} into the kernel definitions \eqref{ICS} yields
exact analytic expressions for $I_c(d,s,x)$ and $I_s(d,s,x)$; however, the resulting  
formulas span several pages and are not reproduced here.\footnote{Exact $I_{c/s}(d,s,x)$ expressions are available in exact analytic expression but are too long to typeset. The full formulas \textit{Mathematica} notebooks are provided in the paper’s repository (see ancillary files or the project URL).} For large $x$, the kernels are decaying as : $I_{c/s}\sim x^{-1}$ (hence $I_{c/s}^2\sim x^{-2}$). 
Consequently $\overline{I^2}$ tends to a constant and the induced energy density  
$\Omega_{\rm GW}(k)$ becomes time independent at late times, in agreement with the summary  
given in Sec.~\ref{summary}. The evolution of the second-order source term is shown in Fig. \ref{fig:NB-source}. The evolution of the corresponding Kernels is shown in Fig.\ref{fig:NB-kernels}.  

For a Dirac–delta isocurvature peak, 
$\mathcal{P}_{S} (k)=\mathcal{A}_{S}\,\delta\!\big(\ln(k/k_{p})\big)$,
we evaluate the spectrum along the delta line in $(d,s)$,
$d=0$ and $s=\tfrac{2}{\sqrt{3}}(k_{p}/k)$, using
Eqs.~\eqref{eq:OmegaGW-dsg2} and \eqref{eq:Ibar2} with the present gauge’s
$I_{c,\infty}(d,s)$ and $I_{s,\infty}(d,s)$.
The resulting $\Omega_{\rm GW}(k)$ (Fig.~\ref{fig:NB-spectrum}) shows the standard
$k^{2}\ln^{2}\!k$ infrared rise, a peak at $k=2c_s k_p$ with $c_s=1/\sqrt{3}$,
and a sharp cutoff at $k=2k_p$. Its evolution is constant in time, matching the late–time
behavior inferred from the kernel scaling.

As elsewhere, we consider modes that reenter during radiation domination and insert the
gauge–specific source into the kernel integrals \eqref{ICS}. The analytic expressions for
$I_c^{(\text{Nb})}(d,s,x)$ and $I_s^{(\text{Nb})}(d,s,x)$ are lengthy; for practical use we
employ their late–time forms $I_{c,\infty}^{(\text{Nb})}(d,s)$ and
$I_{s,\infty}^{(\text{Nb})}(d,s)$ in Eq.~\eqref{eq:OmegaGW-dsg2}. When the analytic expressions extend
over several pages, we omit them from the main text; the figures are based on direct evaluation
of the exact expressions. The procedure applies to any ${\cal P}_S(k)$ provided that the relevant
modes reenter well before the equality.
    
\section{Gauge-independent energy density of SIGWs in isocurvature perturbations}
\label{sec:gauge-independent}

\begin{table*}[t]
\centering
\renewcommand{\arraystretch}{1.35}
\begin{tabular}{|c|c|c|c|p{0.15\textwidth}|}
\hline
\textbf{Gauge} & \textbf{Asymptotic form} &
\textbf{Late-time  } & \textbf{Radiative (luminal)} & \textbf{Remarks} \\
\textbf{} & \textbf{of kernel $I(x)$} &
\textbf{dependence of $\Omega_\text{GW}(x)$} & \textbf{contribution} & \textbf{} \\ 
\hline
Long.   & $\sim x^{-1}$ & $\propto$ (const.) & convergent; standard     & baseline \\
CO      &  $\sim x^{2}$   & $\propto \eta^{6}$            & convergent             & non-radiative pieces removed \\
TT      & $\sim x^{3}$   & $\propto \eta^{8}$            & convergent             & same \\
TM      &  $\sim x$       & $\propto \eta^{4}$            & convergent             & same \\
UC      &  $\sim x$       & $\propto \eta^{4}$            & convergent             & same \\
UD      & $\sim x^{0}$         & $\propto \eta^{2}$            & convergent             & same \\
UE      & $\sim x^{-1}$ & $\propto$ (const.) & convergent; $=$ Long.    & via transform \\
Nm      & $\sim x^{-1}$ & $\propto$ (const.) & convergent; $=$ Long.    & identical to Long.\\ 
N--body & $\sim x^{-1}$ & $\propto$ (const.) & convergent               & standard $x^{-1}$ tail \\ 
\hline 
\end{tabular}\caption{ 
Late–time behavior of the energy density spectra,   $\Omega_\text{GW}(x)\propto x^2 I^2(x)$ as a function of the dimensionless time variable $x\equiv k\eta$, and the corresponding gravitational–wave energy density   
$\Omega_\text{GW}$ during radiation domination, shown before isolating the    
oscillatory (radiative) component. The “Radiative (luminal)” column corresponds to retaining only the  
free–GW pieces $\{\sin x,\cos x\}$ of the tensor solution; this yields the physical,   
gauge–independent spectrum with kernels scaling as $x^{-1}$.} 
\label{tab:gauge_summary}
\end{table*}

\begin{figure*}[t]
\centering
\begin{minipage}[t]{0.48\linewidth}
\centering
\includegraphics[width=\linewidth]{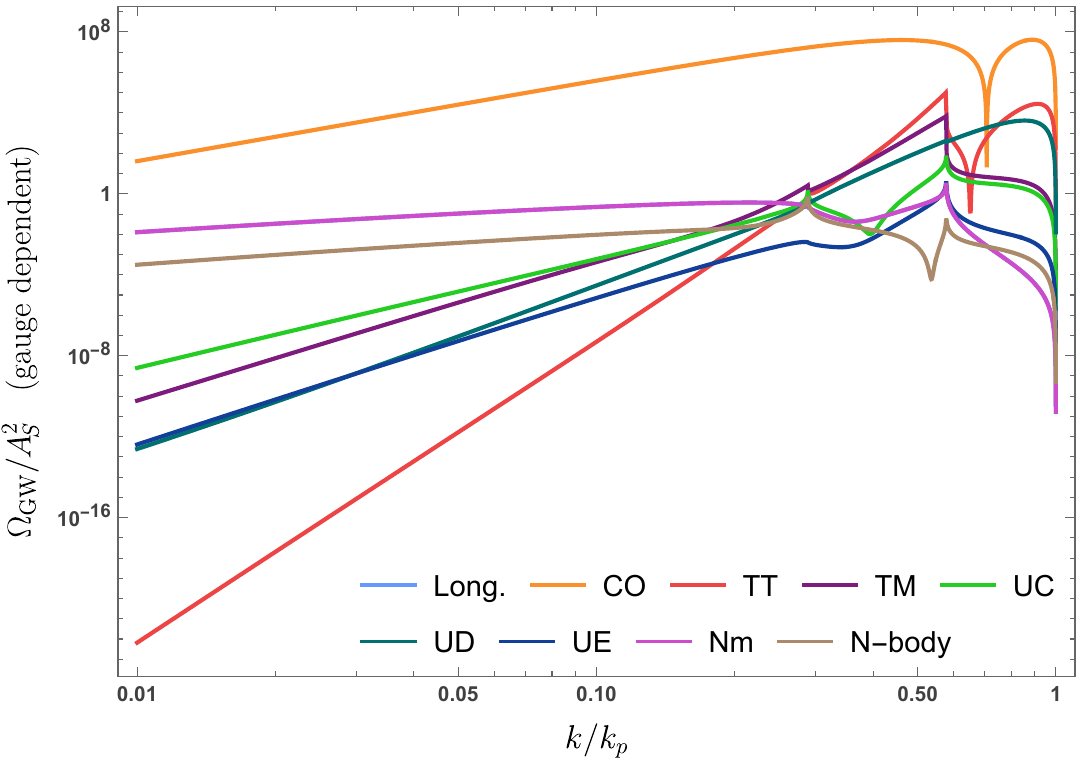} 
\caption{Gauge-dependent (full source).
Scalar–induced GW spectra $\Omega_\text{GW}(k)$ from isocurvature perturbations in nine gauges
(Long., CO, TT, TM, UC, UD, UE, Nm, N–body).  Differences reflect non-radiative
pieces in the kernel that are slicing artifacts at intermediate times.}
\label{fig:gauge_dep_vs_indep01}
\end{minipage}\hfill
\begin{minipage}[t]{0.48\linewidth}
\centering
\includegraphics[width=\linewidth]{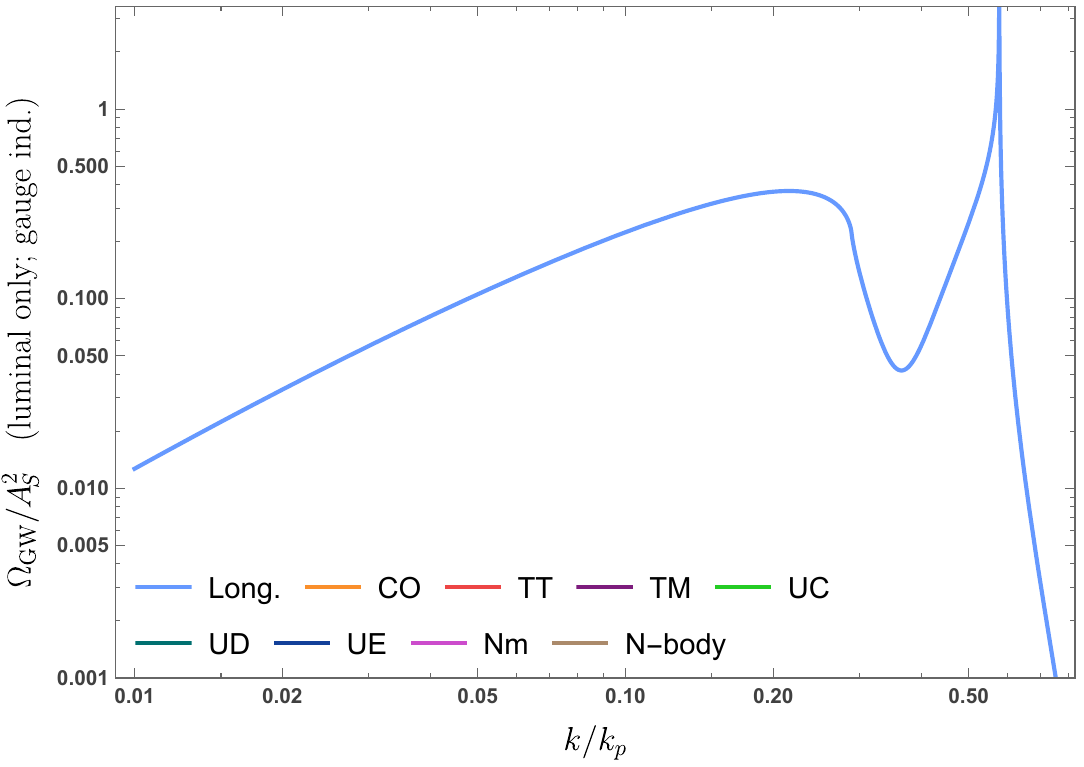}
\caption{Gauge-independent (free GWs or radiative only).
Spectra obtained by keeping the free $\sin x/\cos x$ terms and taking $x\to\infty$. 
All gauges coincide with the longitudinal result; the surviving kernel has the
standard $x^{-1}$ tail, yielding a finite late-time spectrum.  Here $\kappa=1$ for
display; the shape is insensitive for $\kappa\gg1$.} 
\label{fig:gauge_dep_vs_indep02}
\end{minipage}
\end{figure*}

In this section, we first compare the analytic results of the energy spectra computed in all nine gauges. We then isolate the physically relevant propagating, gauge-invariant modes from these spectra by projecting out the unphysical pure-gauge contributions, thereby demonstrating that the SIGWs are gauge invariant.    

At late times during radiation domination, the tensor modes propagate as a superposition of $\sin x$ and $\cos x$, with $x \equiv k \eta$. To isolate the physically relevant, propagating components, we adopt a minimal \emph{radiative projection}, retaining only the $\sin x$ and $\cos x$ terms in Eq.~\eqref{ICS} and discarding the non-radiative contributions whose phases depend on the slicing. These non-luminal terms either average to zero or are power-suppressed as $x \to \infty$. With this selection, the late-time evolution becomes universal across all gauges.

It is convenient to quantify the late-time power of the induced tensor modes by the average
\begin{equation}
\overline{I^2(d,s)}\Big|_{x\to\infty}
  \equiv \tfrac12\!\left(I_{c,\infty}^2(d,s)+I_{s,\infty}^2(d,s)\right),
\end{equation}
constructed from the cosine and sine transfer integrals in Eq.~\eqref{ICS}. After the radiative projection, the surviving kernel exhibits the standard $x^{-1}$ tail, ensuring that the induced fractional energy density remains finite and \emph{gauge independent} at late times.

Several approaches have been proposed in the literature to obtain late-time, gauge-independent SIGW spectra. Early work removed spurious contributions via explicit gauge transformations \cite{Ali:2020sfw}, while later studies refined this procedure systematically \cite{Ali:2023moi}. More recent analyses phrase the selection in terms of the luminal part of the convolution domain (e.g., imposing $u+v=1$),\footnote{In our $(d,s)$ variables, this condition maps to $d \pm s = 1$. In practice, we implement the same physical selection by retaining only the free gravitational-wave pieces, $\sin x$ and $\cos x$, in Eq.~\eqref{ICS}.} which is equivalent to $d \pm s = 1$ \cite{Jiang:2025ysb, Xue:2025yii}. In this work, we implement this principle directly at the level of the Green-function solution by keeping only the $\sin x$ and $\cos x$ terms and discarding non-radiative contributions tied to the slicing. The resulting late-time spectrum is therefore time-independent and, for all gauges considered, coincides with the longitudinal benchmark.

Under the Gaussian isocurvature assumption, the observable energy density is given by Eq.~\eqref{eq:OmegaGW-dsg2}, with the kernel determined by the Green-function solution in Eq.~\eqref{eq:kernel-rad2} and its decomposition in Eq.~\eqref{ICS}. The late-time average is defined in Eq.~\eqref{eq:Ibar2}. As discussed in Sec.~\ref{sec3}, the raw kernels display markedly different time behaviors across gauges before projection: growth in CO/TT/UC/UD gauges, mild drift in TM, near-convergent in UE, and convergent in Long./Nm/Nb. After applying the radiative projection, however, all nine gauges yield the same spectrum as the longitudinal benchmark. Notably, the Nm gauge is already aligned even before projection, while UE and Nb are nearly so.

In practice, we compute $I_{c,\infty}(d,s)$ and $I_{s,\infty}(d,s)$ from Eq.~\eqref{ICS}, retain only the terms multiplying $\sin x$ and $\cos x$, and then use 
\begin{equation}
\overline{I^2} = \tfrac12 \left(I_{c,\infty}^2 + I_{s,\infty}^2\right), 
\end{equation}
in Eq.~\eqref{eq:OmegaGW-dsg2}. This procedure removes gauge-dependent, non-radiative contributions and isolates the unique free-wave content that determines the observable $\Omega_{\rm GW}$. The comparison between the full-source (gauge-dependent) and radiative-only (gauge-independent) spectra is shown in Figs.~\ref{fig:gauge_dep_vs_indep01} and \ref{fig:gauge_dep_vs_indep02}. With this radiative projection, all nine gauges reproduce the longitudinal benchmark; the resulting agreement is shown in 
Fig.~\ref{fig:gauge_dep_vs_indep02}. 
The nine gauges used in our analysis and their late-time behavior of the SIGW energy density are summarized in Table \ref{tab:gauge_summary}, which provides a compact comparison of the gauge classifications discussed above. 
For the Dirac–delta isocurvature peak used in our illustrations, the spectrum
exhibits the familiar IR rise $\propto k^{2}\ln^{2}k$, a peak near $2c_{s}k_{p}$, and a sharp cutoff at $2k_{p}$.
Within RD, the radiative kernels decay as $x^{-1}$ and the induced secondary gravitational waves behave as radiation, yielding a finite, gauge–independent $\Omega_{\rm GW}(k)$.

\section{Discussion and Conclusion}\label{summary} 

In this work we analyzed scalar– induced secondary gravitational waves sourced by primordial
isocurvature perturbations across nine gauges during radiation domination viz. longitudinal (Poisson), 
CO, synchronous/TT, TM, UC,
UD, UE, Nm, and N–body.
We used the $(d,s)$ parametrization of the convolution domain and obtained analytic transfer
kernels $I_{c/s}(d,s,x)$ in each slicing (lengthy formulas are evaluated directly for the figures).
For every gauge we tracked the late–time evolution of the source, the kernels, and the energy–density 
spectrum; the longitudinal result serves as our benchmark.

As detailed in Sec.~\ref{sec.2}, each gauge fixes a slicing (time) and a threading (space) through its
conditions on $(\phi,\psi,B,E)$ and the matter variables. These choices control whether non–radiative 
pieces of the scalar source survive inside the kernel and therefore set the evolution seen at late times
during RD, prior to isolating the free GW. In longitudinal ($B=E=0$; no scalar shear) the non–radiative 
pieces cancel efficiently and the kernel decays as $x^{-1}$, so $\Omega_{\rm GW}$ tends to a constant. 
In CO ($\delta V=0$, $B=0$; comoving, orthogonal slicing) no condition is imposed on the scalar shear $E$, so a non-oscillatory piece of the source survives; the kernel grows $\sim x^{2}$ and hence $\Omega_{\rm GW}\propto x^{6}$.
In TT ($\phi=0$, $B=0$; synchronous slicing) the absence of lapse and shift removes cancellations of non-radiative terms; the time integral then accumulates as $x^{3}$, yielding $\Omega_{\rm GW}\propto x^{8}$. 
In TM ($\delta V=0$, $E=0$; comoving, shear–free threading) and UC ($\psi=0$, $E=0$; flat slices, shear–free) 
the residual lapse/curvature choice leaves a weaker late-time increase, with kernels $\sim x$ and 
$\Omega_{\rm GW}\propto x^{4}$. In UD ($\delta\rho=0$, $E=0$; constant–density slices, shear–free) the kernel 
saturates to $x^{0}$ so that $\Omega_{\rm GW}\propto x^{2}$. By contrast, UE ($E=0$, 
$\nabla^{2}\sigma=3(\mathcal H\phi+\psi')$) enforces uniform expansion and removes the offending pieces, yielding
the same $x^{-1}$ tail as longitudinal. The Nm conditions ($B=0$, $E''=-\mathcal H E'$) supply no intrinsic
oscillatory transfer, and the spectrum overlays the longitudinal curve. The N–body choice
($\psi=\tfrac{1}{3}\nabla^{2}E$, $\delta V=-B$) aligns with $N$–body practice and produces a damped 
evolution with the standard $x^{-1}$ fall–off. 

These gauge-dependent trends were illustrated graphically and tabulated in Sec.~\ref{sec:gauge-independent}. By projecting out the unphysical pure-gauge modes, we retained only the freely propagating tensor radiation, characterized by the oscillatory $\{\sin x, \cos x\}$ components of the Green-function solution, and discarded the non-luminal phases that did not propagate with the gravitational waves. This procedure removed gauge-dependent terms and yielded a universal late-time kernel with an $x^{-1}$ decay. Consequently, the resulting fractional energy density, $\Omega_{\rm GW}(k)$, was finite and fully gauge independent.

Future directions include extending the present analysis beyond a Dirac–delta isocurvature peak to finite-width and multi-feature spectra, as well as incorporating the radiation-to-matter transition within the same $(d,s)$ framework. These extensions will enable more realistic predictions of scalar-induced gravitational waves and facilitate direct comparison with pulsar timing array (PTA) and space-based detector sensitivities. 

A complementary direction concerns black–hole physics
and nonrelativistic gravity. By examining how gauge choices affect the interpretation of energy density spectra, we will assess implications for the semi-classical break-down of black holes and potential  perturbations in evaporation rates. This endeavor will broaden our  understanding of the evolution of black holes and contribute to  foundational theories in quantum  gravity. 

Among various gauge choices, a special one that has attracted considerable interest in the study of non-relativistic gravity is the pre-Newtonian gauge \cite{VandenBleeken:2017rij, Hansen:2019pkl, Hansen:2020pqs}. In this gauge, the $1/c$ expansion of general relativity can be implemented in a cascade structure, with
each order giving rise to a torsional Newton–Cartan gravity theory. A remarkable feature of these Newton–Cartan theories is their consistency with three classical tests of general relativity
\cite{Hansen:2019vqf}: namely, the perihelion precession, the deflection of light, and the gravitational redshift. This suggests an indistinguishability, at this level, between general relativity and the Newton–Cartan type gravities obtained through such an expansion. However, gravitational waves may reveal distinguishing features between the two theories, offering a potential test for the applicability of torsional Newton–Cartan gravity in gravitational wave physics. We therefore aim to revisit our computations in the pre-Newtonian gauge, where both the $1/c$ expansion and Newton–Cartan gravity can be consistently applied.

\begin{acknowledgments}  
We are grateful to the anonymous referees for their constructive comments and suggestions, which helped improve the clarity of the paper.
A. A. is supported by China Postdoctoral Science Foundation grant   number  2023M742547.  Y. L. is  supported by a Project  Funded by the Priority  Academic Program  Development of Jiangsu  Higher Education  Institutions (PAPD) and by  National Natural Science Foundation of China (NSFC) No.12305081 and the international collaboration grant between NSFC and Royal Society No.W2421035. M. S. is supported in part by the National Natural Science Foundation of China (Grant No. 12475105).    
\end{acknowledgments}

%

\end{document}